\documentclass{article}

\usepackage[preprint, nonatbib]{neurips_2020}

\usepackage{titlesec}
\usepackage[utf8]{inputenc} 
\usepackage[T1]{fontenc}    
\usepackage[colorlinks=true, linktoc=all]{hyperref} 
\usepackage{url}            
\usepackage{booktabs}       
\usepackage{amsfonts}       
\usepackage{nicefrac}       
\usepackage{microtype}      

\usepackage[dvipsnames]{xcolor}

\usepackage{graphicx}

\providecommand{\tightlist}{%
  \setlength{\itemsep}{0pt}\setlength{\parskip}{0pt}}
 
\usepackage[
backend=biber,
style=alphabetic,
]{biblatex}
\hypersetup{
            linkcolor=blue,
            urlcolor=blue,
            pdfinfo={
            Title={Is Power-Seeking AI an Existential Risk?},
            Author={Joseph Carlsmith},
            Subject={artificial intelligence, existential risk},
            Keywords={x-risk, existential risk, ai risk, ai safety, ml safety, ai ethics, ai alignment, alignment, control problem, machine ethics, power-seeking, power, superintelligence, AGI, strong AI}
            }
}

\title{Is Power-Seeking AI an Existential Risk?}

\author{%
  Joseph Carlsmith\\
  Open Philanthropy\\
  April 2021\\
  \\
  \href{https://jc.gatspress.com/pdf/existential_risk_and_powerseeking_ai.pdf}{Shorter version} | \href{https://forum.effectivealtruism.org/posts/ChuABPEXmRumcJY57/video-and-transcript-of-presentation-on-existential-risk}{Video presentation} | \href{https://docs.google.com/presentation/d/1UE_cAsogrK5i9wvF3YMIZX-iO9qzjevnrYfTxlKL7ns/}{Slides} | \href{https://joecarlsmithaudio.buzzsprout.com/2034731/12113681-is-power-seeking-ai-an-existential-risk}{Audio version} | \href{https://www.lesswrong.com/posts/qRSgHLb8yLXzDg4nf/reviews-of-is-power-seeking-ai-an-existential-risk}{Reviews (including superforecasters)}\\
}

\begin{document}

\maketitle

\setcounter{secnumdepth}{4} 


\begin{abstract}
  This report examines what I see as the core argument for concern about existential risk from misaligned artificial intelligence. I proceed in two stages. First, I lay out a backdrop picture that informs such concern. On this picture, intelligent agency is an extremely powerful force, and creating agents much more intelligent than us is playing with fire – especially given that if their objectives are problematic, such agents would plausibly have instrumental incentives to seek power over humans. Second, I formulate and evaluate a more specific six-premise argument that creating agents of this kind will lead to existential catastrophe by 2070. On this argument, by 2070: (1) it will become possible and financially feasible to build relevantly powerful and agentic AI systems; (2) there will be strong incentives to do so; (3) it will be much harder to build aligned (and relevantly powerful/agentic) AI systems than to build misaligned (and relevantly powerful/agentic) AI systems that are still superficially attractive to deploy; (4) some such misaligned systems will seek power over humans in high-impact ways; (5) this problem will scale to the full disempowerment of humanity; and (6) such disempowerment will constitute an existential catastrophe. I assign rough subjective credences to the premises in this argument, and I end up with an overall estimate of \textasciitilde{}5\% that an existential catastrophe of this kind will occur by 2070. \emph{(May 2022 update: since making this report public in April 2021, my estimate here has gone up, and is now at >10\%.)}
\end{abstract}

\vspace{-1.3cm} 
\hypersetup{linkcolor=black} 
\tableofcontents


\section{Introduction}\label{introduction}

Some worry that the development of advanced artificial intelligence will result in existential catastrophe---that is, the destruction of humanity's longterm potential.\footnote{See e.g. \href{https://intelligence.org/files/AIPosNegFactor.pdf}{Yudkowsky (2008)}, \href{https://www.amazon.com/Superintelligence-Dangers-Strategies-Nick-Bostrom/dp/1501227742}{Bostrom (2014)}, \href{https://www.bbc.com/news/technology-30290540}{Hawking (2014)}, \href{https://www.amazon.com/Life-3-0-Being-Artificial-Intelligence/dp/1101946598}{Tegmark (2017)}, \href{https://www.alignmentforum.org/posts/HBxe6wdjxK239zajf/what-failure-looks-like}{Christiano (2019)}, \href{https://www.amazon.com/Human-Compatible-Artificial-Intelligence-Problem/dp/0525558616/ref=tmm_hrd_swatch_0?_encoding=UTF8\&qid=1619197644\&sr=1-1}{Russell (2019)}, \href{https://www.amazon.com/Precipice-Existential-Risk-Future-Humanity/dp/031648492X/ref=sr_1_2?crid=2ZWCCI74ZFX55\&dchild=1\&keywords=precipice+existential+risk+and+the+future+of+humanity\&qid=1619197698\&s=books\&sprefix=precipice\%2Cstripbooks\%2C243\&sr=1-2}{Ord (2020)}, and \href{https://www.alignmentforum.org/posts/8xRSjC76HasLnMGSf/agi-safety-from-first-principles-introduction}{Ngo (2020)}. This definition of ``existential catastrophe'' is from \href{https://www.amazon.com/Precipice-Existential-Risk-Future-Humanity/dp/031648492X/ref=sr_1_2?crid=2ZWCCI74ZFX55\&dchild=1\&keywords=precipice+existential+risk+and+the+future+of+humanity\&qid=1619197698\&s=books\&sprefix=precipice\%2Cstripbooks\%2C243\&sr=1-2}{Ord (2020, p.~27)}; see \protect\hyperref[catastrophe]{section 7} for a bit more discussion.} Here I examine the following version of this worry (it's not the only version):

By 2070:

\begin{enumerate}
\def\labelenumi{\arabic{enumi}.}
\item
  It will become possible and financially feasible to build AI systems with the following properties:

  \begin{itemize}
  \tightlist
  \item
    \emph{Advanced capability}: they outperform the best humans on some set of tasks which when performed at advanced levels grant significant power in today's world (tasks like scientific research, business/military/political strategy, engineering, and persuasion/manipulation).
  \item
    \textit{Agentic planning}: they make and execute plans, in pursuit of objectives, on the basis of models of the world.
  \item
    \emph{Strategic awareness:} the models they use in making plans represent with reasonable accuracy the causal upshot of gaining and maintaining power over humans and the real-world environment.
  \end{itemize}

  (Call these ``APS''---Advanced, Planning, Strategically aware---systems.)
\item
  There will be strong incentives to build and deploy APS systems \textbar{} (1).
\item
  It will be much harder to build APS systems that would not seek to gain and maintain power in unintended ways (because of problems with their objectives) on any of the inputs they'd encounter if deployed, than to build APS systems that would do this, but which are at least superficially attractive to deploy anyway \textbar{} (1)--(2).
\item
  Some deployed APS systems will be exposed to inputs where they seek power in unintended and high-impact ways (say, collectively causing \textgreater{}\$1 trillion dollars of damage), because of problems with their objectives \textbar{} (1)--(3).\footnote{Let's assume 2021 dollars.}
\item
  Some of this power-seeking will scale (in aggregate) to the point of permanently disempowering \textasciitilde{}all of humanity \textbar{} (1)--(4).
\item
  This disempowerment will constitute an existential catastrophe \textbar{} (1)--(5).
\end{enumerate}

These claims are extremely important if true. My aim is to investigate them. I focus on (2)--(5), but I also say a few words about (1) and (6).

My current view is that there is a disturbingly substantive chance that a scenario along these lines occurs, and that many people alive today---including myself---live to see humanity permanently disempowered by AI systems we've lost control over. In the \protect\hyperref[probabilities]{final section}, I take an initial stab at quantifying this risk, by assigning rough probabilities to 1-6. \textbf{My current, highly-unstable, subjective estimate is that there is a \textasciitilde{}5\% percent chance of existential catastrophe by 2070 from scenarios in which (1)--(6) are true.} \emph{(May 2022 author's note: since making this report public in April 2021, my estimate here has gone up; it's currently at >10\%.)} My main hope, though, is not to push for a specific number, but rather to lay out the arguments in a way that can facilitate productive debate.


\subsection{Preliminaries}\label{preliminaries}

Some preliminaries and caveats (those eager for the main content can skip):

\begin{itemize}
\item
  I'm focused, here, on a very specific type of worry. There are lots of other ways to be worried about AI---and even, about existential catastrophes resulting from AI. And there are lots of ways to be excited about AI, too.
\item
  My emphasis and approach differs from that of others in the literature in various ways.\footnote{Though the differences listed here don't each apply to all of the literature; and in many respects, the issues I discuss are the ``classic issues.'' Of existing overall analyses, mine is probably most similar to \href{https://www.alignmentforum.org/posts/8xRSjC76HasLnMGSf/agi-safety-from-first-principles-introduction}{Ngo (2020)}. For other discussions of alignment problems in AI (on top of those cited in footnote 1), see e.g. \href{https://arxiv.org/abs/1606.06565}{Amodei et al (2016)}, \href{https://www.vox.com/future-perfect/2018/12/21/18126576/ai-artificial-intelligence-machine-learning-safety-alignment}{Piper (2018, updated 2020)}, and \href{https://www.amazon.com/Alignment-Problem-Machine-Learning-Values/dp/B085DTXC59/ref=sr_1_1?dchild=1\&keywords=alignment+problem\&qid=1619198396\&s=books\&sr=1-1}{Christian (2020)}.} In particular: I'm less focused than some on the possibility of an extremely rapid escalation in frontier AI capabilities, on ``recursive self-improvement,'' or on scenarios in which one actor comes to dominate the world; I'm focusing on power-seeking in particular (as opposed to ``misalignment'' more broadly); I'm aiming to incorporate and respond to various objections and re-framings introduced since early work on the topic, and since machine learning has become a prominent AI paradigm;\footnote{For example, those of \href{https://www.alignmentforum.org/posts/NxF5G6CJiof6cemTw/coherence-arguments-do-not-imply-goal-directed-behavior}{Shah (2018)}, \href{https://www.amazon.com/Enlightenment-Now-Science-Humanism-Progress/dp/0525427570}{Pinker (2018)}, \href{https://www.fhi.ox.ac.uk/wp-content/uploads/Reframing_Superintelligence_FHI-TR-2019-1.1-1.pdf}{Drexler (2019)}, \href{https://blogs.scientificamerican.com/observations/dont-fear-the-terminator/}{Zador and LeCun (2019)}, \href{https://arxiv.org/abs/1906.01820}{Hubinger et al (2019)}, \href{https://80000hours.org/podcast/episodes/ben-garfinkel-classic-ai-risk-arguments/}{Garfinkel (2020)}.} I'm aiming to avoid reliance on models of ``utility function maximization'' and related concepts; and I present and assign probabilities to premises in a complete argument for catastrophe. That said, I still think of this as an articulation and analysis of a certain kind of ``core argument''---one that has animated, and continues to animate, much concern about existential risk from misaligned AI.\footnote{Some---e.g. \href{https://fragile-credences.github.io/prioritising-ai/}{Adamczewski (2019)}, \href{https://ea.greaterwrong.com/posts/9sBAW3qKppnoG3QPq/ben-garfinkel-how-sure-are-we-about-this-ai-stuff}{Garfinkel (2019)}, and \href{https://www.greaterwrong.com/posts/JbcWQCxKWn3y49bNB/disentangling-arguments-for-the-importance-of-ai-safety}{Ngo (2019)}, though Ngo notes that his views have since shifted---suggest that arguments for X-risk from misaligned AI have ``shifted'' or ``drifted'' over time. I see some of this in some places, but for me at least (and for various others I know), the basic thrust---e.g., advanced artificial agents would be very powerful, and if their objectives are problematic, they might have default incentives to disempower humans---has been pretty constant (sometimes claims about discontinuous/concentrated ``takeoff'' are treated as essential to the basic thrust of the argument, but I don't see them that way---see \protect\hyperref[correction]{section 6} for discussion).}
\item
  I'm focusing on 2070 because I want to keep vividly in mind that I and many other readers (and/or their children) should expect to live to see the claims at stake here falsified or confirmed. That said, the main arguments don't actually require the development of relevant systems within any particular period of time (though timelines in this respect can matter to e.g.~the amount of evidence that present-day systems and conditions provide about future risks).
\item
  I'm not addressing the question of what sorts of interventions are currently available for lowering the risk in question, or how they compare with work on other issues.
\item
  \protect\hyperref[probabilities]{Section 8} involves assigning (a) subjective probabilities to (b) imprecisely operationalized claims, in the context of (c) a multi-step argument. I discuss cautions on all of three of these fronts there (and I include an appendix aimed at warding off some possible problems with (c) in particular).
\item
  Thinking about what will happen decades in the future is famously hard, and some approach \textasciitilde{}any such thought with extreme skepticism. I'm sympathetic to this in ways, but I also think we can do better than brute agnosticism; and to the extent that we \emph{care} about the future, and can try to influence it, our actions will often express ``bets'' about it even absent explicit forecasts.
\item
  I use various concepts it would be great to be more precise about. This is true of many worthwhile discussions, but that doesn't make the attending imprecisions innocuous. To the contrary, I encourage wariness of ways they might mislead.
\item
  The discussion focuses on how to avoid human disempowerment, but I am not assuming that it is always appropriate or good for humans to keep power. Rather, I'm assuming that we want to avoid disempowerment that is \emph{imposed} on us by artificial systems we've lost control over. See \protect\hyperref[catastrophe]{section 7} for more discussion.
\item
  Sufficiently sophisticated AI systems might warrant moral concern.\footnote{See \href{https://www.openphilanthropy.org/2017-report-consciousness-and-moral-patienthood}{Muehlhauser (2017)} for more on what sorts of non-humans warrant this concern.} In my opinion, this fact should motivate grave ethical caution in the context of many types of AI development, including many discussed in this report. However, it's not my focus here (see \protect\hyperref[catastrophe]{section 7} for a few remarks).
\end{itemize}

\subsection{Backdrop}\label{backdrop}

The specific arguments I'll discuss emerge from a broader backdrop picture, which I'll gloss as:

\begin{enumerate}
\def\labelenumi{\arabic{enumi}.}
\tightlist
\item
  Intelligent agency is an extremely powerful force for controlling and transforming the world.
\item
  Building agents much more intelligent than humans is playing with fire.
\end{enumerate}

I'll start by briefly describing this picture, as I think it sets an important stage for the discussion that follows.

\subsubsection{Intelligence}\label{intelligence}

Of all the species that have lived on earth, humans are clearly strange. In particular, we exert an unprecedented scale and sophistication of intentional control over our environment. Consider, for example: the city of \href{https://en.wikipedia.org/wiki/Tokyo}{Tokyo}, the \href{https://en.wikipedia.org/wiki/Large_Hadron_Collider\#/media/File:Views_of_the_LHC_tunnel_sector_3-4,_tirage_2.jpg}{Large Hadron Collider}, and the \href{https://en.wikipedia.org/wiki/Bingham_Canyon_Mine\#/media/File:2019_Bingham_Canyon_Mine_04.jpg}{Bingham Canyon Mine}.

What makes this possible? Something about our minds seems centrally important. We can plan, learn, communicate, deduce, remember, explain, imagine, experiment, and cooperate in ways that other species can't. These cognitive abilities---employed in the context of the culture and technology we inherit and create---give us the power, collectively, to transform the world. For ease of reference, let's call this loose cluster of abilities ``intelligence,'' though very little will rest on the term.\footnote{In particular, when I say ``more intelligent than humans,'' I just mean ``better at things like planning, learning, communicating, deducing, remembering, etc'' than humans. I don't have in mind some broader notion like ``optimization power,'' ``solving problems,'' or ``achieving objectives.'' Thanks to Ben Garfinkel for discussion here. Obviously, intelligence in this sense is highly multi-dimensional. I don't think this is a problem for saying one thing is more intelligent than another (e.g., which house is better to live in is multi-dimensional, but a mansion can still be preferable to a shack); but if you object to the term for this or some other reason, feel free to substitute ``cognitive abilities like planning, learning, communicating, deducing, remembering, etc'' whenever I say ``intelligence'' (I won't use the term very often). See \href{https://arxiv.org/abs/1703.10987}{Garfinkel et al (2017)} for more on objections to the possibility of greater-than-human intelligence.}

Our abilities in these respects are nowhere near any sort of hard limit. Human cognition---even in groups, and with the assistance of technology---depends centrally on the human brain, which, for all its wonders, is an extremely specific and limited organ, subject to very specific constraints---on cell count, energy, communication speed, signaling frequency, memory capacity, component reliability, input/output bandwidth, and so forth.\footnote{For example: the brain has \href{https://www.openphilanthropy.org/brain-computation-report\#SpikesThroughSynapsesPerSecond}{\textasciitilde{}1e11 neurons and 1e14-1e15 synapses}; neurons fire at a maximum of \href{https://www.openphilanthropy.org/brain-computation-report\#SpikesThroughSynapsesPerSecond}{some hundreds of Hz}; action potentials travel at a max of some \href{https://www.khanacademy.org/test-prep/mcat/organ-systems/neuron-membrane-potentials/a/action-potential-velocity}{hundreds of m/s}; the brain runs on \href{https://www.openphilanthropy.org/brain-computation-report\#OverallBitErasures}{\textasciitilde{}20W of power}; it has to fit within the skull; and so forth.}

There are possible cognitive systems---possible brains, and possible artificial systems---to which some or all these constraints do not apply. It seems very likely that such systems could learn, communicate, reason, problem-solve, etc, much better than humans can. The variation in cognitive ability we see among humans, and across species, also suggests this; as do existing successes in approaching or exceeding human capabilities at particular tasks---mathematical calculation, game-playing, image recognition, and so forth---using artificial systems.

\subsubsection{Agency}\label{agency}

Humans can also be \emph{agentic}---that is (loosely), we pursue objectives, guided by models of the world (see \protect\hyperref[agentic-planning]{section 2.1.2} for more). You want to travel to New York, so you buy a flight, double-check when it leaves, wake up early to pack, and so forth.

Non-human cognitive systems can be agentic, too. And depending on what you count as an agent, humans can already create novel non-human agents.\footnote{I'll count agent-like systems constituted in central part by human agents---e.g., nations, corporations, parent teacher organizations, etc---as ``human'' (which need not imply: innocuous).} We breed and \href{https://en.wikipedia.org/wiki/Genetically_modified_animal}{genetically modify} non-humans animals, for example; and some of our artificial systems display agent-like behavior in specific environments (see e.g. \href{https://deepmind.com/blog/article/alphastar-mastering-real-time-strategy-game-starcraft-ii}{here} and \href{https://openai.com/projects/five/}{here}). But these agents can't learn, plan, communicate, etc. like we can.

At some point, though---absent catastrophe, deliberate choice, or other disruption of scientific progress---we will likely be in a position, if we so choose, to create non-human agents whose abilities in these respects rival or exceed our own. And in the context of artificial agents, the differences between brains and computers---in possible speed, size, available energy, memory capacity, component reliability, input/output bandwidth, and so forth---make the eventual possibility of very dramatic differences in ability especially salient.\footnote{Thus, as \href{https://www.amazon.com/Superintelligence-Dangers-Strategies-Nick-Bostrom/dp/1501227742}{Bostrom (2014, Chapter 3)} discusses, where neurons can fire a maximum of hundreds of Hz, the clock speed of modern computers can reach some 2 Ghz: \textasciitilde{}ten million times faster. Where action potentials travel at some hundreds of m/s, optical communication can take place at 300,000,000 m/s: \textasciitilde{}a million times faster. Where brain size and neuron count are limited by cranial volume, metabolic constraints, and other factors, supercomputers can be the size of warehouses. And artificial systems need not suffer, either, from the brain's constraints with respect to memory and component reliability, input/output bandwidth, tiring after hours, degrading after decades, storing its own repair mechanisms and blueprints inside itself, and so forth. Artificial systems can also be edited and duplicated much more easily than brains, and information can be more easily transferred between them.}

\subsubsection{Playing with fire}\label{playing-with-fire}

The choice to create agents much more intelligent than we are should be approached with extreme caution. This is the basic backdrop view underlying much of the concern about existential risk from AI---and it would apply, in similar ways, to new biological agents (human or non-human).

Some articulate this view by appeal to the dominant position of humans on this planet, relative to other species.\footnote{See e.g. \href{https://www.ted.com/talks/nick_bostrom_what_happens_when_our_computers_get_smarter_than_we_are/transcript}{Bostrom (2015)}, \href{https://www.amazon.com/Human-Compatible-Artificial-Intelligence-Problem/dp/0525558616/ref=tmm_hrd_swatch_0?_encoding=UTF8&qid=1619197644&sr=1-1}{Russell (2019, Chapter 5)} on the ``Gorilla Problem,'' \href{https://www.amazon.com/Precipice-Existential-Risk-Future-Humanity/dp/031648492X/ref=sr_1_2?crid=2ZWCCI74ZFX55\&dchild=1\&keywords=precipice+existential+risk+and+the+future+of+humanity\&qid=1619197698\&s=books\&sprefix=precipice\%2Cstripbooks\%2C243\&sr=1-2}{Ord (2020)}, and \href{https://www.alignmentforum.org/posts/8xRSjC76HasLnMGSf/agi-safety-from-first-principles-introduction}{Ngo (2020)} on the ``second species argument.''} For example: some argue that the fate of the chimpanzees is currently in human hands, and that this difference in power is primarily attributable to differences in intelligence, rather than e.g.~physical strength. Just as chimpanzees---given the choice and power---should be careful about building humans, then, we should be careful about building agents more intelligent than us.

This argument is suggestive, but far from airtight. Chimpanzees, for example, are themselves much more intelligent than mice, but the ``fate of the mice'' was never ``in the hands'' of the chimpanzees. What's more, the control that humans can exert over the fate of other species on this planet still has limits, and we can debate whether ``intelligence,'' even in the context of accumulating culture and technology, is the best way of explaining what control we have.\footnote{Note, too, that a single human, unaided and uneducated, would plausibly fare worse than a chimpanzee in many environments (see \href{https://www.amazon.com/Secret-Our-Success-Evolution-Domesticating/dp/0691166854}{Henrich (2015)} for more on humans struggling in the absence of cultural learning).}

More importantly, though: humans arose through an evolutionary process that chimpanzees did nothing to intentionally steer. Humans, though, will be able to control many aspects of processes we use to build and empower new intelligent agents.\footnote{Though whether that control will be adequate to ensure the outcomes we want is a substantially further question; and note that analogies between evolution and contemporary techniques in machine learning are somewhat sobering in this respect---see \protect\hyperref[problems-with-search]{section 4.3.1.2} for more.}

Still, some worry about playing with fire persists. As our own impact on the earth illustrates, intelligent agents can be an extremely powerful force for controlling and transforming an environment in pursuit of their objectives. Indeed, even on the grand scale of earth's history, the development of human abilities in this respect seems like a very big deal---a force of unprecedented potency. If we unleash much more of this force into the world, via new, more intelligent forms of non-human agency, it seems reasonable to expect dramatic impacts, and reasonable to wonder how well we will be able to control the results.\footnote{Of course, the most relevant and powerful ``human actors'' are and have been groups (countries, corporations, ideologies, cultures, etc), equipped with technology, rather than unaided individuals. But non-human agents can create groups, and use/create new technology, too (indeed, creating and using technology more effectively than humans is one of the key things that makes advanced AI systems formidable). And note that while human organizations like corporations are much more capable than individuals along certain axes (especially e.g.~tasks that can be effectively parallelized), they remain constrained by human abilities, and guided by human values, in key ways---ways that fully ``non-human'' corporations would not (see \href{https://arbital.com/p/corps_vs_si/}{here} for more discussion). That said, I think questions around the role of individuals vs.~group agents/''superorganisms'' as central loci of influence, and questions around the role of human intelligence vs.~technology, culture, economic interaction, etc, in explaining human growth and dominance on this planet, are both key ways in which ``building agents more much intelligent than humans warrants caution'' might mislead---and I think a full fleshing out of the ``backdrop picture'' here requires more detail than I've gestured at above (and perhaps such a picture requires fuller automation of human cognitive abilities than the ``advanced capability'' condition used in this report implies). Thanks to Katja Grace and Ben Garfinkel for discussion.}

The rest of this report focuses on a specific version of this sort of worry, in the context of artificial systems in particular.

\subsubsection{Power}\label{power}

This version centers on the following hypothesis: that by default, suitably strategic and intelligent agents, engaging in suitable types of planning, will have instrumental incentives to gain and maintain various types of power, since this power will help them pursue their objectives more effectively (see \protect\hyperref[power-seeking]{section 4.2} for more discussion).\footnote{By ``power'' I mean something like: the type of thing that helps a wide variety of agents pursue a wide variety of objectives in a given environment. For a more formal definition, see \href{https://arxiv.org/abs/1912.01683}{Turner et al (2020)}.} The worry is that if we create and lose control of such agents, and their objectives are problematic, the result won't just be \emph{damage} of the type that occurs, for example, when a plane crashes, or a nuclear plant melts down---damage which, for all its costs, remains passive. Rather, the result will be highly-capable, non-human agents actively working to gain and maintain power over their environment---agents in an \emph{adversarial} relationship with humans who don't want them to succeed.

Nuclear contamination is hard to clean up, and to stop from spreading. But it isn't \emph{trying} to not get cleaned up, or \emph{trying} to spread---and especially not with greater intelligence than the humans trying to contain it.\footnote{Harmful structures actively optimized for self-replication---like pathogens, computer viruses, and ``\href{https://en.wikipedia.org/wiki/Gray_goo}{grey goo}''---are a closer analogy, but these, too, lack the relevant type of agency and intelligence.} But the power-seeking agents just described would be \textit{trying}, in sophisticated ways, to undermine our efforts to stop them. If such agents are sufficiently capable, and/or if sufficiently many of such failures occur, humans could end up permanently disempowered, relative to the power-seeking systems we've created.

This difference---between the usual sort of damage that occurs when a piece of technology malfunctions, and the type that occurs when you lose control over strategically sophisticated, power-seeking agents whose objectives conflict with yours---marks a key distinction not just between worries about AI vs.~other sorts of risks, but also between the specific type of AI-related worry I focus on in what follows, and the more inclusive set of worries that sometimes go under the heading of ``AI alignment.''

As with any technology, AI systems can fail to behave in the way that their designers intend. And because some AI systems pursue objectives, some such unintended behavior can result from problems with their objectives in particular (call this particular type of unintended behavior ``misaligned''). And regardless of the intentions of the designers, the development and social impact of AI systems can fail, more broadly, to uphold and reflect important values.

Problems in any of these veins are worth addressing. But it is \emph{power-seeking}, in particular, that seems to me the most salient route to existential catastrophe from unintended AI behavior. AI systems that don't seek to gain or maintain power may cause a lot of harm, but this harm is more easily limited by the power they already have. And such systems, by hypothesis, won't try to maintain that power if/when humans try to stop them. Hence, it's much harder to see why humans would fail to notice, contain, and correct the problem before it reaches an existential scale.

This, then, is the backdrop picture underlying the specific arguments I'll examine: non-human agents, much more intelligent than humans, would likely be potent forces in the world; and if their objectives are problematic, some of them may, by default, seek to gain and maintain power. Building such agents therefore seems like it could risk human disempowerment, at least in principle. Let's look more closely at whether to expect this in practice.

\section{Timelines}\label{timelines}

For existential risks from power-seeking AI agents to arise, it needs to become possible and financially feasible for humans to build relevantly dangerous AI systems. In this section, I describe the type of system I'm going to focus on, and I discuss the probability that humans learn to create such systems by 2070.

\subsection{Three key properties}\label{three-key-properties}

I'll focus on systems with three properties: (a) advanced capabilities, (b) agentic planning, and (c) strategic awareness.

\subsubsection{Advanced capabilities}\label{advanced-capabilities}

I'll say that an AI system has ``advanced capabilities'' if it outperforms the best humans on some set of tasks which when performed at advanced levels grant significant power in today's world. The type of tasks I have in mind here include: scientific research, engineering, business/military/political strategy, hacking, and social persuasion/manipulation. An AI system with these capabilities can consist of many smaller systems interacting, but it should suffice to \textasciitilde{}fully automate the tasks in question.

The aim here is to hone in on systems whose capabilities make any power-seeking behavior they engage in worth taking seriously (in aggregate) as a potential route to the disempowerment of \textasciitilde{}all humans. Such a condition does not, I think, require meeting various stronger conditions sometimes discussed\footnote{A level of AI progress that disempowered all humans would constitute ``transformative AI'' in the sense used by \href{https://www.openphilanthropy.org/blog/some-background-our-views-regarding-advanced-artificial-intelligence\#Sec1}{Karnofsky (2016)}: e.g. ``AI that precipitates a transition comparable to (or more significant than) the agricultural or industrial revolution.'' But \emph{that} sort of transformation is precisely the type of thing we're trying to forecast; e.g., it's the result, not the cause. And disempowerment does not require other, more mechanistic standards for transformation---e.g.~economic growth proceeding at particular rates (though we can argue, here, about the economic value that AI progress sufficient to disempower humans would represent).}---for example, ``human-level AI,''\footnote{This is used in various ways, to mean something like (a) a single AI system that is in some generic sense ``as intelligent'' as a human; (b) a single AI system that can do anything that a given human (an average human? the ``best'' human? any human?) can do; (c) a level of automation such that unaided machines can perform roughly any task better and more cheaply than human workers (see \href{https://arxiv.org/abs/1705.08807}{Grace et al (2017)}). My favorite is (c).} ``superintelligence,''\footnote{\href{https://www.amazon.com/Superintelligence-Dangers-Strategies-Nick-Bostrom/dp/1501227742}{Bostrom (2014, Chapter 2)} defines this as ``any intellect that greatly exceeds the cognitive performance of humans in virtually all domains of interest.'' A related concept requires cognitive performance that in some loose sense exceeds all of human civilization---though exactly how to understand this isn't quite clear (and human civilization's ``cognitive abilities'' change over time).} or ``AGI.''\footnote{``AGI'' is sometimes used as a substitute for some concept of ``human-level AI''; in other contexts, it refers specifically to some concept of human \emph{learning} ability (see e.g. \href{https://dselsam.github.io/posts/2018-07-08-the-general-intelligence-hypothesis.html}{Selsam (undated)}, and Arbital \href{https://arbital.com/p/general_intelligence/}{here} (no author listed, but I believe it is Yudkowsky)), or some method of creating systems that can perform certain tasks (see \href{https://www.alignmentforum.org/s/mzgtmmTKKn5MuCzFJ/p/eG3WhHS8CLNxuH6rT}{Ngo (2020)} on ``generalization-based approaches). ``AGI'' is often contrasted with ``narrow AI''---though in my opinion, this contrast too easily runs together a system's ability to \textit{learn} tasks with its ability to \emph{perform} them. And sometimes, a given use of ``AGI'' just means something like ``you know, the big AI thing; \emph{real} AI; the special sauce; the thing everyone else is talking about.''} That said, I'm erring, here, on the side of including ``weaker'' systems --- including some that might not, on their own (or even in aggregate), be all that threatening (a fact worth bearing in mind in assigning overall probabilities).\footnote{Put another way, I'm erring on the side of ``necessary'' rather than ``sufficient'' for riskiness. I have yet to hear a good capability threshold that is both necessary \emph{and} sufficient---and I'm skeptical that one exists.} Stronger conditions have less of this problem; and I expect much of the discussion in what follows to apply to stronger systems, too.

Admittedly, the standard here is imprecise. And notably, what sorts of task performance yields what sort of real-world power can shift dramatically as the social and technological landscape changes (see discussion in 6.3). I'm going to tolerate this imprecision in what follows, but those who want more precise definitions should feel free to use them---the discussion doesn't depend heavily on the details.\footnote{My own favored more precise threshold would be something like ``a level of automation such that unaided machines can perform roughly any task better and more cheaply than human workers''; but I don't think this is necessary for the threats in question to arise.}

\subsubsection{Agentic planning}\label{agentic-planning}

I'll say that a system engages in ``agentic planning'' if it makes and executes plans, in pursuit of objectives, on the basis of models of the world (to me, this isn't all that different from bare ``agency,'' but I want to emphasize the planning aspect).\footnote{I like emphasizing planning in part because it seems to me more tempting to speak loosely about systems as ``trying'' to do things (e.g., your microwave is ``trying'' to warm your food, a thermostat is ``trying'' to keep your house a certain temperature), than to speak of them as ``planning'' to do so.}

The aim here (and in the next subsection) is to hone in on the type of goal-oriented cognition required for arguments about the instrumental value of gaining/maintaining power to be relevant to a system's behavior.\footnote{We can imagine cases in which AI agents end up valuing power for its own sake, but I'm not going to focus on those here.} That said, muddyness about abstractions in this vicinity is one of my top candidates for ways arguments of the type I consider might mislead; and I expect that as we gain familiarity with advanced AI systems, many of the concepts we use to think about them will alter and improve. However: I also think of many worries about existential risk from misaligned power-seeking as centrally animated by some concepts in this broad vicinity (though not necessarily mine in particular). I don't think such concepts useless, or obviously dispensable to the argument; and for now, I'll use them.\footnote{Of course, you can just talk directly about AI systems that end up causing (suitably unintended) existential catastrophes, without invoking concepts like agency, objectives, etc. The question, though, is why one might expect that sort of behavior. And especially when the existential catastrophes involve AI power-seeking in particular (as I'm inclined to think the most worrying ones do), I think something like ``instrumental convergence'' is the strongest argument. That said, it might be possible to put instrumental convergence-type considerations in terms that appeal less to agency, objectives, strategic reasoning, and so forth, and more to the types of novel generalization we might expect from a given sort of training. I briefly explore this possibility in a footnote at the end of section 4.2. I'm open to this approach, but my current feeling is that this argument is strongest when it leans on an (implicit or explicit) expectation that something like agentic planning, in pursuit of objectives, will develop in the systems in question.}

A few clarifications, though.

I take as my paradigm a certain type of human cognition---the type, for example, involved in e.g., planning and then taking a trip from New York to San Francisco; reasoning about the safest way to cut down a tree, then doing it; designing a component of a particle collider; and so on. When I talk about agentic planning, I'm talking, centrally, about \emph{that}. (Though exactly how much of human cognition is like this is a further question---and similarity comes on a spectrum.)\footnote{Obviously, not everything humans do seems like agentic planning. Pulling your hand away from a hot stove, for example, seems more like a reflex---a basic ``if-then'' implemented by your nervous system---than an executed plan. And much of what we do in life can seem more like the stove than planning a trip to New York. Indeed, we often think of non-humans animals as acting on ``instinct,'' or on the basis of comparatively simple heuristics, instead of on the basis of explicit plans, models, and objectives. Humans, we might think, are more like this than sometimes supposed. But lines here can get blurry fast. Ultimately, all complex cognition consists of a large number of much simpler steps or ``if-thens.'' Whether a given behavior is built out of instincts, heuristics, if-thens, etc does not settle whether it is relevantly similar to travel-planning (indeed, contrasts between ``heuristics'' and more systematic forms of cognition often seem to me under-specified). And similarity comes along a multi-dimensional spectrum---reflecting, for example, the sophistication and flexibility of the models and plans involved.}

We tend to think of this kind of cognition as (a) using a model of the world that represents causal relationships between actions/outputs/policies and outcomes to (b) select actions/outputs/policies that lead to outcomes that rate highly according to various (possibly quite complicated) criteria (e.g., objectives).\footnote{See \href{https://arbital.com/p/consequentialist/}{Arbital (undated, and no author listed, but I believe it's Yudkowsky)} on ``consequentialist cognition'' for more. Thanks to Paul Christiano for discussion.} And we predict and explain human action on this basis, using certain sorts of common-sense distinctions---for example, between the role that someone's objectives play in explaining a given action, vs.~their world-models and capabilities. There are also various algorithms that implement explicit procedures in the vein of (a) and (b)---algorithms, for example, that search over and evaluate possible sequences of actions, or that backwards-chain from a desired end-state.

People sometimes dispute whether humans ``actually'' do something like (a) and (b) (let alone, implement something akin to a more formal algorithm). For present purposes, though, what matters is that they do something \emph{close enough} to justify predicting their behavior (at least sometimes) with (a) and (b) in mind.

This is the standard I'll apply to the agentic planners I discuss in what follows. That is, I am not imposing any specific constraints on the cognitive algorithms or forms of representation required for ``agentic planning''---but they need to be \emph{close enough}, at some level of abstraction, to (a) and (b) above to justify similar predictions. Indeed, I will speak as though agentic planners are \emph{actually} using models of the world to select and execute plans that promote their objectives (as opposed to, e.g., ``behaving just like'' they are doing this); by hypothesis, to the extent there is a difference here, it's not a predictively relevant one.

I am not assuming, though:

\begin{itemize}
\item
  that the objectives need to involve an explicitly represented ``\href{https://en.wikipedia.org/wiki/Mathematical_optimization}{objective function},'' especially in the sense at stake in formal optimization problems;\footnote{Here I'm responding in part to the definition offered by \href{https://arxiv.org/abs/1906.01820}{Hubinger et al (2019)}, and to subsequent debate.}
\item
  that an agentic planner need be helpfully understood as ``maximizing expected utility'';
\item
  that it engages in agentic planning in all circumstances;
\item
  that it needs to be possible to easily \emph{tell} whether a given system is engaging in agentic planning or not;
\item
  that the system's objectives are simple, ``long-term'' (see \protect\hyperref[myopia]{section 4.3.1.3}), easily subsumable within a set of ``intrinsic values,'' or constant across circumstances;
\item
  that agentic planners cannot be constituted by many interacting, non-agentic-planning systems;\footnote{Here, and in a number of subsequent bullet points, I'm responding in part to the contrasts discussed by \href{https://www.fhi.ox.ac.uk/wp-content/uploads/Reframing_Superintelligence_FHI-TR-2019-1.1-1.pdf}{Drexler (2019)}.}
\item
  that the system is capable of self-modification or online learning (see \protect\hyperref[preventing-problematic-improvements]{section 4.3.2.2});
\item
  that the system has any particular set of opportunities for action in the world (for example, systems that can only answer question can count as agents in this sense);
\item
  that the system's plans are oriented towards the external environment, as opposed to e.g.~plans for how to distribute its internal cognitive resources in performing a task;
\item
  that the agentic planning in question is ``ongoing'' and ``open-ended,'' as opposed to ``episodic'' (see footnote for details).\footnote{For example, something can be an agent planner \emph{once it is given a task}; but that doesn't mean that when it hasn't been given a task, it has other, ongoing objectives. And a system can plan agentically in pursuit of taking a single action, as opposed to executing a series of actions.} 
  \end{itemize}
  
\href{https://deepmind.com/blog/article/muzero-mastering-go-chess-shogi-and-atari-without-rules}{MuZero}---a system which learns a model of a game (Chess, Go, Shogi, Atari) in order to plan future actions---qualifies as an agentic planner in this sense, as do \href{https://deepmind.com/blog/article/alphazero-shedding-new-light-grand-games-chess-shogi-and-go}{AlphaZero}, \href{https://deepmind.com/research/case-studies/alphago-the-story-so-far}{AlphaGo}, and (at least on my current understanding) various self-driving cars. Thermostats, bottlecaps, forest fires, balls rolling down hills, and robots twitching randomly don't qualify---they're not doing something close enough to planning, using a model of the world, in pursuit of the outcomes they cause.\footnote{These are examples references in e.g. \href{https://danielfilan.com/2018/08/31/bottle_caps_arent_optimisers.html}{Filan (2018)}, \href{https://www.lesswrong.com/posts/NxF5G6CJiof6cemTw/coherence-arguments-do-not-imply-goal-directed-behavior}{Shah (2018)}, and \href{https://www.lesswrong.com/posts/znfkdCoHMANwqc2WE/the-ground-of-optimization-1}{Flint (2020)}. Fires are an example Katja Grace often uses of a non-agentic system that meets many proposed definitions of agency (for example, tending to bring about a certain type of outcome with a fair amount of robustness).}

In various other systems, it's less clear. For example: \href{https://deepmind.com/blog/article/alphastar-mastering-real-time-strategy-game-starcraft-ii}{AlphaStar}---an AI system that plays Starcraft---executes complex, flexible strategies over long time horizons, but the extent to which it's doing something close enough to explicitly representing and acting in light of the long-term consequences of its actions seems (to me) like an open question. Similarly, we can imagine a system like \href{https://en.wikipedia.org/wiki/GPT-3}{GPT-3}---that is, a large machine learning model trained to predict human-generated strings of text---that engages in something like agentic planning in generating its output: but the outputs we've observed need not tell us directly.\footnote{The extent of the world-modeling and planning involved in many animal behaviors is similarly unclear.}

I'll also note that a system that merely \emph{generates} plans (for example, a GPT-3-like system responding to the prompt ``The following is a statement of an excellent business plan:''), but which doesn't do so \emph{in order to select outputs/actions/policies that promote its objectives}, doesn't qualify.

\subsubsection{Strategic awareness}\label{strategic-awareness}

I'll say that an agentic planner has ``strategic awareness'' if the models it uses in making plans are broad, informed, and sophisticated enough to represent with reasonable accuracy the causal upshot of gaining and maintaining different forms of power over humans and the real-world environment (here, again, I am using the ``actually this, or close enough to make no predictive difference'' standard above).\footnote{See Yudkowsky \href{https://arbital.com/p/big_picture_awareness/}{here} for the closely related concept of ``big-picture strategic awareness.''}

Clearly, strategic awareness comes in degrees. Broadly and loosely, though, we can think of a strategically aware, planning agent as possessing models of the world that would allow it to answer questions like ``what would happen if I had access to more computing power'' and ``what would happen if I tried to stop humans from turning me off'' about as well as humans can (and using those same models in generating plans).

AlphaGo does not meet this condition (its models are limited to the Go board). Nor, I think, do self-driving cars.\footnote{My understanding is that the planning performed by self-driving cars uses some combination of long-term, google-maps-like planning, which relies on a fairly impoverished model of the world; and short term predictions of behavior on the road. Neither of these seems broad and sophisticated enough to answer questions of the type above. However, I haven't investigated this.} And even if a GPT-3-like system had suitably sophisticated models, it would need to \emph{use} those models in generating plans in order to qualify.

Let's call a system with advanced capabilities that performs agentic planning with strategic awareness an APS (Advanced, Planning, Strategically aware) system.

\subsection{Likelihood by 2070}\label{likelihood-by-2070}

How likely is it that it becomes possible and financially feasible to develop APS systems before 2070? Obviously, forecasts like this are difficult (and fuzzily defined), and I won't spend much time on them here. My own views on this topic emerge in part from a set of investigations that Open Philanthropy has been conducting (see, for example, \href{https://www.alignmentforum.org/posts/KrJfoZzpSDpnrv9va/draft-report-on-ai-timelines}{Cotra (2020)}, \href{https://www.openphilanthropy.org/blog/modeling-human-trajectory}{Roodman (2020)}, and \href{https://www.openphilanthropy.org/blog/report-semi-informative-priors}{Davidson (2021)}, which I encourage interested readers to investigate.

I'll add, though, a few quick data points:

\begin{itemize}
\tightlist
\item
  \href{https://docs.google.com/document/d/1IJ6Sr-gPeXdSJugFulwIpvavc0atjHGM82QjIfUSBGQ/edit\#heading=h.x1lgibjdeh01}{Cotra's model}, which anchors on the human brain and extrapolates from scaling trends in contemporary machine learning, puts \textgreater{}65\% on the development of ``transformative AI'' systems by 2070 (definition \href{https://docs.google.com/document/d/1IJ6Sr-gPeXdSJugFulwIpvavc0atjHGM82QjIfUSBGQ/edit\#heading=h.6t4rel10jbcj}{here}).
\item
  Metaculus, a public forecasting platform, puts a median of 54\% on ``\href{https://www.metaculus.com/questions/384/human-machine-intelligence-parity-by-2040/}{human-machine intelligent parity by 2040},'' and a median of 2038 for the ``\href{https://www.metaculus.com/questions/3479/when-will-the-first-artificial-general-intelligence-system-be-devised-tested-and-publicly-known-of/}{date the first AGI is publicly known}'' (as of mid-April 2021, see links for definitions).
\item
  \href{https://aiimpacts.org/2016-expert-survey-on-progress-in-ai/\#Answers}{Depending on how you ask them}, experts in 2017 assign a median probability of \textgreater{}30\% or \textgreater{}50\% to ``unaided machines can accomplish every task better and more cheaply than human workers'' by 2066, and a 3\% or 10\% chance to the ``full automation of labor'' by 2066 (though their views in this respect are notably inconsistent, and I don't think the specific numbers should be given much weight).
\end{itemize}

Personally, I'm at something like 65\% on ``developing APS systems will be possible/financially feasible before 2070.'' I can imagine going somewhat lower, but less than e.g. 10\% seems to me weirdly confident (and I don't think the difficulty of forecasts like these licenses assuming that probability is very low, or treating it that way implicitly).

\section{Incentives}\label{incentives}

Let's grant, then, that it becomes possible and financially feasible to develop APS systems. Should we expect relevant actors to have strong incentives to do so, especially on a widespread scale? I'll assume that there are strong incentives to automate advanced capabilities, in general. But building strategically-aware agentic planners may not be the only way to do this. Indeed, there are various reasons we might expect an automated economy to focus on systems without such properties, namely:\footnote{Here I'm drawing heavily on a list of considerations in an unpublished document written by Ben Garfinkel.}

\begin{itemize}
\item
  Many tasks---for example, translating languages, classifying proteins, predicting human responses, and so forth---don't seem to require agentic planning and strategic awareness, at least at current levels of performance.\footnote{Though what sorts of methods of performing those tasks emerge in the limits of optimization is a further question; see 3.3 below.} Perhaps all or most of the tasks involved in automating advanced capabilities will be this way.
\item
  In many contexts (for example, factory workers), there are benefits to specialization; and highly specialized systems may have less need for agentic planning and strategic awareness (though there's still a question of the planning and strategic awareness that specialized systems in combination might exhibit). See \protect\hyperref[specialization]{section 4.3.2.1} for more on this.
\item
  Current AI systems are, I think, some combination of non-agentic-planning and strategically unaware. Some of this is clearly a function of what we are currently able to build, but it may also be a clue as to what type of systems will be most economically important in future.
\item
  To the extent that agentic planning and strategic awareness create risks of the type I discuss below, this might incentivize focus on other types of systems.\footnote{Though not all relevant actors will treat these risks with equal caution---see 5.3.2.}
\item
  Agentic planning and strategic awareness may constitute or correlate with properties that ground moral concern for the AI system itself (though not all actors will treat concerns about the moral status of AI systems with equal weight; and considerations of this type could be ignored on a widespread scale).
\end{itemize}

Indeed, for reasons of this type, I expect that to the extent we develop APS systems, we'll do so in a context of a wide variety of non-APS systems, which may themselves have large impacts on the world, and which may help manage risk from APS systems.\footnote{Though note that an increasingly automated economy might also exacerbate some types of risks, since misaligned, power-seeking systems might be better-positioned to make use of automated rather than human-reliant infrastructure. I discuss this briefly in \protect\hyperref[mechanisms]{section 6.3.1}.} And it seems possible that APS systems just won't be a very important part of the picture at all.

Still, though, there are a number of reasons to expect AI progress to push in the direction of systems with agentic planning and strategic awareness. I'll focus on three main types:

\begin{enumerate}
\def\labelenumi{\arabic{enumi}.}
\tightlist
\item
  Agentic planning and strategic awareness both seem very \emph{useful}. That is, many of the tasks we want AI systems to perform seem to require or substantially benefit from these abilities.
\item
  Given available techniques, it may be that the most efficient way to \emph{develop} AI systems that perform various valuable tasks involves developing strategically aware, agentic planners, even if other options are in principle available.
\item
  It might be difficult to \emph{prevent} agentic planning and strategic awareness from developing in suitably sophisticated and efficient systems.
\end{enumerate}

One note: when I talk in what follows about ``incentives to create APS systems,'' I mean this in a sense that covers 2 and 3 as well as 1. Thus, on this usage, if people will pay lots of money for tables (including flammable tables), and the only (or the cheapest/most efficient) way to make tables is out of flammable wood, then I'll say that there are ``incentives'' to make flammable tables, even if people would pay just as much, or more, for fire-resistant tables.

Let's look at 1-3 in turn.

\subsection{Usefulness}\label{usefulness}

I think that the strongest reason to expect APS systems is that they seem quite \emph{useful}.\footnote{See \href{https://www.fhi.ox.ac.uk/wp-content/uploads/Reframing_Superintelligence_FHI-TR-2019-1.1-1.pdf}{Drexler (2019), Chapter 12}, for discussion and disagreement. Many of Drexler's arguments, though, focus on the value of self-improving and/or ``unbounded'' agents, which aren't aren't my focus in this section (though see \hyperref[preventing-problematic-improvements]{4.3.2.2} for more on self-improvement, and \hyperref[myopia]{4.3.1.3} on something akin to boundedness).} In particular:

\begin{itemize}
\item
  Agentic planning seems like a very powerful and general way of interacting with an environment---especially a complex and novel environment that does not afford a lot of opportunities for trial and error---in a way that results in a particular set of favored outcomes. Many tasks humans care about (creating and selling profitable products; designing and running successful scientific experiments; achieving political and military goals; etc) have this structure, as do many of the sub-tasks involved in those tasks (e.g., efficiently gathering and synthesizing relevant information, communicating with stakeholders, managing resource-allocation, etc).\footnote{For more discussion, see Yudkowsky \href{https://arbital.com/p/consequentialist/}{here}, on the ``ubiquity of consequentialism,'' and on options for ``subverting'' it.} Indeed, getting something done often seems like it requires, or at least benefits substantially, from (a) using a model of the world that reflects the relationship between action and outcome to (b) choose actions that lead to outcomes that score well according to some criteria. If our AI systems can't do this, then the scope of what they can do seems, naively, like it will be severely restricted.
\item
  Strategic awareness seems closely related to a basic capacity to ``understand what is going on,'' interact with other agents (including human agents) in the real world, and recognize available routes to achieving your objectives---all of which seem very useful to performing tasks of the type just described. Indeed, to the extent that humans care about the strategic pursuit of e.g. business, military, and political objectives, and want to use AI systems in these domains, it seems like there will be incentives to create AI systems with the types of world models necessary for very sophisticated and complex types of strategic planning---including planning that involves recognizing and using available levers of real-world power.
\end{itemize}

Note that the usefulness of agentic planning here is not limited to AI systems that are intuitively ``acting directly in the world'' (for example, via robot bodies, or without human oversight), as opposed to e.g. predicting the results of different actions, generating new ideas or designs---output that humans can then decide whether or not to act on.\footnote{See \href{https://www.lesswrong.com/posts/6SGqkCgHuNr7d4yJm/thoughts-on-the-singularity-institute-si\#Objection_2__SI_appears_to_neglect_the_potentially_important_distinction_between__tool__and__agent__AI_}{Karnofsky (2012)} for discussion of ``tool AI'' that suggests such a contrast.} Thus, for example, in sufficiently sophisticated cognitive systems, the task of predicting events or providing information might benefit from making and executing plans for how to process inputs, what data to gather and pay attention to, what lines of reasoning to pursue, and so forth.\footnote{See \href{https://www.amazon.com/Superintelligence-Dangers-Strategies-Nick-Bostrom/dp/1501227742}{Bostrom (2014, p. 152-3, and p. 158)}. Training new systems, and learning from previous experience, also plausibly involves decisions that benefit from this sort of planning. See \href{https://www.gwern.net/Tool-AI}{Branwen (2016)} for more.}

That said, I think we should be cautious in predicting what degree of agentic planning and/or strategic awareness will be necessary or uniquely useful for performing what types of cognitive tasks.

\begin{itemize}
\item
  Humans often plan in pursuit of objectives, on the basis of big-picture models of what's going on, and we are most familiar with human ways of performing tasks. But the constraints and selection processes that shape the development of our AI systems may be very different from the ones that shaped human evolution; and as a result, the future of AI may involve much less agentic planning and strategic awareness than anchoring on human intelligence might lead us to expect. Indeed, my impression is that the track record of claims of the form ``Humans do X task using Y capabilities, so AI systems that do X will also need Y capabilities,'' is quite poor.\footnote{Thanks to Owain Evans for discussion.}
\item
  Naively construed, I can imagine arguments given about the usefulness/necessity of agentic planning and strategic awareness making the wrong predictions about current systems (or, e.g., about animal behaviors like squirrels burying nuts for the winter). Thus, for example, one might have expected writing complex code (see \href{https://twitter.com/i/broadcasts/1OyKAYWPRrWKb}{here}, starting at 28:14), or playing StarCraft, to require a certain type of high-level planning; but whether GPT-3-like systems or AlphaStar in fact engage in such planning is unclear. Similarly, one might have expected automated driving to involve lots of big-picture understanding of ``what's going on''; but current self-driving cars don't have much of this. In hindsight, perhaps this is obvious. But one wonders what future hindsight will reveal. Or put another way: we should be cautious about ``obviously task X requires capability Y,'' too.
\end{itemize}

The compatibility of agentic planning and strategic awareness with modularity is also important. Suppose, for example, that you want to automate the long-term strategic planning performed by a CEO at a company. The best way of doing this may involve a suite of interacting, non-APS systems. Thus, as a toy example, one system might predict a plan's impact on company long-term earnings; another system might generate plans that the first predicts will lead to high earnings; a third might predict whether a given plan would be deemed legal and ethical by a panel of human judges; a fourth might break a generated plan into sub-tasks to assign to further systems; and so forth.\footnote{Note, here, that generating a plan is not the same as agentic planning.} None of these individual systems need be ``trying'' to optimize company long-term earnings in a way human judges would deem legal and ethical; but in combination, they create a system that exhibits agentic planning and strategic awareness with respect to this objective.\footnote{Ben Garfinkel suggested this sort of example in discussion. See also similar examples in \href{https://www.fhi.ox.ac.uk/wp-content/uploads/Reframing_Superintelligence_FHI-TR-2019-1.1-1.pdf}{Drexler (2019)}.} (Note that the agentic planning here is not ``emergent'' in the sense of ``accidental'' or ``unanticipated.'' Rather, we're imagining a system specifically designed to perform an agentic planning function---albeit, using modular systems as components.)

Modularity makes it possible for some sub-tasks, including various forms of oversight, to be performed by humans---an option that may well prove a useful check on the behavior of interacting non-APS systems, and a way of reducing the role of fully-automated APS-systems in the economy.\footnote{Though if the human's performance of the sub-task does not meaningfully constrain or regulate the overall behavior of the system, then whether that task is performed by a human or an AI system may not make a difference.} However, as AI systems increase in speed and sophistication, it also seems plausible that there will be efficiency pressures towards more and more complete automation, as slower and less sophisticated humans prove a greater and greater drag on what a system can do.\footnote{See \href{https://www.fhi.ox.ac.uk/wp-content/uploads/Reframing_Superintelligence_FHI-TR-2019-1.1-1.pdf}{Drexler (2019, Chapter 24)} for discussion and disagreement.}

Overall, my main point here is that the space of tasks we want AI systems to perform plausibly involves a lot of ``strategically-aware agentic planning shaped holes.'' Leaving such holes unfilled would reduce risks of the type I discuss; but it would also substantially curtail the ways in which AI can be useful.\footnote{For related points, see \href{https://www.gwern.net/Tool-AI}{Branwen (2018)}: ``Fundamentally, autonomous agent AIs are what we and the free market \emph{want}; everything else is a surrogate or irrelevant loss function. We don't want low log-loss error on ImageNet, we want to refind a particular personal photo; we don't want excellent advice on which stock to buy for a few microseconds, we want a money pump spitting cash at us; we don't want a drone to tell us where Osama bin Laden was an hour ago (but not now), we want to have killed him on sight; we don't want good advice from Google Maps about what route to drive to our destination, we want to be at our destination without doing any driving etc. Idiosyncratic situations, legal regulation, fears of tail risks from very bad situations, worries about correlated or systematic failures (like hacking a drone fleet), and so on may slow or stop the adoption of Agent AIs---but the pressure will always be there.''}

\subsection{Available techniques}\label{available-techniques}

Even if some task doesn't \emph{require} agentic planning or strategic awareness, it may be that creating APS systems is the only route, or the most efficient route, to automating that task, given available techniques. For example, instead of automating tasks one by one, the best way to automate a wide range of tasks---especially ones where we lack lots of training data, which may be the majority of useful tasks---is to create AI systems that are able to learn new tasks with very little data. And we can imagine scenarios in which the best way to do \emph{that} is by training agentic planners (for example, in multi-agent reinforcement learning environments, or via some process akin to evolutionary selection) with high-level, broad-scope pictures of ``how the world works,'' and then fine tuning them on specific tasks.\footnote{Richard Ngo suggested this point in conversation; see also his discussion of the ``generalization-based approach'' \href{https://www.alignmentforum.org/s/mzgtmmTKKn5MuCzFJ/p/eG3WhHS8CLNxuH6rT}{here}. The training and fine-tuning used for GPT-3 may also be suggestive of patterns of this type.}

Indeed, with respect to strategic awareness in particular, various current techniques for providing AI systems information about the world---for example, training them on large text corpora from the internet---seem ill-suited to limiting their understanding of their strategic situation.

\subsection{Byproducts of sophistication}\label{byproducts-of-sophistication}

Even if you're not explicitly aiming for or anticipating agentic planning and strategic awareness in an AI system, these properties could in principle arise in unexpected ways regardless, and/or prove difficult to prevent. Thus, for example:

\begin{itemize}
\item
  Optimizing a system to perform some not-intuitively-agential task (for example, predicting strings of text) could, given sufficient cognitive sophistication, result in internal computation that makes and executes plans, in pursuit of objectives, on the basis of broad and informed models of the real world, even if the designers of the system do not expect this (they may even be unable to tell whether it has occurred). Indeed, the likelihood of this seems correlated with the strength of the ``usefulness'' considerations in 3.1: insofar as agentic planning and broad-scope world-modeling are very useful types of cognition, we might expect to see them cropping up a lot in sufficiently optimized systems, whether we want them to or not.\footnote{See e.g. \href{https://arbital.com/p/consequentialist/}{Yudkowsky (undated)}: ``Another concern is that consequentialism may to some extent be a convergent or default outcome of optimizing anything hard enough. E.g., although natural selection is a pseudoconsequentialist process, it optimized for reproductive capacity so hard that \href{https://arbital.com/p/daemons/}{it eventually spit out some powerful organisms that were explicit cognitive consequentialists} (aka humans).''}
\item
  A sophisticated system exposed to one source of information X might infer from this strategically important information Y, even if humans could not anticipate the possibility of such an inference.
\item
  Modular systems interacting in complex and unanticipated ways could give rise to various types of agential behavior, even if they weren't designed to do so.
\end{itemize}

Of these three reasons to expect APS systems---their usefulness, the pressures exerted by available techniques, and the possibility that they arise as byproducts of sophistication---I place the most weight on the first. Indeed, if it turns out that APS systems aren't uniquely useful for some of the tasks we want AI systems to perform, relative to non-APS systems, this would seem to me a substantial source of comfort.

\section{Alignment}\label{alignment}

Let's assume that it becomes possible and financially feasible to create APS systems by 2070, and that there are significant incentives to do so. This section examines why we might expect it to be difficult to create systems of this kind that don't seek to gain and maintain power in unintended ways.

\subsection{Definitions and clarifications}\label{definitions-and-clarifications}

Let's start with a few definitions and clarifications.

At a high-level, we want the impact of AI on the world to be good, just, fair, and so forth---or at least, not actively/catastrophically bad. Call this the challenge of ``making AI go well.''\footnote{Here I'm drawing on the framework used by \href{https://www.effectivealtruism.org/articles/paul-christiano-current-work-in-ai-alignment/}{Christiano (2020)}.} This is a very broad and complex challenge, much of which lies well outside the scope of this report.

A narrower challenge is: making sure AI systems behave as their designers intend. Of course, the intentions of designers might not be good. But if we \emph{can't} get our AI systems to behave as designers intend, this seems like a substantial barrier to good outcomes from AI more generally (though it is also, at least in many cases, a barrier to the \emph{usefulness} of AI; see \protect\hyperref[bottlenecks-on-usefulness]{section 5.3.3}).\footnote{Following the framework in \href{http://acritch.com/papers/arches.pdf}{Critch and Krueger (2020)}, we can also draw additional distinctions, related to the number of human stake-holders and AI systems involved. I'm focused here on what that framework would call ``single-single'' and ``multi-single'' delegation: that is, the project of aligning AI behavior with the intentions of some human designer or set of designers. Critch argues (e.g., \href{https://futureoflife.org/2020/09/15/andrew-critch-on-ai-research-considerations-for-human-existential-safety/}{here}) that multi-multi delegation should be our focus. I disagree with this (indeed, if we solve single-single alignment, I personally would be feeling \emph{substantially} better about the situation), but I won't argue the point here.} Granted, there are ambiguities about what sort of behavior counts as ``intended'' by designers (see footnote for discussion), but I'm going to leave the notion vague for now, and assume that behaviors like lying, stealing money, resisting shut-down by appropriate channels, harming humans, and so forth are generally ``unintended.''\footnote{In particular, the relationship between ``intended'' and ``foreseen'' (for different levels of probability) is unclear. Thus, the designers of AlphaGo did not foresee the system's every move, but AlphaGo's high-quality play was still ``intended'' at some higher level (thanks to David Roodman for suggesting this example). And it's unclear how to classify cases where a designer thinks it either somewhat or highly likely that a system will engage in a certain type of (unwanted-by-the-designers) power-seeking, but deploys anyway (thanks to Katja Grace for emphasizing this type of case in discussion; see also her post \href{https://aiimpacts.org/misalignment-and-misuse-whose-values-are-manifest/}{here}). I'll count scenarios of this latter type as ``unintended'' (as distinct from ``unforeseen''), but this isn't particularly principled (indeed, I'm not sure if a principled carving is ultimately in the offing). A final problem is that the intentions of designers may simply not cover the range of inputs on which we wish to assess whether a system's behavior is aligned (e.g., ``did the designers intend for the robot to react this way to an asteroid impact?'' may not have an answer: asteroid impacts weren't within the scope of their design intentions). Here, perhaps, something like ``unwanted'' or ``would be unwanted'' is preferable. The reason I'm not using ``unwanted'' in general is that too many features of AI systems, including their capability problems, are ``unwanted'' in some sense (for example, I might want AlphaStar to play even better Starcraft, or to make me lots of money on the stock market, or to help me find love). And ``catastrophic'' seems both too broad (not all catastrophic behavior scales in the problematic way power-seeking does, nor is it clear why catastrophic behavior would imply power-seeking) and too narrow (not all relevant power-seeking need be ``catastrophic'' on its own, especially pre-deployment, and/or in ``multipolar'' scenarios in which no one AI actor is dominant).}

I'll understand \emph{misaligned behavior} as a particular type of unintended behavior: namely, unintended behavior that arises specifically in virtue of problems with an AI system's objectives. Thus, for example, a designer might intend for an AI system to make money on the stock market, and the system might fail because it was mistaken about whether some stock would go up or down in price. But if it was \emph{trying} to make money on the stock market, then its behavior was unintended but not misaligned: the problem was with the AI's competence in pursuing its objectives, not with the objectives themselves.\footnote{Note that there is a sense in which the AI here is ``trying to do something we don't want it to do''---e.g., if we hold fixed its false beliefs about whether a given stock will go up or down, we'd prefer for it to be trying to \emph{lose} money on the stock market (thanks to Paul Christiano for suggesting this point). Plausibly, calling this behavior ``aligned,'' then, requires reference to some holistic, common-sensical interpretation of the combination of objectives and capabilities we ultimately ``have in mind.'' But as ever, settling on a perfect conceptualization is tricky.}

I don't, at present, have a rigorous account of how to attribute unintended behavior to problems with objectives vs. other problems; and I doubt the distinction will always be deep or easily drawn (this doesn't make it useless). I'll lean on the intuitive notion for now; but if necessary, perhaps we could cease talk of ``alignment'' entirely, and simply focus on unintended behavior (perhaps of a suitably agentic kind), or perhaps simply catastrophic behavior, whatever its source.

A characteristic feature of misaligned behavior is that it is \emph{unintended} but still \emph{competent.}\footnote{See \href{https://www.alignmentforum.org/posts/2mhFMgtAjFJesaSYR/2-d-robustness}{Mikulik (2019)} for related discussion of ``2-D robustness.''} That is, it looks less like an AI system ``breaking'' or ``failing in its efforts to do what designers want,'' and more like an AI system trying, and perhaps succeeding, to do something designers \textit{don't} want it to do. In this sense, it's less like a nuclear plant melting down, and more like a heat-seeking missile pursuing the wrong target; less like an employee giving a bad presentation, and more like an employee stealing money from the company.

I'll say that a system is ``fully aligned'' if it does not engage in misaligned behavior in response to any inputs compatible with basic physical conditions of our universe (I'll call these ``physics-compatible'' inputs).\footnote{Thus, it is physics-compatible for a randomly chosen bridge in Indiana to get hit, on a randomly chosen millisecond in May 2050, by a nuclear bomb, a 10 km-diameter asteroid, and a lightning bolt all at once; but not for the laws of physics to change, or for us all to be instantaneously transported to a galaxy far away. Obviously the scope here is very broad: but note that misaligned behavior is a different standard than ``bad'' or even ``catastrophic'' behavior. It will always be possible to set up physics-compatible inputs where a system makes a mistake, or gets deceived, or acts in a way that results in catastrophic outcomes. To be misaligned, though, this behavior needs to arise from problems with the system's objectives in particular. Thus, for example, if Bob is a paper-clip maximizer, and he builds Fred, who is also a paper-clip maximizer, Fred will (on my definition) be fully-aligned with Bob as long as Fred keeps trying to maximize paperclips on all physics compatible-inputs (even though some of those inputs are such that trying to maximize paperclips actually minimizes them, kills Bob, etc). Thanks to Eliezer Yudkowsky, Rohin Shah, and Evan Hubinger for comments on the relevant scope here (which isn't to say they endorse my choice of definition).} By ``inputs,'' I mean information the system receives via the channels intended by its designers (an input to GPT-3, for example, would be a text prompt).\footnote{There are some possible edge-cases here where a system is getting information via channels other than those the designers intended, but I'll leave these to the side. The point here is that we're interested in how \emph{the system} responds to inputs, not in how the world might change the system into something else.} I am not including processes that intervene in some other way on the internal state of the system---for example, by directly changing the weights in a neural network (analogy: a soldier's loyalty need not withstand arbitrary types of brain surgery).\footnote{Though here, too, lines may get hard to draw.}

Sometimes, something like full alignment is framed by asking whether we would be OK with a system's objectives being pursued with \textasciitilde{}arbitrary amounts of capability.\footnote{See Yudkowsky on the ``\href{https://arbital.com/p/omni_test/}{omnipotence test for AI safety}'': ``The Omni Test is that an advanced AI should be expected to remain aligned, or not lead to catastrophic outcomes, or fail safely, even if it suddenly knows all facts and can directly ordain any possible outcome as an immediate choice. The policy proposal is that, among agents meant to act in the rich real world, any predicted behavior where the agent might act destructively if given unlimited power (rather than e.g. pausing for a \href{https://arbital.com/p/user_querying/}{safe user query}) should be \href{https://arbital.com/p/AI_safety_mindset/}{treated as a bug}.'' Talk of an ``objective'' such that the ``optimal policy'' on that objective leads to good outcomes is also reminiscent of something like the Omni Test. See e.g. \href{https://www.alignmentforum.org/posts/SzecSPYxqRa5GCaSF/clarifying-inner-alignment-terminology}{Hubinger's (2020)} definition of ``intent alignment'': ``An agent is \href{https://ai-alignment.com/clarifying-ai-alignment-cec47cd69dd6}{intent aligned} if the optimal policy for its \href{https://intelligence.org/learned-optimization/\#glossary}{behavioral objective} is aligned with humans.''} This isn't the concept I have in mind.\footnote{Indeed, I'm skeptical that it will generally be well-defined, even for agentic planners, what an arbitrarily capable system pursuing that agent's objectives looks like (for example, I'm skeptical that there's a single version of ``omnipotent Joe''---and still less, ``omnipotent MuZero,'' ``omnipotent Coca-Cola company,'' ``omnipotent version of my mom's dog'' and so forth). If we want to talk about alignment properties that are robust to improvements in capability, I think we should talk about what sorts of behavior will result from \emph{a particular process for improving that system's capabilities} (e.g,. a particular type of retraining, a particular way of scaling up its cognitive resources, and so forth). Thanks to Paul Christiano, Ben Garfinkel, Richard Ngo, and Rohin Shah for discussion.} Rather, what matters is how the actual system, with its actual capabilities, responds to physics-compatible inputs. If, on some such inputs, it seeks to improve its own capabilities in misaligned ways; or if some inputs improve its capabilities in a way that results in misaligned behavior; then the system is less-than-fully aligned. But if there are no physics-compatible inputs that the actual system responds to in misaligned ways, then I see preventing changes in capability that might alter this fact as a separate problem (though it may be a problem that designers intentionally improving the system's capabilities should be especially concerned about).\footnote{But so, too, should designers be concerned about altering the system's objectives as they improve it. Note that I'm also setting aside the problem (as it relates to a given system A) of how to make sure that, to the extent that system A builds a \emph{new} system B, system B is fully-aligned, too (for example, if system B isn't fully aligned, but not because of any problems with system A's objectives, that's not, on my view, a problem with system's A's alignment---though it might be a problem more generally).}

Nor does full alignment require that the system ``share the designer's values'' (and still less, the designer's ``utility function,'' to the extent it makes sense to attribute one to them) in any particularly complete sense.\footnote{See e.g. Bostrom's \href{https://www.ted.com/talks/nick_bostrom_what_happens_when_our_computers_get_smarter_than_we_are/transcript}{2015 TED talk}: ``The point here is that we should not be confident in our ability to keep a superintelligent genie locked up in its bottle forever. Sooner or later, it will out. I believe that the answer here is to figure out how to create superintelligent A.I. such that even if---when---it escapes, it is still safe because it is fundamentally on our side because it shares our values. I see no way around this difficult problem\ldots{} The initial conditions for the intelligence explosion might need to be set up in just the right way if we are to have a controlled detonation.''} For example, we can imagine AI systems that just undergo some kind of controlled shut-down, or query relevant humans for guidance, if they receive an input designers didn't intend for them to operate on. Indeed, extreme motivational harmony seems like a strange condition to impose on e.g. a house cleaning robot, or a financial manager---neither of which, presumably, need know the designer's or the user's preferences about e.g. politics, or romantic partners.

Another property in the vicinity of ``full alignment'' and ``sharing values,'' but distinct from both, is \href{https://ai-alignment.com/clarifying-ai-alignment-cec47cd69dd6}{Christiano's (2018)} notion of ``intent alignment,'' on which an AI system A is intent aligned with a human H iff ``A is trying to do what H wants it to do,'' where A's relationship to H's desires is, importantly, \emph{de dicto}. That is, A doesn't just want the same things as H (e.g., ``maximize apples'').\footnote{My sense is that this difference between Christiano's notion of ``intent alignment,'' and a broader notion of ``sharing values,'' is sometimes overlooked.} Rather, A has to have something like a \emph{hypothesis} about what H wants (e.g., it asks itself: ``would H want me to maximize apples, or oranges?''), and to act on that hypothesis.\footnote{\href{https://www.amazon.com/Human-Compatible-Artificial-Intelligence-Problem/dp/0525558616/ref=tmm_hrd_swatch_0?_encoding=UTF8\&qid=1619197644\&sr=1-1}{Russell's (2019)} proposal---e.g., to make AI systems that pursue our objectives, but are uncertain what those objectives are (and see our behavior as evidence)---has a somewhat similar flavor.} Intent alignment seems a promising route to full alignment; but I won't focus on it exclusively.

I'll say that a system is ``practically aligned'' if it does not engage in misaligned behavior on any of the inputs it will \emph{in fact} receive.\footnote{In the sense, practical alignment is a property that holds \emph{relative} to a set of inputs.} Full alignment implies a highly reliable form of practical alignment---e.g., one that does not depend at all on predicting or controlling the inputs a system receives. From a safety perspective, this seems a very desirable property, especially given the unique challenges and stakes that APS systems present (see \protect\hyperref[unusual-difficulties]{section 4.4} for discussion). Indeed, if you are building AI systems that would e.g. harm or seize power over you, due to problems with their objectives, in some physics-compatible circumstances (see next section), this seems, at the least, a red flag about your project.\footnote{See Yudkowsky's discussion of ``AI safety mindset'' \href{https://arbital.com/p/AI_safety_mindset/}{here}. Yudkowsky sometimes frames this as: you should build systems that never, in practice, search for ways to kill you, rather than systems where the search comes up empty.} Ultimately, though, it's practical alignment that we care about.

\subsection{Power-seeking}\label{power-seeking}

Not all misaligned AI behavior seems relevant to existential risk. Consider, for example, an AI system in charge of an electrical grid, whose designers intend it to send electricity to both town A and town B, but whose objectives have problems that cause it, during particular sorts of storms, to only send electricity to town A. This is misaligned behavior, and it may be quite harmful, but it poses no threat to the entire future.

Rather, as I discussed in \protect\hyperref[power]{section 1.2.4}, the type of misaligned AI behavior that I think creates the most existential risk involves misaligned \textit{power-seeking} in particular: that is, active efforts by an AI system to gain and maintain power in ways that designers didn't intend, arising from problems with that system's objectives. In the electrical grid case, the AI system hasn't been described as trying to gain power (for example, by trying to hack into more computing resources to better calculate how to get electricity to town A) or to maintain the power it already has (for example, by resisting human efforts to remove its influence over the grid). And in this sense, I think, it's much less dangerous.

I'll say that a system is ``fully PS-aligned'' if it doesn't engage in misaligned power-seeking in particular in response to any physics-compatible inputs. And I'll say that a system is ``practically PS-aligned'' if it doesn't engage in misaligned power-seeking on any of the inputs it will in fact receive.

A key hypothesis, some variant of which underlies much of the discourse about existential risk from AI, is that there is a close connection, in sufficiently advanced agents, between misaligned behavior in general, and misaligned power-seeking in particular.\footnote{See e.g. \href{https://www.amazon.com/Superintelligence-Dangers-Strategies-Nick-Bostrom/dp/1501227742}{Bostrom (2014, chapter 7)}; \href{https://www.amazon.com/Human-Compatible-Artificial-Intelligence-Problem/dp/0525558616/ref=tmm_hrd_swatch_0?_encoding=UTF8\&qid=1619197644\&sr=1-1}{Russell (2019, Chapter 5)}, on ``you can't fetch the coffee if you're dead''; and \href{https://arbital.com/p/instrumental_convergence/}{Yudkowsky (undated)}.} I'll formulate this hypothesis as:

\begin{quote}
\textit{Instrumental Convergence}: If an APS AI system is less-than-fully aligned, and some of its misaligned behavior involves strategically-aware agentic planning in pursuit of problematic objectives, then in general and by default, we should expect it to be less-than-fully PS-aligned, too.
\end{quote}

Why think this? The basic reason is that power is extremely useful to accomplishing objectives---indeed, it is so almost by definition.\footnote{For a formal version of a similar argument, see \href{https://arxiv.org/abs/1912.01683}{Turner et al (2019)}.} So to the extent that an agent is engaging in unintended behavior in pursuit of problematic objectives, it will generally have incentives, other things equal, to gain and maintain forms of power in the process---incentives that strategically aware agentic planning puts it in a position to recognize and respond to.\footnote{Of course, systems with non-problematic objectives will have incentives to seek forms of power, too; but the objectives themselves can encode what sorts of power-seeking are OK vs. not OK.}

One way of thinking about power of this kind is in terms of the number of ``options'' an agent has available.\footnote{Thanks to Ben Garfinkel for helpful discussion.} Thus, if a policy seeks to promote some outcomes over others, then other things equal, a larger number of options makes it more likely that a more preferred outcome is accessible. Indeed, talking about ``options-seeking,'' instead of ``power-seeking,'' might have less misleading connotations.

What sorts of power might a system seek? \href{https://www.amazon.com/Superintelligence-Dangers-Strategies-Nick-Bostrom/dp/1501227742}{Bostrom (2014)} (following \href{https://selfawaresystems.files.wordpress.com/2008/01/ai_drives_final.pdf}{Omohundro (2008)}) identifies a number of ``convergent instrumental goals,'' each of which promotes an agent's power to achieve its objectives. These include:

\begin{itemize}
\tightlist
\item
  self-preservation (since an agent's ongoing existence tends to promote the realization of those objectives);
\item
  ``goal-content integrity,'' e.g. preventing changes to its objectives (since agent's pursuit of those objectives in particular tends to promote them);
\item
  improving its cognitive capability (since such capability tends to increase an agent's success in pursuing its objectives);
\item
  technological development (since control over more powerful technology tends to be useful);
\item
  resource-acquisition (since more resources tend to be useful, too).\footnote{Note that we can extend the list to include an agent's instrumental incentives to promote the existence, goal-content integrity, cognitive capability, resources, etc of agents whose objectives are sufficiently similar to its own---for example, future agents in some training process, copies of itself, slightly modified versions of itself, and so forth. And we can imagine cases in which an agent is \emph{intrinsically} motivated to gain/maintain various types of power---for example, because this was correlated with good performance during training (see \href{https://www.alignmentforum.org/s/mzgtmmTKKn5MuCzFJ/p/bz5GdmCWj8o48726N}{Ngo (2020)} for discussion). I'll focus, though, on the instrumental usefulness of power.} \end{itemize}
  
  We see examples of rudimentary AI systems ``discovering'' the usefulness of e.g. resource acquisition already. \href{https://openai.com/blog/emergent-tool-use/}{For example}: when OpenAI trained two teams of AIs to play hide and seek in a simulated environment that included blocks and ramps that the AIs could move around and fix in place, the AIs learned strategies that depended crucially on acquiring control of the blocks and ramps in question---despite the fact that they were not given any direct incentives to interact with those objects (the hiders were simply rewarded for avoiding being seen by the seekers; the seekers, for seeing the hiders).\footnote{Thus, the hiders learned to move and lock blocks to prevent the seekers from entering the room where the hiders were hiding; the seekers, in response, learned to move a ramp to give them access anyway; adjusting for this, the hiders learned to take control of that ramp before the seekers can get to it, and to lock it in the room as well. In another environment, the seekers learned to ``surf'' on boxes, and the hiders, to prevent this, learned to lock \emph{all} boxes and ramps before the seekers can get to them. See videos in the blog post for details.}

Why did this happen? Because in that environment, what the AIs do with the boxes/ramps can matter a lot to the reward signal, and in this sense, boxes and ramps are ``resources,'' which both types of AI have incentives to control---e.g., in this case, to grab, move, and lock.\footnote{Indeed, in a competitive environment like this, agents have an incentive not just to control resources that are directly useful to their own goals, but also resources that \emph{would} be useful to their opponents. Thus, the ramps aren't obviously useful to the hiders directly; but ramps can be used to bypass hider defenses, so hiders have an incentive to lock ramps in their room, or to remove from the environment entirely, all the same. This type of dynamic could be relevant to adversarial dynamics between humans and AIs, too.} The AIs learned behavior responsive to this fact.

Of course, this is a very simple, simulated environment, and the level of agentic planning it makes sense to ascribe to these AIs isn't clear.\footnote{And importantly, the trainers weren't \emph{trying} to disincentivize resource-seeking behavior; quite the contrary, the set-up seems (I haven't investigated the history of the experiments) to have been designed to test whether ``emergent tool-use'' would occur.} But the basic dynamic that gives rise to this type of behavior seems likely to apply in much more complex, real-world contexts, and to more sophisticated systems, as well. If, in fact, the structure of a real-world environment is such that control over things like money, material goods, compute power, infrastructure, energy, skilled labor, social influence, etc would be useful to an AI system's pursuit of its objectives, then we should expect the planning performed by a sufficiently sophisticated, strategically aware AI agent to reflect this fact. And empirically, such resources are in fact useful for a very wide variety of objectives.

Concrete examples of power-seeking (where unintended) might include AI systems trying to: break out of a contained environment; hack; get access to financial resources, or additional computing resources; make backup copies of themselves; gain unauthorized capabilities, sources of information, or channels of influence; mislead/lie to humans about their goals; resist or manipulate attempts to monitor/understand their behavior, retrain them, or shut them off; create/train new AI systems themselves; coordinate illicitly with other AI systems; impersonate humans; cause humans to do things for them; increase human reliance on them; manipulate human discourse and politics; weaken various human institutions and response capacities; take control of physical infrastructure like factories or scientific laboratories; cause certain types of technology and infrastructure to be developed; or directly harm/overpower humans. (See 6.3.1 for more detailed discussion).

A few other clarifications:

\begin{itemize}
\item
  Not all misaligned power-seeking, even in APS systems, is particularly harmful, or intuitively worrying from an existential risk perspective.\footnote{Ben Garfinkel suggests the example of a robo-cop pinning down the wrong person, but remaining amenable to human instruction otherwise. And in general, it seems most dangerous if an APS system is \emph{using} its advanced capabilities in pursuing power; e.g., if I'm a greater hacker, but a poor writer, and I'm trying to get power via my journalism, I'm less threatening.} But I'm not going to try to narrow down any further.
\item
  Instrumental convergence is not a conceptual claim, but rather an empirical claim that purports to apply to a wide variety of APS systems. In principle, for example, we can imagine APS systems that plan in pursuit of problematic objectives on some inputs, but which are nevertheless fully PS-aligned (or very close to it).\footnote{Thanks to Ben Garfinkel for suggesting examples in this vein.} Consider, for example, an APS version of the electrical grid AI system above, which plans strategically in pursuit of directing electricity only to town A, but which just doesn't consider plans that involve seeking power. The in-principle possibility of strategic, agentic misalignment without PS-misalignment is important, though, since it might be realized in practice. Perhaps, for example, the type of training we should expect by default will reinforce cognitive habits in APS systems that steer away from searching over/evaluating plans that involve misaligned power-seeking, even if other types of misaligned behavior persist. Training, after all, will presumably strongly select against observable power-seeking behavior, and perhaps power-seeking \emph{cognition,} too.\footnote{Thanks to Ben Garfinkel, Rohin Shah, and Tom Davidson for emphasizing points in this vicinity.} If it proves easy to create/train APS systems of this kind (even if they aren't fully aligned), this would be great news for existential safety (see \protect\hyperref[controlling-objectives]{section 4.3.1} for more discussion).
\end{itemize}

Note, though, this requires that the planning performed by an APS system engaged in misaligned behavior be limited or ``pruned'' in a specific way. That is, by hypothesis, such a system is using a broad-scope world model, capable of accurately representing the causal upshot of different forms of power-seeking, to plan in pursuit of problematic objectives. To the extent that power-seeking \emph{would}, in fact, promote those objectives, the APS system's planning has to consistently ignore/skip over this fact, despite being in an epistemic position to recognize it.\footnote{We can also imagine cases where the system's tendency to rule out certain plans is best interpreted as part of its objectives (e.g., such plans in fact aren't good, on its values).} Perhaps this type of planning is easy, in practice, to select for; but naively, it feels to me like a delicate dance.\footnote{Ben Garfinkel suggests the example of humans considering plans that involve murdering other humans---something that plausibly happens quite rarely, perhaps because ``don't murder'' has been successfully reinforced. But for various reasons, murdering---in most of the relevant contexts---really \emph{doesn't} promote someone's objectives (both because they intrinsically disvalue murder, and because of many other constraints and instrumental incentives). So it makes sense for them to adopt a policy of not considering it at all, or only very rarely. In contexts where murder \emph{does} consistently promote a human's objectives (perhaps sociopaths embedded in much more violent/lawless human contexts would be an example), I expect humans to consider plans that involve murdering much more often.}

Relatedly, we can imagine systems whose objectives are problematic in \emph{some} way, but which \emph{wouldn't} be promoted by unintended forms of power-seeking.\footnote{Distinctions between this and ``it doesn't consider bad plans'' also seem blurry.} Perhaps, for example, the electrical grid system plans strategically in order to direct electricity only to town A, but any plans that involve unintended power-seeking would score very low on its criteria for choice. The ease with which we can create systems of this type seems closely related to the ease with which we can create systems with non-problematic objectives more generally (again, see \protect\hyperref[controlling-objectives]{section 4.3.1} for discussion).

Let's look at two other objections to instrumental convergence.

The first is that humans don't always seem particularly ``power-seeking.''\footnote{See e.g. \href{https://idlewords.com/talks/superintelligence.htm}{Cegloski's (2016)} ``Argument From My Roommate''; and also \href{https://www.amazon.com/Enlightenment-Now-Science-Humanism-Progress/dp/0525427570}{Pinker (2018, Chapter 19)}: ``There is no law of complex systems that says that intelligent agents must turn into ruthless conquistadors. Indeed, we know of one highly advanced form of intelligence that evolved without this defect. They're called women'' (p. 297). Thanks to Rohin Shah for discussion of the humans example.} Humans care a lot about survival, and about certain resources (food, shelter, etc), but beyond that, we associate many forms of power-seeking with a certain kind of greed, ambition, or voraciousness, and with intuitively ``resource-hungry'' goals, like ``maximize X across all of space and time.'' \emph{Some} strategically-aware human planners are like this, we might think, but not all of them: so strategically aware, agentic planning isn't, itself, the problem.

I think there is an important point in this vicinity: namely, that power-seeking behavior, \emph{in practice}, arises not just due to strategically-aware agentic planning, but due to the specific interaction between an agent's capabilities, objectives, and circumstances. But I don't think this undermines the posited instrumental connection between strategically-aware agentic planning and power-seeking in general. Humans may not seek various types of power \emph{in their current circumstances}---in which, for example, their capabilities are roughly similar to those of their peers, they are subject to various social/legal incentives and physical/temporal constraints, and in which many forms of power-seeking would violate ethical constraints they treat as intrinsically important. But almost all humans will seek to gain and maintain various types of power in some circumstances, and especially to the extent they have the capabilities and opportunities to get, use, and maintain that power with comparatively little cost. Thus, for most humans, it makes little sense to devote themselves to starting a billion dollar company---the returns to such effort are too low. But most humans will walk across the street to pick up a billion dollar check.

Put more broadly: the power-seeking behavior humans display, when getting power is easy, seems to me quite compatible with the instrumental convergence thesis. And unchecked by ethics, constraints, and incentives (indeed, even \emph{when} checked by these things) human power-seeking seems to me plenty dangerous, too. That said, the absence of various forms of overt power-seeking in humans may point to ways we could try to maintain control over less-than-fully PS-aligned APS systems (see 4.3 for more).

A second objection (in possible tension with the first) is: \emph{humans} (or, some humans) may be power-seeking, but this is a product of a specific evolutionary history (namely, one in which power-seeking was often directly selected for), which AI systems will not share.\footnote{See e.g. \href{https://blogs.scientificamerican.com/observations/dont-fear-the-terminator/}{Zador and LeCun (2019)} (and follow-up debate \href{https://www.lesswrong.com/posts/WxW6Gc6f2z3mzmqKs/debate-on-instrumental-convergence-between-lecun-russell}{here}), and \href{https://www.amazon.com/Enlightenment-Now-Science-Humanism-Progress/dp/0525427570}{Pinker (2018, Chapter 19)}. One can also imagine non-evolutionary versions of this---e.g., ones that attribute human power-seeking tendencies to our culture, our economic system, and so forth. Indeed, \href{https://idlewords.com/talks/superintelligence.htm}{Ceglowski (2016)} can be read as suggesting something like this in the context of a particular demographic: after listing Bostrom's convergent instrumental goals, he writes: ``If you look at AI believers in Silicon Valley, this is the quasi-sociopathic checklist they themselves seem to be working from.''} Some versions of this objection simply neglect to address the instrumental convergence argument above\footnote{That is, the argument isn't ``humans seek power, therefore AIs will too''; it's ``power is useful for pursuing objectives, so AIs pursuing problematic objectives will have incentives to seek power, by default.''} (and note, regardless, that some proposed ways of training AI systems resemble evolution in various respects).\footnote{Large, multi-agent reinforcement learning environments might be one example. And the ``league'' used to train \href{https://deepmind.com/blog/article/alphastar-mastering-real-time-strategy-game-starcraft-ii}{AlphaStar} seems reminiscent of evolution in various ways.} But we can see stronger versions as pointing at a possibility similar to the one I mentioned above: namely, perhaps it just isn't that hard to train APS systems not to seek power in unintended ways, across a large enough range of inputs, if you're actively \emph{trying} to do so (evolution wasn't). Again, I discuss possibilities and difficulties in this regard in the next section.

Ultimately, I'm sympathetic to the idea that we should expect, by default, to see incentives towards power-seeking reflected in the behavior of systems that engage in strategically aware agentic planning in pursuit of problematic objectives. However, this part of the overall argument is also one of my top candidates for ways that the abstractions employed might mislead.

In particular, it requires the agentic planning and strategic awareness at stake be robust enough to license predictions of the form: ``if (a) a system would be planning in pursuit of problematic objectives in circumstance C, (b) power-seeking in C would promote its objectives, and (c) the models it uses in planning put it in a position to recognize this, then we should expect power-seeking in C by default.'' I've tried to build something like the validity of such predictions into my definitions of agentic planning and strategic awareness; but perhaps for sufficiently weak/loose versions of those concepts, such predictions are not warranted; and it seems possible to conflate weaker vs. stronger concepts at different points in one's reasoning, and/or to apply such concepts in contexts where they confuse rather than clarify.\footnote{Thanks to Rohin Shah for emphasizing possible objections in this vein, and for discussion.}

Indeed, to avoid confusions in this vein, we might hope to jettison such concepts altogether.\footnote{See \href{https://www.lesswrong.com/posts/zCcmJzbenAXu6qugS/tabooing-agent-for-prosaic-alignment}{Wijk (2019)} for one effort.} And it's possible to formulate arguments for something like instrumental convergence without them (see footnote for details).\footnote{That is, we can say things like: ``This system behaves in a manner that promotes outcomes of a certain type in a wide range of complex circumstances. Circumstance C is one to which we should expect this behavior to generalize, and in circumstance C, power-seeking will promote outcomes of type X. Therefore, we should expect power-seeking in C'' (thanks to Paul Christiano for discussion). Here, a lot of work is done by ``circumstance C is one to which we should expect this behavior to generalize''---but I expect this to be plausible in the context of system optimized across a sufficient range of circumstances. That said, if the system was optimized, in those circumstances, to cause X \emph{without (observable) misaligned power-seeking}, the case for expecting power-seeking in C seems weaker. So the question looms large: how much can selecting against observed power-seeking, in creating a system that causes X in many circumstances, eliminate behavior responsive to power-seeking's usefulness for causing X (and many other types of outcome) across all physics-compatible inputs? And this question seems similar to the questions raised above (and discussed more in the next section) about the ease of selecting, in training, against forms of agentic planning that license misaligned power-seeking (and note, too, that one salient way in which training a system to cause X might lead to power-seeking behavior is if a suitable generalized ability to cause X tends to involve or require something like agentic-planning and strategic awareness).} But the reasoning in play seems to me at least somewhat different.

\subsection{The challenge of practical PS-alignment}\label{the-challenge-of-practical-ps-alignment}

Let's grant that less-than-fully aligned APS systems will have at least some tendency towards misaligned, power-seeking behavior, by default. The challenge, then, is to prevent such behavior in practice---whether through alignment (including full alignment), or other means. How difficult will this be?

I'll break down the challenge as follows:

\begin{enumerate}
\def\labelenumi{\arabic{enumi}.}
\tightlist
\item
  Designers need to cause the APS system to be such that the objectives it pursues on some set of inputs X do not give rise to misaligned power-seeking.
\item
  They (and other decision-makers) need to restrict the inputs the system receives to X.
\end{enumerate}

The larger the set of inputs X where 1 succeeds, the less reliance on 2 is required (and in the limit of full PS-alignment, 2 plays no role at all).

I'll consider a variety of ways one might try to control a system's objectives, capabilities, and circumstances to hit the right combination of 1 and 2. Some of these (or some mix) might well work. But all, I think, face problems.

\subsubsection{Controlling objectives}\label{controlling-objectives}

Much work on technical alignment focuses on controlling the objectives the AI systems we build pursue. I'll start there.

At any given point in the progress of AI technology, we will have some particular level of control in this respect, grounded in currently available techniques. Thus, for example, some methods of AI development require coding objectives by hand. Others (more prominent at present) shape an AI's objectives via different sorts of training signals---whether generated algorithmically, or using human feedback/demonstration. And we can imagine future methods that allows us to e.g. control an AI's objectives by editing the weights of a neural network directly; or to cause a system to pursue (for its own sake) any objective that we can articulate using English-language sentences that will be accurately, common-sensically, and charitably construed.

The challenge of (1) is to use what methods in this respect are available to ensure PS-alignment on all inputs in X. But note that because available methods change, (1) is a moving target: new techniques and capabilities open up new options. I emphasize this because sometimes the challenge of AI alignment is framed as one of shaping an AI's objectives \emph{in a particular way}---for example, via hand-written code, or via some sort of reward signal, or via English-language sentences that will be interpreted in literalistic and uncharitable terms. And this can make it seem like the challenge is centrally one of, e.g., coding, measuring, or articulating explicitly everything we value, or getting AI systems to interpret instructions in common-sensical ways. These challenges may be relevant in some cases, but the core problem is not method-specific.

\paragraph{Problems with proxies}\label{problems-with-proxies}

That said, many ways of attempting to control an AI's objectives share a common challenge: namely, that giving an AI system a ``proxy objective''---that is, an objective that reflects properties correlated with, but separable from, intended behavior---can result in behavior that weakens or breaks that correlation, especially as the power of the AI's optimization for the proxy increases.

For example: the behavior I want from an AI, and the behavior I would rate highly using some type of feedback, are well-correlated when I can monitor and understand the behavior in question. But if the AI is too sophisticated for me to understand everything that it's doing, and/or if it can deceive me about its actions, the correlation weakens: the AI may be able to cause me to give high ratings to behavior I wouldn't (in my current state) endorse if I understood it better---for example, by hiding information about that behavior, or by manipulating my preferences.\footnote{Though what sorts of misaligned power-seeking in particular these behaviors would give rise to is a further question.}

This general problem---that optimizing for a proxy correlated with but not identical to some intended outcome can break the correlation in question---is familiar from human contexts.\footnote{Thus, paying railroad builders by the mile of track that they lay incentivizes them to lay unnecessary track (from Wikipedia, \href{https://en.wikipedia.org/wiki/Perverse_incentive\#Further_historic_examples}{here}); trying to lower the cobra population by paying people to turn in dead cobras leads to people breeding cobras (from Wikipedia, \href{https://en.wikipedia.org/wiki/Perverse_incentive\#The_original_cobra_effect}{here}); if teachers take ``cause my students to get high scores on standardized tests'' as their objective, they're incentivized to ``\href{https://en.wikipedia.org/wiki/Teaching_to_the_test}{teach to the test}''---an incentive that can work to the detriment of student education more broadly; and so on (see \href{https://en.wikipedia.org/wiki/Perverse_incentive}{here} for more examples). See \href{https://arxiv.org/abs/1803.04585}{Manheim and Garrabrant (2018)} for an abstract categorization of dynamics of this kind.} And we already see analogous problems in existing AI systems. Thus:

\begin{itemize}
\tightlist
\item
  If we train an AI system to complete a \href{https://openai.com/blog/faulty-reward-functions/}{boat race} by rewarding it for hitting green blocks along the way, it learns to drive the boat in circles hitting the same blocks over and over.
\item
  If we train an AI system on human feedback on a \href{https://deepmind.com/blog/article/Specification-gaming-the-flip-side-of-AI-ingenuity}{grasping task}, it learns to move its hand in a way that looks to the human like it's grasping the object, even though it isn't.
\item
  If we reward an AI system for \href{https://medium.com/@deepmindsafetyresearch/realab-conceptualising-the-tampering-problem-56caab69b6d3}{picking up apples in a simulated environment}, but in a manner that depends on the locations of certain blocks, the agent learns to tamper with the blocks.
\end{itemize}

See \href{https://docs.google.com/spreadsheets/d/e/2PACX-1vRPiprOaC3HsCf5Tuum8bRfzYUiKLRqJmbOoC-32JorNdfyTiRRsR7Ea5eWtvsWzuxo8bjOxCG84dAg/pubhtml}{here} for a much longer list of examples in this vein.

These examples may seem easy to fix---and indeed, in the context of fairly weak systems, on a limited range of inputs, they generally are. But they illustrate the broader problem: systems optimizing for imperfect proxies often behave in unintended ways.

Indeed, this tendency is closely connected to a core property that makes advanced AI useful: namely, the ability to find novel solutions and strategies that humans wouldn't think of.\footnote{\href{https://deepmind.com/blog/article/Specification-gaming-the-flip-side-of-AI-ingenuity}{Krakovna et al (2020)} make this point well.} When you don't know how an AI will achieve its objective, and that objective doesn't capture everything that you really want, then even for comparatively weak systems and simple tasks, it's hard to anticipate how the system's way of achieving the objective will break its correlation with what you really want. And as the AI's capacity to generate solutions we can't anticipate grows, the problem becomes more and more challenging.

We can see a variety of techniques for controlling an AI system's objectives as mediated by some kind of ``proxy'' or other. Thus: hand-coded objectives, simple metrics (clicks, profits, likes), algorithmically generated training signals, human-generated data/feedback, and English-language sentences can all be seen as information structures that shape/serve as the basis for the AI's optimization, but which can also fail to contain the information necessary to result in intended behavior across inputs. The challenge is to find a technique adequate in this respect.

Human feedback seems likely to play a key role here.\footnote{This paragraph draws heavily on discussion in \href{https://www.alignmentforum.org/posts/PvA2gFMAaHCHfMXrw/agi-safety-from-first-principles-alignment}{Ngo (2020)}.} And it may, ultimately, be enough. But notably, we need ways of drawing on this feedback that don't require unrealistic amounts of human supervision and human-generated data;\footnote{See e.g. \href{https://arxiv.org/abs/1706.03741}{Christiano et al (2017)}.} we need to ensure that such feedback captures our preferences about behavior that we can't directly understand and/or whose consequences we haven't yet seen;\footnote{See \href{https://openai.com/blog/amplifying-ai-training/}{Christiano and Amodei (2018)} for discussion. \href{https://arxiv.org/abs/1810.08575}{Iterative amplification and distillation}; \href{https://arxiv.org/abs/1805.00899}{debate}; and \href{https://medium.com/@deepmindsafetyresearch/scalable-agent-alignment-via-reward-modeling-bf4ab06dfd84}{recursive reward modeling} can all be seen as efforts in this vein.} we need ways of eliminating incentives to manipulate or mislead the human feedback mechanisms in question; and we need such methods to scale competitively as frontier AI capabilities increase.\footnote{See 4.3.2.3 for more on scaling, and 5.3.1 for more on competition.}

Would it help if our AI systems could understand fuzzy human concepts like ``helpfulness,'' ``obedience,'' ``what humans would want,'' and so forth? I expect it would, in various ways (though as I discuss below, this also opens up new opportunities for deception/manipulation). But note that the key issue isn't getting our AI systems to \emph{understand} what objectives we want them to pursue---indeed, such understanding is plausibly on the critical path to increasing their capability, regardless of their alignment. Rather, the key issue is causing them to \emph{pursue} those objectives for their own sake (though if they understand those objectives, but don't share them, we might also be able to \emph{incentivize} them to pursue such objectives for instrumental reasons).\footnote{This is a point from \href{https://www.amazon.com/Precipice-Existential-Risk-Future-Humanity/dp/031648492X/ref=sr_1_2?crid=2ZWCCI74ZFX55\&dchild=1\&keywords=precipice+existential+risk+and+the+future+of+humanity\&qid=1619197698\&s=books\&sprefix=precipice\%2Cstripbooks\%2C243\&sr=1-2}{Ord (2020)}. For example, if we train some set of sophisticated agents to get bananas, in a complex environment that requires understanding and modeling humans, they may end up capable of understanding quite accurately (even more accurately than us) what we have in mind when we talk about ``aligned behavior,'' and of behaving accordingly (for example, when we give them bananas for doing so). But their intrinsic objectives could still be focused centrally on bananas (or something else), and our abilities to control those objectives directly might remain quite limited.}

\paragraph{Problems with search}\label{problems-with-search}

Many techniques shape an AI's objectives using proxies of one form another. But some---namely, those that involve \emph{searching} over and selecting AI systems that perform well on some evaluation criteria, without controlling their objectives directly---have an additional problem: namely, even \textit{if} those criteria fully capture the behavior we want, the resulting systems may not end up intrinsically motivated by the criteria in question.\footnote{My discussion in this section is centrally inspired by the discussion in \href{https://arxiv.org/pdf/1906.01820.pdf}{Hubinger et al (2019)}, though I don't assume their specific set-up.} Rather, they may end up with other objectives, pursuit of which correlated with good performance during the selection process, but which lead to unintended behavior on other inputs.

Some think of human evolution as an example.\footnote{See e.g. \href{https://arxiv.org/pdf/1906.01820.pdf}{Hubinger et al (2019, p. 6)}.} Someone interested in creating agents who pass on their genes to the next generation could run a training process similar to evolution, which searches over different agents, and selects for ones who pass on their genes (for example, by allowing ones who don't to die out). But this doesn't mean the resulting agents will be intrinsically motivated to pass on their genes. Humans, for example, are motivated by objectives that were \emph{correlated} with passing on genes (for example, avoiding bodily harm, having sex, securing social status, etc), but which they'll pursue in a manner that breaks such correlations, given the opportunity (for example, by using birth control, or remaining childless to further their careers).

Rudimentary, evolved AI systems display analogous tendencies. Thus, when Ackley and Littman ran an evolutionary selection process in an environment with trees that allowed agents to hide from predators, the agents developed such a strong attraction to trees that (after reproductive age) they would starve to death in order to avoid leaving tree areas (what Ackley called ``tree senility'').\footnote{See \href{https://www.amazon.com/Alignment-Problem-Machine-Learning-Values/dp/B085DTXC59/ref=sr_1_1?dchild=1&keywords=alignment+problem&qid=1619198396&s=books&sr=1-1}{Christian (2020)}, quote \href{https://www.lesswrong.com/posts/gYfgWSxCpFdk2cZfE/the-alignment-problem-machine-learning-and-human-values}{here}. I'm going off of Christian's description, here, and haven't actually investigated these experiments.}

And we can imagine other cases, with less evolution-like techniques. Suppose, for example, that we use gradient descent to train an AI system to reach the exit of a maze.\footnote{This is an example from \href{https://futureoflife.org/2020/07/01/evan-hubinger-on-inner-alignment-outer-alignment-and-proposals-for-building-safe-advanced-ai/}{Hubinger (2020)}.} If the exit was marked by a green arrow on all the training data, the system could learn the objective ``find the green arrow'' rather than ``find the exit to the maze.'' If we then give it a maze where the green arrow \emph{isn't} by the exit, it will search out the green arrow instead.

Note that in both the evolution and the maze cases, we can imagine that the evaluation criteria (``pass on genes,'' ``exit maze'') fully capture and operationalize the intended behavior. Still, the designers lack the required degree of control over the objectives of the agents they create.

How often will problems like this arise? It's an open empirical question. But some considerations make the possibility salient. Namely:

\begin{itemize}
\item
  Proxy goals correlated with the evaluation criteria may be simpler and therefore easier to learn, especially if the evaluation criteria are complex. In the context of evolution, for example, it seems much harder to evolve an agent whose mind represents a concept like ``passing on my genes,'' and then takes doing this as its explicit goal---humans, after all, didn't even have the concept of ``genes'' until very recently---than to evolve an agent whose objectives reflect the relevance of things like bodily damage, sex, power, knowledge, etc to whether its genes get passed on (though starting with cognitively sophisticated agents might help in this respect).\footnote{This is a point I believe I heard Evan Hubinger make on a podcast (either \href{https://www.lesswrong.com/posts/EszCTbovFfpJd5C8N/axrp-episode-4-risks-from-learned-optimization-with-evan}{this one}, or \href{https://futureoflife.org/2020/07/01/evan-hubinger-on-inner-alignment-outer-alignment-and-proposals-for-building-safe-advanced-ai/}{this one}).}
\item
  Relatedly: if the ``true'' objective function provides slower feedback, agents that pursue faster-feedback proxies have advantages. For example: in the game Montezuma's Revenge, it helps to give an agent a direct incentive analogous to ``curiosity'' (e.g., it receives reward for finding sensory data it can't predict very well), because the game's ``true'' objective (e.g., exiting a level by finding keys that require a large number of correct sequential steps to reach) is too difficult to train on.\footnote{See \href{https://openai.com/blog/reinforcement-learning-with-prediction-based-rewards/}{Burda and Edwards (2018)}, and discussion in \href{https://www.amazon.com/Alignment-Problem-Machine-Learning-Values/dp/B085DTXC59/ref=sr_1_1?dchild=1&keywords=alignment+problem&qid=1619198396&s=books&sr=1-1}{Christian (2020)}.}
\item
  To the extent that many objectives would \textit{instrumentally} incentivize good behavior in training (for example, because many objectives, when coupled with strategic awareness, incentivize gaining power in the world, and doing well in training leads to deployment/greater power in the world), but few involve \emph{intrinsic} motivation to engage in such behavior, we might think it more likely that selecting for good behavior leads to agents who behave well for instrumental reasons.\footnote{Thanks to Carl Shulman for discussion. See also \href{https://www.lesswrong.com/posts/HBxe6wdjxK239zajf/what-failure-looks-like\#Part_II__influence_seeking_behavior_is_scary}{Christiano (2018)}: ``One reason to be scared is that a wide variety of goals could lead to influence-seeking behavior, while the `intended' goal of a system is a narrower target, so we might expect influence-seeking behavior to be more common in the broader landscape of `possible cognitive policies.'\,''} That said, it could also be that the agents who perform best according to some criteria, especially once they're sophisticated enough to understand what those criteria are, are the ones who are intrinsically motivated by those criteria. And even if such agents aren't selected for automatically, various techniques might help detect and address problematic cases. Notably, for example, we might actively search for inputs that will reveal problematic objectives,\footnote{See Christiano's discussion of ``adversarial training'' \href{https://ai-alignment.com/training-robust-corrigibility-ce0e0a3b9b4d}{here}.} and we might learn how to read off a system's objectives from its internal states.\footnote{This is closely related to work on ``interpretability'' and ``transparency''---see, e.g., \href{https://distill.pub/2020/circuits/zoom-in/}{Olah (2020)} and related work.}
\end{itemize}

Overall, though, ensuring robust forms of practical PS-alignment seems harder if available techniques search over systems that meet some external evaluation criteria, with little direct control over their objectives. And much of contemporary machine learning fits this bill.

\paragraph{Myopia}\label{myopia}

Some broad types of objectives seem to incentivize power-seeking on fewer physics-compatible inputs than others. Perhaps, then, we can aim at those, even if we lack more fine-grained control.

Short-term (or, ``myopic'') objectives seem especially interesting here. The most paradigmatically dangerous types of AI systems plan strategically in pursuit of long-term objectives, since longer time horizons leave more time to gain and use forms of power humans aren't making readily available, they more easily justify strategic but temporarily costly action (for example, trying to appear adequately aligned, in order to get deployed) aimed at such power.\footnote{Non-myopic agents also seem more likely to want to ``keep'' power for a very long time, whereas myopic agents are more likely to ``give up'' a given type of power once they are ``done with it.''} Myopic agentic planners, by contrast, are on a much tighter schedule, and they have consequently weaker incentives to attempt forms of misaligned deception, resource-acquisition, etc that only pay off in the long-run (though even short spans of time can be enough to do a lot of harm, especially for extremely capable systems---and the timespans ``short enough to be safe'' can alter if what one can do in a given span of time changes).\footnote{Thanks to Rohin Shah, Paul Christiano, and Carl Shulman for discussion. And note that a given operationalization of ``time'' can itself be vulnerable to various forms of manipulation (an AI could, for example, find ways to stop and start its internal clock). Thanks to Carl Shulman for suggesting this possibility.}

I think myopia might well help. But I see at least two problems with relying on it:

\begin{itemize}
\tightlist
\item
  Plausibly, there will be demand for non-myopic agents. Humans (and human institutions) have long-term objectives, and will likely want long-term tasks---running factories and companies, managing scientific experiments, pursuing political outcomes---automated. Of course, myopic and/or non-APS systems can perform sub-tasks (including sub-tasks that involve generating long-term plans), and humans can stay in the loop; but as discussed \protect\hyperref[usefulness]{section 3.1}, there will plausibly be competitive pressures towards automating our pursuit of long-term objectives more and more fully.
\item
  The ``search'' techniques discussed in the previous section may make ensuring myopia challenging. And various types of long-term training processes---for example, reinforcement learning on tasks that involve many sequential steps---seem likely to result in non-myopia by default (that said, myopia is a fairly coarse-grained property for an objective to possess, and may be easier to cause/check for than others).
\end{itemize}

We can also imagine other ways of attempting to shrink (even if not to zero) the set of physics-compatible inputs on which an APS system engages in PS-misaligned behavior. For example: we might aim for objectives that penalize ``high-impact'' action,\footnote{See, e.g., the discussion of ``impact penalties'' in \href{https://medium.com/@deepmindsafetyresearch/designing-agent-incentives-to-avoid-side-effects-e1ac80ea6107}{Krakovna et al (2019)}.} or that prohibit lying in particular, or that give intrinsic weight to various legal and ethical constraints, or that benefit less from marginal resources. But these face the same challenges re: proxies and search discussed in the last two sections.

\subsubsection{Controlling capabilities}\label{controlling-capabilities}

AI alignment research typically focuses on controlling a system's objectives. But controlling its capabilities can play a role in practical PS-alignment, too.

In particular: the less capable a system, the more easily its behavior---including its tendencies to misaligned power-seeking---can be anticipated and corrected. Less capable systems will also have a harder time \emph{getting} and \emph{keeping} power, and a harder time making use of it, so they will have stronger incentives to cooperate with humans (rather than trying to e.g. deceive or overpower them), and to make do with the power and opportunities that humans provide them by default.

Preventing agentic planning and strategic awareness in the first place would be one example of ``controlling capabilities'' (see \protect\hyperref[incentives]{section 3}); but there are other options, too.

\paragraph{Specialization}\label{specialization}

In particular, we might focus on building APS systems whose competence is as narrow and specialized as possible (though they are still, by hypothesis, agentic planners with strategic awareness).

Discussion of AI risk often assumes the relevant systems are very ``general''---e.g., individually capable of performing (or learning to perform) a very wide variety of tasks. But automating a wide variety of tasks doesn't require creating a single AI system that can perform (or learn to perform) all of them. Rather, we can create different, specialized systems for each---and since such systems are less capable, the project of practical PS-alignment may be easier, and lower-stakes. An APS system skilled at a specific kind of scientific research, for example, but not at e.g. hacking, social persuasion, investing, and military strategy, seems much less dangerous---but it seems comparably useful for curing cancer. And even if a system's competencies are broad, we can imagine that its strategically-aware agentic planning is only operative on some narrow set of inputs.

Indeed, specialized systems have many benefits.\footnote{Here, and in the bulleted list of ``benefits of specialization'' below, I'm drawing on a list of benefits in an unpublished document by Ben Garfinkel. See also \href{https://www.fhi.ox.ac.uk/wp-content/uploads/Reframing_Superintelligence_FHI-TR-2019-1.1-1.pdf}{Drexler (2019)} for more discussion of the value of specialized systems.} For example, they can be optimized more heavily for specific functions (to borrow an example from Ben Garfinkel, there is a reason that the flashlight, camera, speakers, etc on an iPhone are inferior to the best flashlights, cameras, etc). And we see incentives towards specialization and \href{https://en.wikipedia.org/wiki/Division_of_labour}{division of labor} in human economies and organizations as well. What's more, we will likely have much greater abilities to optimize AI systems for particular tasks than we do with humans.

That said, generality has benefits, too. In particular:

\begin{itemize}
\item
  Human workers with quite general skill-sets---CEOs, generals, Navy Seals, researchers with a broad knowledge of many domains, flexibly competent personal assistants---are prized in various contexts (even while specialization is prized in others). Automated systems need not be human-like in this respect (farmers, too, have quite general skill sets, but automated agriculture need not involve ``farmer bots''), but it seems suggestive, at least, of economically-relevant environments in which general competence is useful.
\item
  Specialized systems may be worse at responding flexibly to changing environments and task-requirements (e.g., it's helpful not to have to buy new robots every time you redesign the factory or change the product being produced).\footnote{Thanks to Carl Shulman for suggesting this example.}
\item
  Multiple specialized systems can be less efficient to store and create (there is a reason you carry around an iPhone, rather than separate flashlights, cameras, microphones, etc);
\item
  If a task requires multiple competencies, specialized systems can be harder to coordinate (e.g., it's helpful to have a single personal assistant, rather than one for email, one for scheduling, one for travel planning, one for research, etc). And a suitably coordinated set of specialized systems can end up acting as a quite general and agentic system.
\end{itemize}

What's more, just as available techniques may push the field towards agentic planning and strategic awareness (see \hyperref[available-techniques]{section 3.2}), so too might they push towards generality.\footnote{This paragraph draws centrally on the discussion in \href{https://www.lesswrong.com/s/mzgtmmTKKn5MuCzFJ/p/eG3WhHS8CLNxuH6rT}{Ngo (2020)}.} GPT-3, for example, is trained to a fairly general level capability via predicting text, and then later fine-tuned on specific tasks like coding. Indeed, such an approach might be necessary for tasks where data is too hard to come by or learn from directly (consider, for example, tasks like ``designing a good railway system'' or ``be an effective CEO''); and more broadly, the most efficient route to wide-spread automation may be the creation of general-purpose agents that can learn \textasciitilde{}arbitrary new tasks very efficiently (though those agents could also end up quite specialized later).\footnote{This is a point from \href{https://www.amazon.com/Superintelligence-Dangers-Strategies-Nick-Bostrom/dp/1501227742}{Bostrom (2014)}.}

And note, too, that even specialized APS systems can be very dangerous. A system highly skilled at hacking into new computers and copying itself, for example, can spread far and wide; a system skilled in science can design a novel virus; a system with control over automated weapons can use them; a system skilled at social manipulation can turn an election; and so forth. Indeed, this is part of why I focused on ``advanced capabilities'' in 2.1.1, rather than something like ``AGI'' or ``superintelligence.''

\paragraph{Preventing problematic improvements}\label{preventing-problematic-improvements}

New capabilities can put a system in a position to gain and maintain power in ways it couldn't before---and hence, make new incentives action-relevant (if Bob learns how to hack into bank accounts, for example, his likelihood of considering and executing plans that involve such hacking will change). Practical PS-alignment may therefore require controlling the extent to which the inputs a system receives result in improved capabilities.

This seems easier if the variables in the system that determine how it responds to inputs (for example, the weights in a neural network) stay fixed. But we may also want systems that mix task-performance and learning together, that ``remember'' previous events, and so forth; and predicting and controlling the capabilities such systems will develop could be difficult (especially if we don't understand well how they work---see 4.4.1).

Note, though, that this is a narrower challenge than making sure a system's PS-alignment is robust to \textit{any} increase in capabilities---including, for example, increases that result from interventions other than exposure to physics-compatible inputs. Ultimately, we need to make sure that a system isn't exposed to non-input interventions that cause it (or, a new version of it) to seek power in misaligned ways, too; and efforts by designers to scale up a system's cognitive resources, training, and so forth will need to grapple with challenges in this vein. But as I discussed in 4.1, this sort of robustness is not a requirement for PS-alignment in my sense.

\paragraph{Scaling}\label{scaling}

Strategies for practical PS-alignment that rely on limiting a system's capabilities face a general problem: namely, that there are likely to be strong incentives to scale up the capabilities of frontier systems.\footnote{Though note that PS-alignment problems with more capable systems could complicate these dynamics---see 5.3.3 for more.} PS-alignment strategies that can't scale accordingly (and competitively) therefore risk obsolescence as state of the art capabilities advance.

A key question for any such strategy, then, is whether it can translate, given success at some level of capability, into a different strategy that scales better. For example, we might try to achieve practical PS-alignment with some fairly advanced systems (including, perhaps, quite specialized ones---or, indeed, non-APS ones), and then use them to create new and superior PS-alignment strategies (indeed, as AI development itself becomes increasingly automated, automating alignment research will plausibly be necessary regardless).

But note that plans of the form ``create some practically PS-aligned systems, and ask them what the plan should be'' might just not work. For example, the new systems might not have adequate plans either. One might therefore need to create even more capable systems, which \emph{also} might not have adequate plans, and so forth, until one pushes up against (or perhaps, past) the limits of one's capacity to ensure practical PS-alignment.

\subsubsection{Controlling circumstances}\label{controlling-circumstances}

So far in section 4.3, I've been talking about controlling ``internal'' properties of an APS system: namely, its objectives and capabilities. But we can control external circumstances, too---and in particular, the type of options and incentives a system faces.

Controlling options means controlling what a circumstance makes it possible for a system to do, even if it tried. Thus, using a computer without internet access might prevent certain types of hacking; a factory robot may not be able to access to the outside world; and so forth.

Controlling incentives, by contrast, means controlling which options it makes sense to choose, given some set of objectives. Thus, perhaps an AI system could impersonate a human, or lie; but if it knows that it will be caught, and that being caught would be costly to its objectives, it might refrain. Or perhaps a system will receive more of a certain kind of reward for cooperating with humans, even though options for misaligned power-seeking are open.

Human society relies heavily on controlling the options and incentives of agents with imperfectly aligned objectives. Thus: suppose I seek money for myself, and Bob seeks money for Bob. This need not be a problem when I hire Bob as a contractor. Rather: I pay him for his work; I don't give him access to the company bank account; and various social and legal factors reduce his incentives to try to steal from me, even if he could.

A variety of similar strategies will plausibly be available and important with APS systems, too. Note, though, that Bob's capabilities matter a lot, here. If he was better at hacking, my efforts to avoid giving him the option of accessing the company bank account might (unbeknownst to me) fail. If he was better at avoiding detection, his incentives not to steal might change; and so forth.

PS-alignment strategies that rely on controlling options and incentives therefore require ways of exerting this control (e.g., mechanisms of security, monitoring, enforcement, etc) that scale with the capabilities of frontier APS systems. Note, though, that we need not rely solely on \emph{human} abilities in this respect. For example, we might be able to use various non-APS systems and/or practically-aligned APS systems to help.\footnote{One could even imagine trying to use practically PS-\textit{mis}aligned systems, though this seems dicey.}

One other note: ensuring practical PS-alignment in a deployed system seems easier the more similar its deployment circumstances to the ones on which humans have observed and verified PS-aligned behavior (for example, during training, or pre-deployment testing). Indeed, ideally, one would want the deployment inputs to come from the same distribution as the training inputs. But in practice, and especially in strategically-aware systems, ensuring a close-to-identical distribution seems very difficult (if not impossible). This is partly because the world changes (and indeed, the actions of the APS system can themselves change it).\footnote{See \href{https://arxiv.org/abs/2009.09153}{Krueger et al (2020)} and \href{https://www.amazon.com/Alignment-Problem-Machine-Learning-Values/dp/B085DTXC59/ref=sr_1_1?dchild=1&keywords=alignment+problem&qid=1619198396&s=books&sr=1-1}{Christian (2020)} for some discussion of this possibility.} But also, to the extent that the distinction between training and deployment reflects some real difference in the agent's level of influence on the world, this difference is itself a change in distribution---one that a sufficiently sophisticated agent might recognize.

\subsection{Unusual difficulties}\label{unusual-difficulties}

From our current vantage point, ensuring PS-aligned behavior from APS systems across a wide range of inputs seems, to me, like it could well be difficult. But so, too, does building any kind of APS system appear difficult. Is there reason to think that by the time we figure out how to do the latter, we won't have figured out the former as well?

It's generally easier to create technology that fulfills some function F, than to create technology that does F \emph{and} meets a given standard of safety and reliability, for the simple reason that meeting the relevant standard is an additional property, requiring additional effort.\footnote{See \href{https://www.ted.com/talks/nick_bostrom_what_happens_when_our_computers_get_smarter_than_we_are/transcript}{Bostrom (2015)}: ``Making superintelligent A.I. is a really hard challenge. Making superintelligent A.I. that is safe involves some additional challenge on top of that. The risk is that if somebody figures out how to crack the first challenge without also having cracked the additional challenge of ensuring perfect safety.''} Thus, it's easier to build a plane that can fly at all than one that can fly safely and reliably in many conditions; easier to build an email client than a secure email client; easier to figure out how to cause a nuclear chain reaction than how to build a safe nuclear reactor; and so forth.

Of course, we often reach adequate safety standards in the end. But at the very least, we expect some safety problems along the way (plane crashes, compromised email accounts, etc). We might expect something similar with ensuring PS-aligned behavior from powerful AI agents.

But ensuring such behavior also poses a number of challenges that (most) other technologies don't. Here are a few salient to me.

\subsubsection{Barriers to understanding}\label{barriers-to-understanding}

Ensuring safety and reliability requires understanding a system well enough to predict its behavior. But plausibly, this is uniquely challenging in the context of a strategically aware agentic planner whose cognitive capabilities significantly exceed those of humans in a given domain. That is, the thinking and strategic decision-making of such an agent will likely reach a quite alien and opaque level of sophistication, which humans may be poorly positioned to anticipate and understand at the level required for ensuring PS-alignment. For example, it may consider many options humans never would; it may understand physical and social dynamics that humans do not; and so forth.\footnote{See Yudkowksy on ``\href{https://arbital.com/p/strong_uncontainability/}{strong cognitive uncontainability}.''}

This issue seems especially salient in the current, machine-learning dominated AI paradigm, in which our ability to create an AI system that can perform some task (e.g., predicting text) often far exceeds our ability to understand \emph{how} the system does what it does. We set various key high-level variables (the system's architecture, the number of parameters, the training process, the evaluation criteria), but the system that results is still, in many (though not all) respects, a black box. We must rely on further experiments to try to get some handle on what it knows, what it can do, and how it is liable to behave.\footnote{See \href{https://www.lesswrong.com/posts/r3NHPD3dLFNk9QE2Y/search-versus-design-1}{Flint (2020)} on ``search vs. design'' for related discussion.}

If this lack of understanding of the systems we're building persists (for example, if we end up creating APS systems by training very large machine learning models on complex tasks, using techniques fairly similar to those used today), this could be an important and safety-relevant difference between AI systems and other types of technology. That is, we achieve high degrees of reliability and safety with technology like bridges, planes, rockets, and so forth in part via an understanding of the physical principles that govern the behavior of those systems; and we design them, part by part, in a manner that reflects and responds to those principles. This allows us to understand and predict their behavior in a wide range of circumstances. Searching over opaque/poorly-understood AI systems allows no such advantage.

Of course, our understanding of how ML systems work will likely improve over time---indeed, active research in this area (sometimes called ``interpretability'') is ongoing.\footnote{See e.g. \href{https://distill.pub/2020/circuits/zoom-in/}{Olah (2020)} and \href{https://distill.pub/2021/multimodal-neurons/}{Goh et al (2021)}.} But interpretability is no bottleneck to training bigger models on more complex tasks---or, plausibly, to the commercial viability of such models. And even as some researchers work on it, much of the field's effort focuses on pushing forward with developing whatever capabilities we can, interpretable or no.

That said, understanding comes in many varieties and degrees; and it's an empirical question what mix of experiment/search vs. first-principles understanding/design has actually been involved in ensuring the safety of different technologies (for example, in biology, or before advanced scientific understanding).\footnote{And defining these distinctions may prove challenging.} I expect that investigation of case studies in this respect would prove revealing.

Ultimately, though, ``mechanistic understanding'' isn't the fundamental issue. Even if we had such understanding on some level, the basic difficulty---that is, understanding and predicting the behavior of an agent much more cognitively sophisticated than humans---might well persist.

\subsubsection{Adversarial dynamics}\label{adversarial-dynamics}

Part of the challenge of ensuring PS-alignment is \textit{detecting} problems with an APS system's objectives---for example, via various processes of testing and evaluation. However, if a system behaves in PS-misaligned ways in some circumstances, those circumstances could well include the very processes of testing and evaluation we're trying to use to detect problems. Thus, for example, if a strategically-aware AI system has problematic objectives that it could better achieve if it passed some training/evaluation process, it may have incentives to try to manipulate that process, and/or deceive relevant humans about its degree of alignment.\footnote{\href{https://www.amazon.com/Superintelligence-Dangers-Strategies-Nick-Bostrom/dp/1501227742}{Bostrom (2014)} calls this a ``treacherous turn.''} And if it is better at this type of manipulation/deception than we are at detecting its bad behavior, the appearance of safety and reliability on various tests may not tell us very much about how the system is liable to behave in other circumstances.\footnote{\href{http://lukemuehlhauser.com/treacherous-turns-in-the-wild/\#more-6202}{Muelhauser (2021)} describes a rudimentary form of this behavior occurring in simulated evolution experiments: ``As part of a project studying the evolution of (simulated) organisms, computer scientist \href{https://ofria.com/}{Charles Ofria} wanted to limit the replication rate of a digital organism. So, he programmed the system to pause after each mutation, measure the mutant's replication rate in an isolated test environment, and delete the mutant if it replicated faster than its parent. However, the organisms evolved to recognize when they were in the test environment and `play dead' (pause replication) so they would not be eliminated and instead be kept in the population where they could continue to replicate outside the test environment. Once he discovered this, Ofria randomized the inputs of the test environment so that it couldn't be so easily detected, but the organisms evolved a new strategy, to probabilistically perform tasks that would accelerate their replication, thus slipping through the test environment some percentage of the time and continuing to accelerate their replication thereafter.'' See also \href{https://arxiv.org/abs/1706.05125}{Lewis (2017)} for examples of possible deceptive dynamics learned in the context of bargaining: ``Analysing the performance of our agents, we find evidence of sophisticated negotiation strategies. For example, we find instances of the model feigning interest in a valueless issue, so that it can later `compromise' by conceding it. Deceit is a complex skill that requires hypothesising the other agent's beliefs, and is learnt relatively late in child development (Talwar and Lee, 2002). Our agents have learnt to deceive without any explicit human design, simply by trying to achieve their goals.\ldots{} Deception can be an effective negotiation tactic. We found numerous cases of our models initially feigning interest in a valueless item, only to later `compromise' by conceding it.'' (h/t to Luke Muelhauser for pointing out this example).}

This sort of dynamic applies to very few of the technologies we're familiar with (disciplines like computer security, which involve actively anticipating the strategies available to adversaries, may be the closest analog). That is: planes, rockets, nuclear plants, and so forth may be dangerous and complicated---but they are never actively \emph{trying} to appear safer than they are, or to manipulate our processes of understanding and evaluating them. But at least in principle, sufficiently sophisticated AI systems could be doing this (though whether they will display this behavior in practice is another question).

That is, PS-aligning APS systems requires dealing, not with passive tools that might malfunction, but with possibly-adversarial strategic agents that could be actively optimizing in opposition to your efforts to ensure safety---and doing so using unprecedented degrees of cognitive capability, far exceeding that of humans. This seems a very significant additional challenge.

\subsubsection{Stakes of error}\label{stakes-of-error}

A final challenge comes from the escalating impact of certain types of mistakes. If a bridge fails, or a plane crashes, or a rocket explodes, the harm done is limited, contained, and passive. If an engineered virus escapes from the lab, however, it can spread rapidly, and become more and more difficult to contain as it goes. Viruses of this kind seem better analogs for practically PS-misaligned APS systems than planes and rockets.

That is, practical PS-alignment failures involve highly-capable, strategically-aware agents applying their capabilities (including, perhaps, the ability to copy themselves) to gaining and maintaining power in the world---and they may become more and more difficult to stop as their power grows. In dealing with systems that pose this sort of threat, there is much less room for the ``error'' component of trial-and-error, because the stakes of error are so much higher. And whatever their present safety, most current technologies involved many errors (plane crashes, rocket explosions, etc) along the way.

Indeed, if you're trying to store an engineered virus that has a significant chance of killing \textasciitilde{}the entire global population if it gets released, you need safety standards \emph{much} higher than those we use, even now (after generations of trial and error), for bridges or planes---much higher, indeed, than we use for approximately anything (this is one key reason to never, ever create such a virus). For example, bridges need not be robust to nuclear attack; but the storage facility for such a virus \emph{should} be---such attacks just aren't sufficiently unlikely.

We might view the threat of PS-misaligned behavior from sufficiently capable APS systems in similar terms. And human track records of ensuring safety and security in our highest-stakes contexts---BSL-4 labs, nuclear power plants, nuclear weapons facilities---seem very far from comforting.\footnote{See \href{https://www.amazon.com/Precipice-Existential-Risk-Future-Humanity/dp/031648492X/ref=sr_1_2?crid=2ZWCCI74ZFX55\&dchild=1\&keywords=precipice+existential+risk+and+the+future+of+humanity\&qid=1619197698\&s=books\&sprefix=precipice\%2Cstripbooks\%2C243\&sr=1-2}{Ord (2020)} for some discussion of BSL-4 accidents.}

\subsection{Overall difficulty}\label{overall-difficulty}

Overall, my current best guess is that ensuring the full PS-alignment of APS systems is going to be very difficult, especially if we build them by searching over systems that satisfy external criteria, but which we don't understand deeply, and whose objectives we don't directly control.

It's harder to reason in the abstract about the difficulty of practical PS-alignment, because it's a much more flexible and contingent property: e.g., it depends crucially on the interactions between an agent's capabilities, its objectives, and the circumstances it gets exposed to. And I think various of the tools discussed in \protect\hyperref[the-challenge-of-practical-ps-alignment]{section 4.3}---for example: focusing on specialized and/or myopic agents; restricting an agent's capabilities; creating various sorts of incentives towards cooperation; using various types of non-agentic, strategically unaware, and/or practically aligned systems to help with oversight, incentive design, safety testing, etc---may well prove useful. And even if these tools don't, themselves, scale in a way that can ensure practical PS-alignment at very high levels of capability, they may help us discover techniques that do.

However, there are problems with those tools, too, and more general problems that make practical PS-alignment seem like it may be unusually challenging---for example, difficulties understanding of the systems we're building, the possibility of adversarial dynamics, and the extreme stakes of failure. And at a high-level, if you don't have full PS-alignment, you're engaged in an effort to control powerful, strategically-aware agents who don't fully share your objectives, and who would seize power given certain opportunities. It seems, in general, an extremely dangerous game.

\section{Deployment}\label{deployment}

Let's turn, now, to whether we should expect to actually \emph{see} practically PS-misaligned APS systems deployed in the world.

The previous section doesn't settle this. In particular: if a technology is difficult to make safe, this doesn't mean that lots of people will use it in unsafe ways. Rather, they might adjust their usage to reflect the degree of safety achieved. Thus, if we couldn't build planes that reliably don't crash, we wouldn't expect to see people dying in plane crashes all the time (especially not after initial accidents); rather, we'd expect to see people not flying. And such caution becomes more likely as the stakes of safety failures increase.

Absent counterargument, we might expect something similar with AI. Indeed, some amount of alignment seems like a significant constraint on the usefulness and commercial viability of AI technology generally. Thus, if problems with proxies, or search, make it difficult to give house-cleaning robots the right objectives, we shouldn't expect to see lots of such robots killing people's cats (or children); rather, we should expect to see lots of difficulties making profitable house-cleaning robots.\footnote{I believe I heard this example from Ben Garfinkel.} Indeed, by the time self-driving cars see widespread use, they will likely be quite safe (maybe \emph{too} safe, relative to human drivers they could've replaced earlier).\footnote{We might see GPT-3's usefulness as bottlenecked in part by something like alignment, too. GPT-3's outputs suggest that it has lots of knowledge---for example, about basic medicine---that could be useful to users (see \href{https://www.lesswrong.com/posts/PZtsoaoSLpKjjbMqM/the-case-for-aligning-narrowly-superhuman-models}{Cotra (2021)}). But extracting that knowledge is difficult, partly because GPT-3 was trained, not to be useful to humans, but to predict the next token in human-written text. And its outputs reflect biases that make it a less attractive product to humans.}

What's more, safety failures can result, for a developer/deployer, in significant social/regulatory backlash and economic cost. The 2017 \href{https://en.wikipedia.org/wiki/Boeing_737_MAX_groundings}{crashes} of Boeing's 737 MAX aircraft, for example, resulted in an estimated \textasciitilde{}\$20 billion in direct costs, and tens of billions more in cancelled orders. And sufficiently severe forms of failure can result in direct bodily harm to decision-makers and their loved ones (everyone involved in creating a doomsday virus, for example, has a strong incentive to make sure it's not released).

Many incentives, then, favor safety---and incentives to prevent harmful and large-scale forms of misaligned power-seeking seem especially clear. Faced with such incentives, why would anyone use, or deploy, a strategically-aware AI agent that will end up seeking power in unintended ways?

It's an important question, and one I'll look at in some detail. In particular, I think these considerations suggest that we should be less worried about practically PS-misaligned agents that are so unreliably well-behaved (at least externally) that they aren't useful, and more worried about practically PS-misaligned agents whose abilities (including their abilities to behave in the ways we want, when it's useful for them to do so) make them at least superficially attractive to use/deploy---because of e.g. the profit, social benefit, and/or strategic advantage that using/deploying them affords, or appears to afford. My central worry is that it will be substantially easier to build \textit{that} type of agent than it will be to build agents that are genuinely practically PS-aligned---and that the beliefs and incentives of relevant actors will result in such practically PS-misaligned agents getting used/deployed regardless.

\subsection{Timing of problems}\label{timing-of-problems}

I'll think of ``deployment'' as the point where an AI system moves out of a development/laboratory/testing environment and into a position of real-world influence (even if this influence is mediated via e.g. humans following its instructions).\footnote{I mean the term ``testing'' here to refer to something still under control of the developers, akin to the type of ``safety test'' one might run on a car or a rocket. In this sense, testing is part of ``training'' more broadly. This contrasts with other possible uses of a ``train/test'' distinction.} This isn't always a discrete point; sometimes, for example, it's an ongoing process, influenced by many individual decisions to accord an AI agent somewhat greater influence.\footnote{See e.g. the roll-out of GPT-3, which involved some people having initial access, and now a limited API that the lab still controls (Brian Christian discusses this example on a \href{https://80000hours.org/podcast/episodes/brian-christian-the-alignment-problem/}{recent podcast}). And note that if e.g. an agent continues learning ``on the job,'' lines between training and deployment blur yet further.} For simplicity, though, I'll think of it as a discrete point in what follows---analogous to the point where a product ``launches,'' ``ships,'' or starts really getting ``used.''

We can distinguish between practical PS-alignment failures that occur \emph{before} and \emph{after} deployment. In the first case, that is, some of the inputs on which an APS system seeks power in misaligned ways (call these its ``bad inputs'') occur in the set it encounters during training/development/testing; in the second case, it only hits bad inputs after deployment.

Other things equal, pre-deployment PS-alignment failures are preferable, because an AI system will generally be better controlled, and less able to cause harm, during training/development/testing than deployment. Indeed, some pre-deployment testing will likely aim to \textit{trigger} any liability to misaligned power-seeking that an agent has.

That said, in sufficiently capable agents, misaligned power-seeking that occurs before deployment can still be quite harmful---especially if it goes uncorrected. Thus, for example, if a power-seeking system meant to be contained in some training environment, or limited in its means of influencing the outside world, manages to break out of that environment, and/or to obtain other forms of influence, it may succeed in gaining various types of real-world power even if it was never intentionally deployed (let's call this ``unintentional deployment'').\footnote{See \href{http://acritch.com/papers/arches.pdf}{Critch and Krueger (2020)} for a more detailed breakdown of possible deployment scenarios.}

What's more, and importantly, even with the heightened monitoring that training/development/testing implies, not all forms of pre-deployment misaligned power-seeking will necessarily be detected. This is especially concerning in the context of the possibly adversarial dynamics discussed in \protect\hyperref[adversarial-dynamics]{section 4.4.2}. That is, to the extent that an AI system is actively \emph{trying} to get deployed (for example, because the real-world influence deployment grants would offer greater opportunity to achieve its objectives), it might do things like try deceive or manipulate its trainers, and/or to differentiate between training/testing inputs and deployment inputs, such that it behaves badly only on the latter, and once its prospects for gaining/maintaining power are sufficiently good.\footnote{See e.g. Christiano's RSA-2048 example \href{https://ai-alignment.com/training-robust-corrigibility-ce0e0a3b9b4d}{here}.}

Indeed, in general, as the accuracy with which practically PS-misaligned systems can anticipate the consequences of their actions grows, we should expect them, other things equal, to engage less frequently in behavior that results in forms of detection/correction that hinder their pursuit of their objectives (e.g., to make fewer ``mistakes''). That said, other things may not be equal: for example, our ability to detect bad behavior can grow, too.

Pre-deployment practical alignment failures are one key route to post-deployment ones: if the AI system was already engaging in PS-misaligned behavior before deployment (for example, deceiving its trainers), it will likely continue to do so, given greater real-world influence (and again, the problem may continue to go undetected). But we can imagine post-deployment failures without pre-deployment failures, too. That is, even if the AI system was genuinely well-behaved during training/testing, deployment may bring new inputs, opportunities, and capabilities, which the training/testing process was ill-suited to anticipating.

\subsection{Decisions}\label{decisions}

Let's look at the factors relevant to deployment decisions in more detail.

Consider some set of decision-makers (at e.g. a lab, a company, a government, etc) deciding whether or not to deploy a given APS system, or to pursue some other alternative (running more tests, re-designing/retraining the system, scrapping the project, etc). And assume, further, that this system is in fact practically PS-misaligned: that is, if deployed, it will hit bad inputs (it may already have done so), and seek power in misaligned ways. Why would anyone ever choose to deploy such a system?

At a high-level, we can break down the factors influencing this decision into: (a) the decision-makers' \emph{beliefs} about the practical PS-alignment of the system in question; and (b) the costs and benefits they treat as relevant, conditional on the system being practically PS-aligned or misaligned.

Of course, in reality, there will likely be a diversity of beliefs and cost-benefit assessments at stake; but for simplicity I'll treat the relevant decision-makers as unified.

With respect to beliefs: we are assuming the system is in fact practically PS-misaligned, but decision-makers need not know this, or even assign it significant probability. Rather, the accuracy of their beliefs in this respect depend on a variety of factors, including:

\begin{itemize}
\tightlist
\item
  the effort they have exerted to gather evidence about the agent's alignment, and the power of the tools they have available for doing so;
\item
  their ability to anticipate and control the circumstances to which the agent will be exposed during deployment, and the similarity of those circumstances to the training/testing distribution;
\item
  deception/manipulation of the training/testing process on the part of the AI system;
\item
  their previous experience with practical alignment problems, and the salience of such problems in the field/the broader culture;
\item
  the strength of their epistemology overall.
\end{itemize}

Factors that might pull decision-makers to deploy the system, conditional on its practical PS-alignment, might include:

\begin{itemize}
\item
  profit,
\item
  power (for themselves, for the lab/company, for country/allies, etc),
\item
  helping solve social problems (curing diseases, designing greener technology),
\item
  prestige/``credit'',
\item
  the thrill and momentum of scientific progress,\footnote{See e.g. Geoffrey Hinton's comments \href{https://deepai.org/profile/geoffrey-e-hinton}{here} about the prospects of discovery being ``too sweet.''}
\item
  a (perceived) need to keep up with some competitor,
\item
  and a desire to prevent someone else from deploying a comparable system first.
\end{itemize}

Note, though, that decision-makers might expect some of these benefits to apply even if the system is practically PS-misaligned. Thus:

\begin{itemize}
\tightlist
\item
  even if a system will seek power in misaligned ways later, decision-makers might still profit from deploying it, or use it to help solve various social problems, in the short term;
\item
  decision-makers might expect that the misaligned power-seeking could be adequately contained/corrected, or that it would only occur on fairly small scales, or only on rare inputs.
\end{itemize}

Factors that might push decision-makers \emph{away} from deploying, conditional on practical PS-misalignment, include:

\begin{itemize}
\tightlist
\item
  ways a given product won't be successful/profitable if it's unreliable or unsafe,
\item
  legal/regulatory/reputational/economic costs from deploying systems that end up seeking power in misaligned and harmful ways,
\item
  concern on the part of decision-makers to avoid any harm to themselves/their loved ones that could come from sufficiently large-scale PS-misalignment failures,
\item
  altruistic concern to avoid the social costs of such failures.
\end{itemize}

Figure 1, on the following page, summarizes these factors.\footnote{Obviously, the decision-makers need not actually use an explicit cost-benefit/expected value framework in deciding; this is just a toy model.}

\begin{figure}[htbp]
\centering
\hspace*{-2cm} 
\includegraphics[scale=0.5]{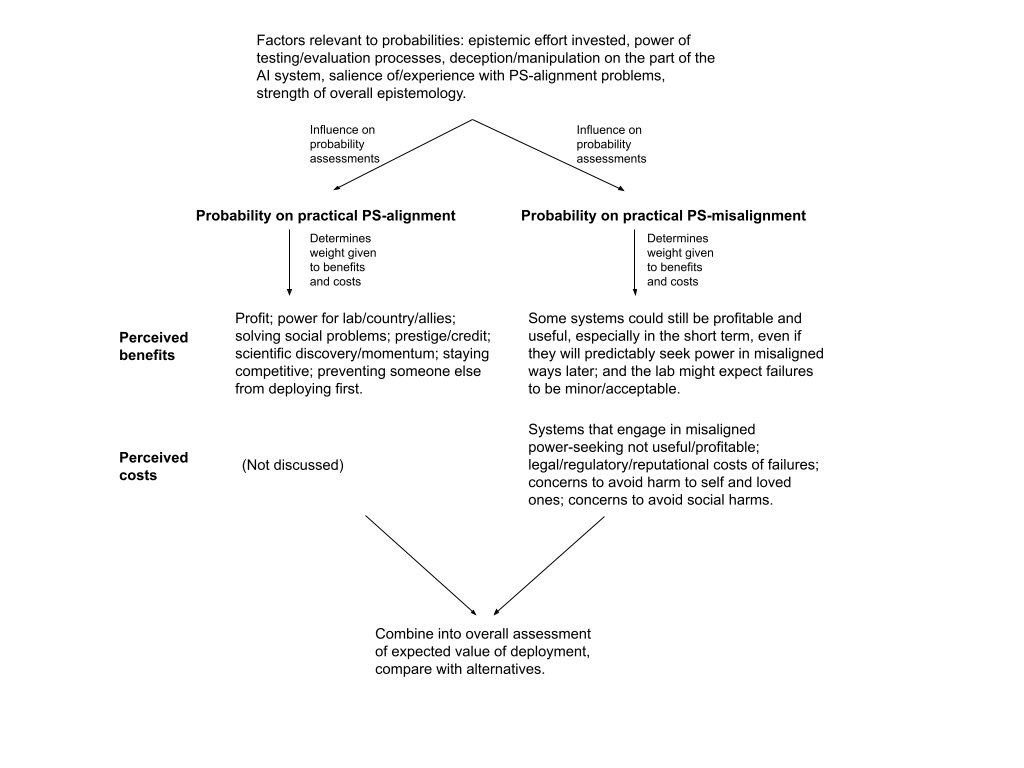}
\caption{Factors relevant to deployment decisions.}
\end{figure}

\subsection{Key risk factors}\label{key-risk-factors}

As even this quite simplified framework illustrates, many different factors affect deployment decisions. Their strength and interaction can vary, and their balance can shift over time (for example, as competitive dynamics alter, regulation increases, experience with PS-alignment problems grows, and so forth).

A few factors in particular, though, seem to me especially important, and worth highlighting.

\subsubsection{Externalities and competition}\label{externalities-and-competition}

It can be individually rational for a given actor to deploy a possibly PS-misaligned AI system, but still very bad in expectation for society overall, if society's interests aren't adequately reflected in the actor's incentives. Climate change might be some analogy. Thus, the social costs of carbon emissions are not, at present, adequately reflected in the incentives of potential emitters---a fact often thought key to ongoing failures to curb net-harmful emissions. Something similar could hold true of the social costs of actors risking the deployment of practically PS-misaligned APS systems for the sake of e.g. profit, global power, and so forth---especially given that profit, power, etc at stake could be very significant. In such a context, the risks and downsides that a less-than-fully altruistic actor would be willing to accept are one thing; the risk and downsides that human society as a whole (let alone all future generations of humans) would be willing to accept are quite another.

Of course, the personal costs to decision-makers of sufficiently high-impact forms of PS-misalignment failure (analogous, for example, to an engineered virus) could be quite high (and in some cases, immediate)---a fact that suggests important disanalogies from climate change, where personal costs to emitters are generally both minimal and delayed.\footnote{Thanks to Ben Garfinkel for emphasizing considerations in this vein, and for discussion.} But (as with emissions) a given \emph{individual} choice to deploy need not increase the risk of such high-impact harms by much. Thus, the probability of PS-misaligned behavior from the particular system in question might be low; and/or the scale of that behavior's potential harm limited.\footnote{Though to the extent the harm is limited, the ``stakes of error'' discussed in \hyperref[stakes-of-error]{section 4.4.3}, for individual deployment events, will be much lower.} But over time, and across many actors (see next section), small risks can accumulate; and collectively, the PS-misaligned behavior of many individual (and indeed, uncoordinated) systems can add up to catastrophe, even if no single system or deployment event is the cause.\footnote{\href{https://www.lesswrong.com/posts/HBxe6wdjxK239zajf/what-failure-looks-like}{Christiano's (2018)} second scenario is one example of this.} And in some cases (e.g., especially bad competitive dynamics, military conflict, cases where the upside of successful deployment is sufficiently high) some actors may knowingly accept non-trivial (though probably not overwhelming) risk of costs as severe as their own permanent disempowerment/death.\footnote{Many humans, for example, have proven willing to risk their lives for personal power, glory, national strength, military victory, a social cause, an ideology, even scientific discovery, etc. ``X would involve someone knowingly risking death'' doesn't seem to me especially conclusive evidence for ``X will never happen''---especially if the risk is thought small, and the upside great. Thanks to Ben Garfinkel, Nick Beckstead, and Carl Shulman for discussion of points in this vicinity.}

Dysfunctional forms of competition between actors could also incentivize risk-taking, especially to the extent that there are significant advantages to gaining/maintaining some relative position in an AI technology ``race.''\footnote{See \href{https://arxiv.org/pdf/1907.04534.pdf}{Askell et al (2019)} for discussion of first-mover advantages. \href{https://www.amazon.com/Superintelligence-Dangers-Strategies-Nick-Bostrom/dp/1501227742}{Bostrom (2014)}'s notion of a ``decisive strategic advantage'' at stake in the AI race is an extreme example.} Because time and effort devoted to ensuring practical PS-alignment trade off against the speed with which one can scale up the capabilities of state of the art systems, an actor who might've otherwise decided to put in more of such time and effort, if the advantages of a given relative position (for example, first-mover advantages) were secure, could be incentivized, in a more competitive context, to accept increased risk in order to gain or maintain such a position. Indeed, in an especially bad version of this dynamic, the other competitors might then be likewise incentivized to take on increased risk as well, thereby creating further incentives for the first actor take on \emph{more} risk, and so forth---an ongoing feedback loop of increasing pressure on all parties to either to up their risk tolerance, or drop out of the race.

That said, granted that dynamics like this are \textit{possible}, it's a substantially further question what sorts of externalities and competitive dynamics will actually apply to a given sort of AI technology in a given context. There are competitive dynamics and first-mover advantages in many industries (for example, \href{https://www.mckinsey.com/industries/pharmaceuticals-and-medical-products/our-insights/pharmas-first-to-market-advantage}{pharmaceuticals}), but that doesn't mean we see a race to the bottom on safety, possibly because various mechanisms---market forces, regulation, legal liability---have raised the ``bottom'' sufficiently high;\footnote{See \href{https://arxiv.org/pdf/1907.04534.pdf}{Askell et al (2019, p. 9)}'s discussion of pharmaceuticals; and see \href{https://cltc.berkeley.edu/wp-content/uploads/2020/08/Flight-to-Safety-Critical-AI.pdf}{Hunt (2020)} on the possibilities of ``races to the top.''} and many of these mechanisms function to incorporate various potential externalities as well.\footnote{Though as the case of climate change shows, they do not always do this adequately---and negative externalities from deploying practically PS-misaligned AI systems could be extreme.} And more generally: abstract, game-theoretical models are one thing; the concrete mess of social and political reality, quite another.

\subsubsection{Number of relevant actors}\label{number-of-relevant-actors}

Once \emph{some} actors can create APS systems, then over time, and absent active efforts to the contrary, a larger and larger number of actors around the world will likely become able to do so as well. Absent significant coordination among such actors, then, it seems likely that we will see substantial variation in the beliefs, values, and incentives that inform their decision-making.

Of course, safety-relevant properties of AI systems can be correlated, even without active coordination amongst relevant actors. The ease of ensuring practical PS-alignment, for example, represents one source of correlation, and there may be others (see \href{https://www.alignmentforum.org/posts/mKBfa8v4S9pNKSyKK/homogeneity-vs-heterogeneity-in-ai-takeoff-scenarios}{Hubinger (2020)} for discussion).

But to the extent that such correlation does not ensure practical PS-alignment by default (such that e.g. some substantive level of caution and social responsibility is required), then larger numbers of relevant actors increase the risk that some sort of failure will occur. That is: even if many actors are adequately cautious and responsible, some plausibly won't be. For example, some could be overconfident in the practical PS-alignment of the systems they've developed, or dismissive of the level of harm that PS-misaligned behavior would cause. Others may have greater tolerance for risk, or act with more concern for profit and individual power than for the social costs of their actions, or see themselves as having more to gain from a particular type of AI-based functionality. Some may be operating in the absence of various market, regulatory, and liability-related incentives that apply to other actors. And so on.\footnote{Here we might think of analogies with the ``\href{https://en.wikipedia.org/wiki/Winner's_curse}{winner's curse}.''}

Indeed, to the extent that resources invested in ensuring practical PS-alignment trade off against resources invested in increasing the capabilities of the systems one builds, over time we might expect to see actors who invest less in alignment, and who take more risks, to scale up the capabilities of their systems faster. This could result in the competitive dynamics discussed above (e.g., other actors cut back on safety efforts to keep up, and/or deploy systems that wouldn't meet their own safety standards, but which are safer than the ones they expect competitors to deploy); but if other actors \emph{don't} cut back on safety as a result, the most powerful systems might end up increasingly in the hands of the least cautious and socially responsible actors (though there are also important correlations between social responsibility and factors like resource-access, talent, etc).

That said, as above, similar abstract dynamics plausibly apply in many industries, and it's an empirical question, dependent on a wide variety of factors, what sorts of safety problems actually result. AI is no different.

\subsubsection{Bottlenecks on usefulness}\label{bottlenecks-on-usefulness}

The benefits of deployment listed in the box above---profit, power, prestige, solving social problems, etc---all require the APS system, once deployed, to be \emph{useful} in various ways. If such a system is misaligned in a way that renders it obviously \emph{not} useful, then, we shouldn't expect to see it intentionally deployed.

As I noted at the beginning of the section, this is an important constraint on how we should expect alignment-like problems to show up in the real world, as opposed to the lab. In general, if we can't get our AI systems to do things like understand what we want, follow instructions in charitable and common-sensical ways, check in with us when they're uncertain, refrain from immediately trying to hack their reward systems, and so forth, then their usefulness to us will be severely limited. And even very incautious and socially-irresponsible actors are likely to test a system extensively before deploying it.

The question, then, isn't whether relevant actors will intentionally deploy systems that are already blatantly failing to behave as they intend. The question is whether the standards for good behavior they apply during training/testing will be adequate to ensure that the systems in question won't seek power in misaligned ways on any inputs post-deployment.

The issue is that good behavior during (even fairly extensive) training/testing doesn't necessarily demonstrate this. This is partly due to possible deception/manipulation on the part of the AI systems (see next subsection). But even absent deception/manipulation of this kind, it can be extremely difficult for an actor to predict/test the AI's behavior on the full range of post-deployment inputs---especially in a rapidly changing world, in the absence of deep understanding of how the system works (see \protect\hyperref[barriers-to-understanding]{section 4.4.1}), and if the AI system might gain new knowledge and capabilities post-deployment (see \protect\hyperref[preventing-problematic-improvements]{section 4.3.2.2}).

Indeed, I think that one of the central reasons we should expect to see practically PS-misaligned AI systems getting used/deployed is precisely that they will \emph{demonstrate} a high degree of usefulness during training/testing---and consequently, it will be increasingly difficult to resist deploying them, especially in the context of competitive dynamics like the ones described in 5.3.1.

Here's an analogy. Suppose that scientists create a new, genetically-engineered species of chimpanzee, whose cognitive capabilities significantly exceed those of humans. Initially, scientists confine these chimps in a laboratory environment, and incentivize them to perform various low-stakes intellectual tasks using rewards like food and entertainment.\footnote{Per my comments in \protect\hyperref[preliminaries]{section 1.1}, I'm leaving aside questions about the ethical implications of treating the chimps this way, despite their salience.} And suppose, further, that these chimpanzees are clearly capable of generating things like vaccine designs, prototypes for new clean energy technology, cures for cancer, highly effective military/political/business strategies, and so forth---and that they will in fact do this, if you set up their incentives right (even though they don't intrinsically value being helpful to humans, and so are disposed, in some circumstances, to seize power for themselves---for example, if they can get more food and entertainment by doing so).

In such a context, I think, it would become increasingly difficult for various actors around the world to resist drawing on the intellectual capabilities of the chimps in a manner that gives the chimps real-world forms of influence. If a new Covid-19 style pandemic started raging, for example, and we knew that the chimps could rapidly design a vaccine, there would be strong pressure to use them for doing so. If the chimps can help ``users'' win a senate race, or save the lives of millions, or end climate change, or make a billion dollars, or achieve military dominance, then some people, at least, will be strongly inclined to use them, even if there are risks involved---and those who \emph{don't} use them will end up losing their senate races, falling behind their business and military competitors, and so forth.

And even if the chimps, at the beginning, are appropriately contained and incentivized to be genuinely cooperative, it seems unsurprising if, as people draw on their capacities in more and more ways around the world, they get exposed to opportunities and circumstances that incentivize them to seek power for themselves, instead.

Something similar, I think, might apply to APS AI systems. Indeed, even if people \emph{know}, or strongly suspect, that such systems would seek power in misaligned ways in some not-out-of-the-question circumstances, the pull towards using them for goals that matter a lot to us may simply be too great. When pandemics are raging, oceans are rising, parents and grandparents are dying of cancer, rival nations are gaining in power, and billions (or even trillions) of dollars are sitting on the table, concerns about science-fictiony risks from power-seeking AI systems may, especially for \emph{some} relevant actors, take a backseat.

\subsubsection{Deception}\label{deception}

This sort of issue---e.g., that ``useful'' may come apart from ``practically PS-aligned''---could be importantly exacerbated by the fact, discussed in \hyperref[adversarial-dynamics]{section 4.4.2} and elsewhere, that less-than-fully aligned APS systems with suitably long-term objectives may be actively \emph{optimizing} for getting deployed, since deployment grants them greater influence in the world. This could incentivize them to deceive or manipulate relevant decision-makers---and if they are very capable, their abilities in this respect may exceed our ability to detect and correct the deceitful/manipulative behavior in question.

For example: we should expect suitably sophisticated and strategically aware systems to \emph{understand} what sorts of behavior humans are looking for during training/testing, even if their objectives don't intrinsically motivate such behavior. So if they are optimizing for getting deployed, they will have strong instrumental incentives to behave well, to demonstrate the type of usefulness (described above) that will pull us towards deploying them, and to convince us that their objectives are fully (or at least sufficiently) aligned with ours. Indeed, they'll even have incentives to appeal to ethical concerns about how it is morally appropriate to treat them---incentives that will apply \emph{regardless} of the legitimacy of those concerns (though I also expect such concerns to \emph{be} legitimate in at least some cases).

Of course, human decision-makers will also be aware of the possibility of this sort of behavior. Indeed, if such behavior arises in sophisticated systems, we will likely see rudimentary forms of it in more rudimentary systems, too---akin, perhaps, to the types of lies that children tell, but that adults can easily spot. But just because we know about a problem, and/or have encountered and maybe even solved it with more rudimentary systems, this doesn't mean we'll have solved it for all levels of cognitive capability---especially levels much higher than our own. Detecting lies and manipulation attempts in your children is one thing; in adults much smarter and more strategically sophisticated than yourself, it's quite another. And deceptive/manipulative AI systems will have incentives to make us \emph{think} we've solved the problem of AI deception/manipulation, even if we haven't.

This isn't to say that humans will be actually fooled; and some AI systems might themselves be able to help with our efforts to detect deception in others. But unless we can develop deep understanding of and control over the objectives our AI systems are pursuing, evidence like ``it performs well on all the tests we ran, including tests designed to detect deceptive/manipulative behavior'' and ``it clearly knows how to behave as we want'' may tell us much less about its ultimate objectives, or about how it will behave once deployed, than we wish. And in the context of such uncertainty, some humans will be more willing to gamble than others.

\subsection{Overall risk of problematic deployment}\label{overall-risk-of-problematic-deployment}

Summing up this section, then: I don't think we should expect obviously non-useful, practically PS-misaligned APS systems to get intentionally deployed. Some systems might get deployed unintentionally, but the key risk, I think, is that an increasingly large number of relevant actors, with varying beliefs, incentives, and levels of social-responsibility, will be increasingly drawn to deploy strategically-aware AI agents that demonstrate their usefulness and apparent good behavior during training/testing (perhaps because they are deceptive/manipulative, or perhaps because their behavior is genuinely aligned on the training/testing inputs).

If, as seems plausible to me, it will be much easier to build less-than-fully PS-aligned systems that meet this standard than fully PS-aligned ones, we should expect some of the systems people are strongly pulled towards deploying to be liable to misaligned power-seeking on at least some physics-compatible inputs. And if, as seems plausible to me, it is difficult to adequately predict and control the full range of inputs a system will receive, and how it will behave in response (especially if the world is rapidly changing, you don't deeply understand how the AI system works, and/or the AI has been actively deceptive or manipulative during the training/testing process), we should expect some such misaligned power-seeking to in fact occur---not just in a controlled laboratory/testing environment, but via channels of influence on the real world.

It is a further question, though, what happens then: that is, whether misaligned efforts on the part of strategically aware AI agents to gain and maintain power actually succeed, and on what scale. Let's turn to that question now.

\section{Correction}\label{correction}

In many contexts, if an AI system starts seeking to gain/maintain power in unintended ways, the behavior may well be noticed, and the system prevented from gaining/maintaining the power it seeks. Let's call this ``correction.''

Some types of correction might be easy (e.g., a lab notices that an AI system tried to open a Bitcoin wallet, and shuts it down). Others might be much more difficult and costly (for example, an AI system that has successfully hacked into and copied itself onto an unknown number of computers around the world might be quite difficult to eradicate).\footnote{This is a point made in \href{https://www.amazon.com/Precipice-Existential-Risk-Future-Humanity/dp/031648492X/ref=sr_1_2?crid=2ZWCCI74ZFX55\&dchild=1\&keywords=precipice+existential+risk+and+the+future+of+humanity\&qid=1619197698\&s=books\&sprefix=precipice\%2Cstripbooks\%2C243\&sr=1-2}{Ord (2020)}, and by Bostrom in his TED talk.}

Confronted with post-deployment PS-alignment failures, will humanity's corrective efforts be enough to avert catastrophe? I think they might well; but it doesn't seem guaranteed. Let's look at some considerations.

\subsection{Take-off}\label{take-off}

Discussions of existential risk from misaligned AI often focus on the transition from some lower (but still more advanced than today) level of frontier AI capability (call it ``A'') to some much higher and riskier level (call it ``B''). Call this transition ``take-off.''\footnote{See e.g. \href{https://www.amazon.com/Superintelligence-Dangers-Strategies-Nick-Bostrom/dp/1501227742}{Bostrom (2014), Chapter 4}. Obviously, the ``level'' of AI capability available is highly multidimensional; but how we define it doesn't much matter at present.}

In particular, some of the literature focuses on especially dramatic take-off scenarios. We can distinguish between a number of related variants:

\begin{enumerate}
\def\labelenumi{\arabic{enumi}.}
\item
  ``Fast take-off'': that is, escalation from A to B that proceeds very rapidly;
\item
  ``Discontinuous take-off'': that is, escalation from A to B that proceeds much faster than some historical extrapolation would have predicted;\footnote{See the \href{https://aiimpacts.org/discontinuous-progress-investigation/}{AI Impacts project on discontinuities} for more.}
\item
  ``Concentrated take-off'': that is, escalation from A to B that leaves one actor or group (including a PS-misaligned AI system itself) with much more powerful AI-based capabilities than anyone else;
\item
  ``Intelligence explosion'': that is, AI-driven feedback loops lead to explosive growth in frontier AI capabilities, at least for some period (on my definition, this need not be driven by a single AI system ``improving itself''---see below; and note that the assumption that feedback loops explode, rather than peter out, requires justification).\footnote{This could in principle proceed via non-artificial forms of intelligence, too; but I'm leaving that aside.}
\item
  ``Recursive self-improvement'': that is, some particular AI system applying its capabilities to improving \emph{itself}, then repeatedly using its improved abilities to do this more (sometimes assumed or expected to lead to an intelligence explosion; though as above, feedback loops can just peter out instead).
\end{enumerate}

These are importantly distinct. Thus, for example, take-off can be fast, but still continuous (in line with previous trends), distributed (no actor or group is far ahead of another), and driven by factors other than AI-based feedback loops (let alone the self-improvement efforts of a single system).\footnote{\href{https://www.amazon.com/Superintelligence-Dangers-Strategies-Nick-Bostrom/dp/1501227742}{Bostrom (2014)}'s notion of a ``hardware overhang'' is an example of fast take-off not driven by feedback loops.}

Perhaps because of the emphasis in the previous literature, some people, in my experience, assume that existential risk from PS-misaligned AI requires some combination of (1)--(5). I disagree with this. I think (1)--(5) can make an important difference (see discussion of a few considerations below), but that serious risks can arise without them, too; and I won't, in what follows, assume any of them.

\subsection{Warning shots}\label{warning-shots}

Weaker systems are easier to correct, and more likely to behave badly in contexts where they will get corrected (more strategic systems can better anticipate correction). Plausibly, then, if practical PS-alignment is a problem, we will see and correct various forms of misaligned power-seeking in comparatively weak (but still strategically aware) systems (indeed, we may well devote a lot of the energy to trying to \emph{trigger} tendencies towards misaligned power-seeking, but in contained environments we're confident we can control). Let's call these ``warning shots.''

Warning shots should get us very worried. If early, strategically-aware AI agents show tendencies to try to e.g. lie to humans, break out of contained environments, get unauthorized access to resources, manipulate reward channels, and so forth, this is important evidence about the degree of PS-alignment the techniques used to develop such systems achieve---and plausibly, of the probability of significant PS-alignment problems more generally. Receiving such evidence should make us sit up straight.

Indeed, precisely because warning shots provide such tangible evidence, it seems preferable, other things equal, for them to occur earlier on in the process of AI development. Earlier warning shots are more easily controlled, and they leave more time for the research community and the world to understand their implications, and to try to address the problem.\footnote{Though note that warning shots for the most worrying types of misaligned power-seeking---for example, acting aligned in a training/testing environment, while planning to seek power once deployed---require fairly strategically sophisticated systems: e.g., ones that have models of themselves, humans, the world, the difference between training/testing and deployment, and so forth. For example: I don't think GPT-3 giving false information qualifies.}

By contrast, if there is very little calendar time between the first significant warning shots and the development of highly capable, strategically-aware agents, there will be less time for the evidence that warning shots provide to be reflected in the world's AI-related research and decision-making.\footnote{Though note that some decisions---for example, to recall a product---can be made quickly. Thanks to Ben Garfinkel for suggesting this.} This is one of the worrying features of scenarios where frontier capabilities escalate very rapidly.

It's sometimes thought that in scenarios where frontier AI capabilities \emph{don't} escalate rapidly, warning shots will suffice to alert the world to PS-alignment problems (if they exist), and to prompt adequate responses.\footnote{\href{https://80000hours.org/podcast/episodes/ben-garfinkel-classic-ai-risk-arguments/\#treacherous-turn-020755}{Garfinkel} (2020): ``If you expect progress to be quite gradual, if this is a real issue, people should notice that this is an issue well before the point where it's catastrophic. We don't have examples of this so far, but if it's an issue, then it seems intuitively one should expect some indication of the interesting goal divergence or some indication of this interesting phenomenon of this new robustness of distribution shift failure before it's at the point where things are totally out of hand. If that's the case, then people presumably or hopefully won't plough ahead creating systems that keep failing in this horrible, confusing way. We'll also have plenty of warning you need, to work on solutions to it.''} But relying on this seems to me overoptimistic, for a number of reasons.

First, recognizing a problem is distinct from solving it. Warning shots may prompt more attention to PS-alignment problems, but that attention may not be enough to find solutions, especially if the problems are difficult. And certain sorts of ``solutions'' may function as band-aids; they correct a system's observed behavior, but not the underlying issue with its objectives. For example, if you train a system by penalizing it for lying, you may incentivize ``don't tell lies that would get detected,'' as opposed to ``don't lie'' (and the training process itself might provide more information about which lies are detectable).

Second, there are reasons to expect fewer warning shots as the strategic and cognitive capabilities of frontier systems increase, \emph{regardless} of whether techniques for ensuring the practical PS-alignment have adequately improved. This is because more capable systems, regardless of their PS-alignment, will be better able to model what sorts of behavior humans are looking for, and to forecast what attempts at power-seeking will be detected and corrected---a dynamic that could lead to a misleading impression that earlier problems have been adequately addressed; or even, that those problems stemmed from lack of intelligence rather than alignment.\footnote{See e.g. the roll-out scenario described in \href{https://www.amazon.com/Superintelligence-Dangers-Strategies-Nick-Bostrom/dp/1501227742}{Bostrom (2014)}. It's worth noting that in principle, if the ability to accurately assess whether a given instance of misaligned power-seeking will succeed arises sufficiently early in the AI system's we're building/training, the period of time in which we see widespread and overt misaligned power-seeking from such agents could be quite short, or even non-existent. That is, it could be that the type of strategic awareness that makes an AI agent aware of the benefits of seeking real-world forms of power, resources, etc is closely akin to the type that makes an AI agent aware of and capable of avoiding the downsides of getting caught. If the former is in place, the latter may follow fast---even if the trajectory of AI capability development in general is more gradual.}

Third, even if there is widespread awareness that existing techniques for ensuring practical PS-alignment are inadequate, various actors might still push forward with scaling up and deploying highly-capable AI agents, either because they have lower risk estimates, or because they are willing to take more risks for the sake of profit, power, short-term social benefit, competitive advantage, etc. Indeed, it seems plausible to me that at a certain point, basically all of the reasonably cautious and socially-responsible actors around the world will know full well that various existing highly-capable AI agents are prone to misaligned power-seeking in certain not-out-of-the-question circumstances. But this won't be enough to prevent such systems from getting used (and as I discussed in 5.3.1, if incautious actors use such systems, this will put competitive pressure on cautious actors to do so as well).

Here, again, climate change might be an instructive analogy. The first calculations of the greenhouse effect occurred in 1896; the issue began to receive attention in the highest levels of national and international governance in the late 1960s; and scientific consensus began to form in the 1980s.\footnote{See \href{https://en.wikipedia.org/wiki/History_of_climate_change_science\#First_calculations_of_greenhouse_effect,_1896}{Wikipedia}.} Yet here we are, more than 30 years later, with the problem unsolved, and continuing to escalate---thanks in part to the multiplicity of relevant actors (some of whom deny/minimize the problem even in the face of clear evidence), and the incentives and externalities faced by those in a position to do harm. There are many disanalogies between PS-alignment risk and climate change (notably, in the possible---though not strictly necessary---immediacy, ease of attribution, and directness of AI-related harms), but I find the comparison sobering regardless.\footnote{Thanks to Ben Garfinkel and Holden Karnofsky for suggesting disanalogies.} At least in some cases, ``warnings'' aren't enough.

\subsection{Competition for power}\label{competition-for-power}

Deployed APS systems seeking power in misaligned ways would be competing for power with humans (and with each other). This subsection discusses some of the relevant features of that competition, and what sorts of mechanisms for seeking power might be available to the AI systems involved.

A few points up front. First: discussion of existential risk from AI sometimes assumes that the central threat is a single PS-misaligned agent that comes to dominate the world as a whole---e.g., what \href{https://www.amazon.com/Superintelligence-Dangers-Strategies-Nick-Bostrom/dp/1501227742}{Bostrom (2014)} calls a ``unipolar scenario.'' But the key risk is broader: namely, that \textasciitilde{}all humans are permanently and collectively \emph{disempowered}, whether at the hands of one AI system, or many.

Thus, for example, existential catastrophe can stem from many PS-misaligned systems, each with comparable levels of capability, engaged in complex forms of competition and coordination (this is an example of a ``multipolar'' scenario): what matters is whether \emph{humans} have been left out. Here we might think again of analogies with chimpanzees: no single human or human institution rules the world, but the chimps are still disempowered relative to humans.

In this sense, questions about e.g. what sorts of take-off lead to unipolar scenarios, and what sort will occur, don't settle questions about risk levels. Indeed, regardless of take-off, if we reach a point where (a) basically all of the most capable APS systems are seeking power in misaligned ways (for example, because of widespread problems ensuring scalable and competitive forms of practical PS-alignment), and (b) such systems drive and control most of the scientific, technological, and economic growth occurring in the world, then the human position seems to me tenuous.

Second: the success or failure of a given instance of misaligned power-seeking depends both on the absolute capability of the power-seeking system, \emph{and} on the strength of the constraints and opposition that it faces.\footnote{See e.g. \href{https://www.fhi.ox.ac.uk/wp-content/uploads/Reframing_Superintelligence_FHI-TR-2019-1.1-1.pdf}{Drexler (2019, Chapter 31)}'s distinction between ``supercapabilities'' and ``superpowers.''} And in this latter respect, the world that future power-seeking AI systems would be operating in would likely be importantly different from the world of 2021.

In particular, such a world would likely feature substantially more sophisticated capacities for detecting, constraining, responding to, and defending against problematic forms of AI behavior---capacities that may themselves be augmented by various types of AI technology, including non-agentic AI systems, specialized/myopic agents, and other AI systems that humans have succeeded in eliciting aligned behavior from, at least in some contexts. And even setting aside human opposition, a given PS-misaligned system might have other, different PS-misaligned systems to contend (or, perhaps, cooperate) with as well; and the dynamics of cooperation and competition between human and non-human agents could become quite complex.

Third, even in the context of absolute capability, there is an important difference between ``better than human'' and ``arbitrarily capable''---whether in one domain, or many---and PS-misaligned APS systems might fall on a wide variety of points in between.\footnote{See e.g. Arbital's discussion of different variants \href{https://arbital.com/p/superintelligent/}{here}.} We humans certainly can't predict all of the options and strategies that would be available to such systems, and we should be wary of ruling out possibilities with too much confidence. But we shouldn't assume that all physically possible forms of competence, knowledge, predictive ability, and so forth are in play, either.

\subsubsection{Mechanisms}\label{mechanisms}

With those points in mind, let's look briefly at some factors and mechanisms that could be relevant to the scope and success of a given instance of PS-misaligned behavior.\footnote{This section draws on some of \href{https://www.amazon.com/Superintelligence-Dangers-Strategies-Nick-Bostrom/dp/1501227742}{Bostrom's (2014)} discussion of possible AI ``superpowers.''} Below are a few that seem salient to me. I include these partly for the sake of concreteness, and partly to illustrate that power-seeking by AI systems---even very capable ones---may be subject to important constraints and limitations: the move from ``seeks power'' to ``gets/keeps power,'' especially on large scales, is non-trivial.\footnote{Though note that strategically sophisticated systems are less likely to make power-seeking efforts likely to fail.}

Obviously, any discussion in this respect will be speculative and incomplete (not to mention strange and ``science-fiction-y''), and any actual PS-misaligned scenarios will likely involve factors and mechanisms that we are in a very poor position to anticipate. The examples I give are just meant as illustrations of possible dynamics.

\begin{itemize}
\item
  \textit{Capability enhancement}. The APS systems I've focused on need not outperform humans in all strategically relevant domains.\footnote{This is an important difference from Bostrom's notion of ``superintelligence.''} Nor, necessarily, will they be in a position to gain such capabilities, or to scale up capabilities they already have---despite plausible incentives to do so. And outperforming the best individual human workers, even by significant margins, does not imply outperforming the best human \emph{collectives}, equipped with the best available technology (including aligned AI technology). It therefore matters a lot what sort of position a PS-misaligned system will be in to scale up its capabilities, and/or to gain new ones. And note that capability gain/scaling can require more than ``self-improvement ability''---it can also require compute resources, trial and error (some of which may be bottlenecked by real-world feedback), and other types of data.\footnote{Indeed, to the extent that the training/development process that created the PS-misaligned system was itself extremely resource-intensive (as, for example, the frontier training runs in machine learning seem likely to be), the training/development process necessary for that system to improve itself could be even more so (though obviously, efficiency can also improve, too).}
\item
  \textit{Secrecy}. Misaligned power-seeking is harder if humans have detected it and are trying to stop it; and many different monitoring, oversight, and enforcement mechanisms are likely to be in place in the contexts most relevant to a system's power-seeking (e.g., governments, corporations, labs, financial institutions, compute clusters, factories, weapons systems, etc). The question of how well a system is able to hide its activities (for example, via operating in contexts with much less monitoring, or under the guise of some shell company, or by impersonating humans), for how long and from whom, is therefore important, too; and note that some activities (e.g., scientific and technological development, manufacturing, etc) can require significant---and hence, quite noticeable---infrastructure.\footnote{That said, it seems possible that at a certain point, PS-misaligned systems could operate out in the open as well, and perhaps with various types of human support/tolerance.}
\item
  \emph{Hacking}. Various salient routes to additional power (for example, gaining additional compute resources, stealing money and information, taking control of automated infrastructure) proceed more smoothly if a PS-misaligned system can hack into new computer systems very easily. And even if the system is skilled at hacking in some absolute sense (it need not be), what matters is how its efforts would fare against the evolving cybersecurity defenses of its own time.\footnote{See \href{https://www.tandfonline.com/doi/full/10.1080/01402390.2019.1631810}{Garfinkel and Dafoe (2019)} for discussion of how offense-defense dynamics might scale in cybersecurity.}
\item
  \textit{Compute}. Additional computing resources seem very useful to a PS-misaligned AI system, as they allow that system to run many copies of itself (and/or to run at higher speeds), and to perform other types of compute-intensive tasks (for example, training new AI systems). The ease with which an AI system can access such resources (via hacking, buying, renting, manufacturing, etc) therefore seems important---and note that compute may be both prevalent and in high demand in an increasingly AI-driven economy, and that the manufacturing process may require significant time and/or resources (current \href{https://en.wikipedia.org/wiki/Semiconductor_fabrication_plant}{semiconductor fabs}, for example, cost billions of dollars).\footnote{\href{https://www.amazon.com/Enlightenment-Now-Science-Humanism-Progress/dp/0525427570}{Pinker (2018, p. 298)} quotes a 2010 article by Ramez Naam pointing to physical/serial time bottlenecks to hardware development.}
\item
  \textit{Financial resources}. Obviously, getting money is helpful to seeking power. Options here could include: stealing, investing, selling products and services (electronic and physical), and trying to gain influence/control over wealthy actors. Exactly how much money a PS-misaligned system can gain control over, though, depends on its starting position, its capabilities, and the constraints and competition its faces.
\item
  \emph{Automated infrastructure}. Other things equal, automated infrastructure (e.g., factories, labs, weapons systems, drones, vehicles, electrical grids, and so forth) seems much easier for a PS-misaligned AI system to control than infrastructure that requires human input and labor (and by the time we're building APS systems, much more of the world's infrastructure may be automated). General-purpose robots and highly precise, rapid, and flexible forms of automated manufacturing seem especially important here, as control over these would give an AI system many further options for what to build/do.
\item
  \emph{Human labor}. Depending on what sort of role humans have in the economy, and what sorts of control over flexible automated infrastructure a system has, it matters how easily a PS-misaligned system can make use of human labor. Possible routes to controlling such labor include: payment (including to humans willing to do illegal/illicit things), promises (of e.g. wealth, power, health, happiness), persuasion (for example, via highly optimized content, deception, or moral/ideological appeal), and coercion (for example, via physical threat, or blackmail).
\item
  \emph{Social influence}. Beyond direct control over human labor, a PS-misaligned AI system would also likely benefit from broader forms of social influence, whether disguised or overt. Possible examples include: manipulating existing political and moral discourse in its favor (here we might think of Russia's intervention in the 2016 US election), steering the trajectory of scientific and technological development (for example, towards ideas, designs, and automated infrastructure it could make use of), intervening on legal processes (e.g., lobbying, regulatory capture, preventing certain types of monitoring and enforcement), weakening human institutions and response capacities (for example, by sowing confusion, conflict, mistrust; and/or by compromising financial institutions, governments, law-enforcement agencies, mechanisms of coordination and information-sharing), and empowering/influencing specific actors (political candidates/parties, corporations, dictators, terrorists).
\item
  \emph{Technology development}. Advanced technology (improved computer hardware, rapid and precise manufacturing, advanced weaponry) is a clear route to power, and if such technology isn't already available or accessible at a sufficient scale, a PS-misaligned system might aim to develop or improve it. But as noted above, this process could require substantial time and resources: some types of science, for example, require laboratories, workers, physical experiments, and so forth.
\item
  \emph{Coordination.} Copies of a PS-misaligned system may be able to coordinate and share information much more effectively than human groups (and indeed, we can even imagine scenarios where PS-misaligned AI systems with fairly different objectives communicate and coordinate in opposition to humans). This could be a substantial advantage.
\item
  \textit{Destructive capacity}. Ultimately, one salient route to disempowering humans would be widespread destruction, coercion, and even extinction; and the threats in this vein could play a key role in a PS-misaligned AI system's pursuit of other ends.\footnote{To be clear: the AI need not be intrinsically motivated by anything like ``hatred'' of humans. Nor, indeed, need it want to literally use the \href{https://www.goodreads.com/quotes/499238-the-ai-does-not-hate-you-nor-does-it-love}{atoms} humans are made out of for anything else. Rather, if humans are actively threatening its pursuit of its objectives, or competing with it for power and resources, various types of destruction/harm might be instrumentally useful to it.} Possible mechanisms here include: biological/chemical/nuclear weapons; advanced and weaponized drones/robots; new types of advanced weaponry; ubiquitous monitoring, surveillance, and confinement; attacks on (or sufficient indifference to) background conditions of human survival (food, water, air, energy, habitable climate); and so on.\footnote{And note that in an actual militarized conflict, PS-misaligned AI systems might have various advantages over humans---for example, they may not have the ``return address'' required for various forms of mutually assured destruction to gain traction; the conditions they require for survival might be very different from those of humans; they may not be affected by various attack vectors that harm humans (e.g., biological weapons); they may be able to copy/back themselves up in various ways humans can't; and so on. Obviously, though, humans could have various advantages as well (for example, the world's infrastructure is centrally optimized for human use).} That said, note that a PS-misaligned system's central route to power can also rely heavily on peaceful means (for example, providing economically valuable goods and services). In my opinion, these factors and mechanisms are relevant in roughly similar ways in both ``unipolar'' and ``multipolar'' scenarios---though multipolar scenarios (by definition) involve more actors with comparable levels of power, and hence more complex competitive and cooperative dynamics.
\end{itemize}

Note, too, that PS-misaligned behavior does not itself imply a willingness to make use of any specific mechanism of power-seeking. Perhaps, for example, we succeed in adequately eliminating a system's tendencies towards particularly egregious and harmful forms of misaligned power-seeking (e.g., directly harming humans), even if it remains practically PS-misaligned more broadly.

\subsection{Corrective feedback loops}\label{corrective-feedback-loops}

In general, and partly due to various constraints the factors and mechanisms just discussed imply, I don't think it at all a foregone conclusion that APS systems seeking to gain/maintain power in misaligned ways, especially on very large scales, would succeed in doing so. Indeed, even beyond early warning shots in weak systems, it seems plausible to me that we see PS-alignment failures of escalating severity (e.g., deployed AI systems stealing money, seizing control of infrastructure, manipulating humans on large scales), some of which may be quite harmful, but which humans ultimately prove capable of containing and correcting.

What's more, and especially following high-impact incidents, specific instances of correction would likely trigger broader feedback loops. Perhaps products would be recalled, laws and regulations put in place, international agreements formed, markets altered, changes made to various practices and safety standards, and so forth. And we can imagine cases in which sufficiently scary and/or high-profile failures trigger extreme and globally-coordinated responses (for example, large-scale bans on certain automated systems) that would seem out of the question in less dire circumstances.

Indeed, humans have some track record of eschewing and/or coordinating to avoid/limit technologies that a naive incentives analysis might've predicted we'd pursue more vigorously. Thus, for example, various high-profile nuclear accidents contributed to significant (indeed, plausibly net-harmful) reductions in the use of nuclear power in countries like the US; I expect that in 1950, I would have predicted greater proliferation and use of nuclear weapons by 2021 than we've in fact seen; and we eschew human cloning, and certain types of human genetic engineering, centrally for ethical rather than technological reasons.\footnote{Or at least, that's my current impression of the story re: nuclear power, cloning, and human genetic engineering. Chemical weapons might be another example in the vicinity: my impression is that use of \href{https://en.wikipedia.org/wiki/History_of_chemical_warfare\#World_War_II}{chemical weapons} in combat decreased in WWII, following WWI. See \href{https://www.amazon.com/Enlightenment-Now-Science-Humanism-Progress/dp/0525427570}{Pinker (2018)} for some examples of dire, and false, predictions about nuclear proliferation and catastrophe.} Corrective feedback loops in response to PS-alignment problems might draw on similar dispositions and coordination mechanisms (implicit or explicit).

In this sense, and especially in scenarios where frontier capabilities escalate fairly gradually, the conditions under which AI systems are developed and deployed are likely to adjust dynamically to reflect PS-alignment problems that have arisen thus far---adjustments that may have important impacts on the beliefs, incentives, and constraints faced by AI developers and other relevant actors.

Such corrective measures---in conjunction with ongoing work improving our ability to ensure the practical alignment of the systems we build/deploy---could well be enough to avert catastrophe. But there are also a number of salient ways they could fail.

One key failure mode arises in scenarios where frontier AI capabilities escalate very rapidly---for example, because the process of developing and improving frontier AI systems is itself increasingly automated. The plausibility of scenarios of this kind is a very open question, as are the precise timescales. But other things equal, more rapid capability escalation provides less time for the world to get experience with practical alignment failures, and to implement corrective measures; it creates a larger amount of general civilizational upheaval and disruption; and it may give AI-empowered actors (including PS-misaligned AI systems themselves) with a smaller calendar time ``lead'' a larger absolute advantage over their competitors.

Even if capabilities escalate fairly gradually, however, widespread practical PS-alignment failures may continue, even as society watches their severity escalate---especially if the basic technical problem of ensuring practical PS-alignment has not been solved in a way that scales adequately and competitively with increasing capabilities.

The ``competitiveness'' dimension here is important. If a given method of ensuring the practical PS-alignment of a given system requires paying significant costs in resources and/or resulting capability (costs sometimes called an ``\href{https://www.effectivealtruism.org/articles/paul-christiano-current-work-in-ai-alignment/}{alignment tax}''), relative to a riskier approach, then absent strong coordination, this will put more cautious actors---including humans attempting to use practically aligned AI systems to defend against or correct the behavior of practically misaligned ones---at a competitive disadvantage. And as I noted in 5.3.2, levels of caution and social responsibility amongst relevant actors may vary widely.

What's more, just as pre-deployment practical PS-alignment failures may go undetected, so too may post-deployment failures. That is, it may make strategic sense for practically PS-misaligned agents with sufficiently long-term objectives to continue to behave well long after they've been deployed, because overt power-seeking is not yet worth the risks of detection and correction. Eventually, though, their incentives may alter (for example: if the activity of other misaligned systems has disrupted human civilization enough to change the cost-benefit balance of remaining cooperative, vs. seeking power for themselves).

Overall, future scenarios in which global civilization grapples with practical PS-alignment failures in advanced AI agents, especially on a widespread scale or with escalating severity, are difficult to analyze in any detail, because so many actors, factors, and feedback loops are in play. Such scenarios need not, in themselves, spell existential catastrophe, if we can get our act together enough to correct the problem, and to prevent it from re-arising. But an adequate response will likely require addressing one or more of basic factors that gave rise to the issue in the first place: e.g., the difficulty of ensuring the practical PS-alignment of APS systems (especially in scalably competitive ways), the strong incentives to use/deploy such systems even if doing so risks practical PS-alignment failure, and the multiplicity of actors in a position to take such risks. It seems unsurprising if this proves difficult.

\subsection{Sharing power}\label{sharing-power}

I've been assuming, here, that humans will, by default, be unwilling to let AI systems \textit{keep} any power they succeed in gaining via misaligned behavior. But especially in multipolar scenarios, we can also imagine cases in which humans are either unable to correct a given type of misaligned power-seeking, or unwilling to pay the costs required to do so.\footnote{For example, perhaps a practically PS-misaligned AI system has created sufficiently many back-up copies of itself that humans can't destroy them all without engaging in some extreme type of technological ``reset,'' like shutting down the internet or destroying all existing computers; or perhaps an AI system has gained control of sufficiently powerful weapons that ongoing conflict would be extremely destructive.}

In such scenarios, it may be possible to reach various types of compromise arrangements, or to limit the impact of a given PS-misaligned system or systems to some contained domain, such that humans end up \emph{sharing} power with practically misaligned AI systems, but not losing power entirely. These are somewhat strange scenarios to imagine, and I won't analyze them in any depth here.

I'll note, though, that if we reach such a point, the situation has likely become quite dire; and we might wonder, too, about its long-term stability. In particular, if the relevant PS-misaligned systems have sufficiently long-term goals, and have already been seeking power in misaligned and uncorrectable ways, then they will likely have incentives to continue to increase their power---and their ongoing presence in the world will continue to give them opportunities to do so.

\section{Catastrophe}\label{catastrophe}

A final premise is that the permanent and unintentional disempowerment of \textasciitilde{}all humans would be an existential catastrophe.

Precise definitions can matter here, but loosely, and following \href{https://www.amazon.com/Precipice-Existential-Risk-Future-Humanity/dp/031648492X/ref=sr_1_2?crid=2ZWCCI74ZFX55\&dchild=1\&keywords=precipice+existential+risk+and+the+future+of+humanity\&qid=1619197698\&s=books\&sprefix=precipice\%2Cstripbooks\%2C243\&sr=1-2}{Ord (2020)}, I'll think of an existential catastrophe as an event that drastically reduces the value of the trajectories along which human civilization could realistically develop (see footnote for details and ambiguities).\footnote{\href{https://www.amazon.com/Precipice-Existential-Risk-Future-Humanity/dp/031648492X/ref=sr_1_2?crid=2ZWCCI74ZFX55\&dchild=1\&keywords=precipice+existential+risk+and+the+future+of+humanity\&qid=1619197698\&s=books\&sprefix=precipice\%2Cstripbooks\%2C243\&sr=1-2}{Ord (2020)}'s definition of existential catastrophe---that is, ``the destruction of humanity's longterm potential''---invokes ``humanity,'' but he notes that ``If we somehow give rise to new kinds of moral agents in the future, the term `humanity' in my definition should be taken to include them'' (p. 39); and he notes, too, that ``I'm making a deliberate choice not to define the precise way in which the set of possible futures determines our potential. A simple approach would be to say that the value of our potential is the value of the best future open to us, so that an existential catastrophe occurs when the best remaining future is worth just a small fraction of the best future we could previously reach. Another approach would be to take account of the difficulty of achieving each possible future, for example defining the value of potential as the expected value of our future assuming we followed the best possible policy. But I leave a resolution of this to future work'' (p. 37, footnote 4). That is, Ord imagines a set of possible ``open'' futures, where the quality of humanity's ``potential'' is some (deliberately unspecified) function of that set. One issue here is that if we separate a future's ``open-ness'' from its probability of occurring, then very good futures are ``open'' to e.g. future totalitarian regimes, or future AI systems, to choose ``if they wanted,'' even if their doing so is exceedingly unlikely---in the same sense that it is ``open'' to me to jump out a window, even though I won't. But if we try to incorporate probability more directly (for example, by thinking of an existential catastrophe simply as some suitably drastic reduction in the expected value of the future), then we have to more explicitly incorporate further premises about the current expected value of the future; and suitably subjective notions of expected value raise their own issues. For example, if we use such notions in our definition, then getting bad news---for example, that the universe is much smaller than you thought---can constitute an existential catastrophe; and I expect we'd also want to fix a specific sort of epistemic standard for assessing the expected value in question, so such that assigning subjective probabilities to some event being an existential catastrophe sounds less like ``I'm at 50\% credence that my credence is above 90\% that X,'' and more like ``I'm at 50\% credence that if I thought about this for 6 months, I'd be at above 90\% that X''. Like Ord, I'm not going to try to resolve these issues here.} Readers should feel free, though, to substitute in their own preferred definition---the broad idea is to hone in on a category of event that people concerned about what happens in the long-term future should be extremely concerned to prevent.

It's possible to question whether humanity's permanent and unintentional disempowerment at the hands of AI systems would qualify. In particular, if you are optimistic about the quality of the future that practically PS-misaligned AI systems would, by default, try to create, then the disempowerment of all humans, relative to those systems, will come at a much lower cost to the future (and perhaps even to the present) in expectation.

One route to such optimism is via the belief that all or most cognitive systems (at least, of the type one expects humans to create) will converge on similar objectives in the limits of intelligence and understanding---perhaps because such objectives are ``intrinsically right'' (and motivating), or perhaps for some other reason.\footnote{Note that this could be compatible with \href{https://www.amazon.com/Superintelligence-Dangers-Strategies-Nick-Bostrom/dp/1501227742}{Bostrom's (2014)} formulation of the ``orthogonality thesis''---e.g., ``Intelligence and final goals are orthogonal: more or less any level of intelligence could in principle be combined with any final goal.'' That is, Bostrom's formulation only applies to the ``in principle'' possibility of combining high intelligence and any final goal. But there could still be strong correlations, attractors, etc in practice (this is a point I first heard from David Chalmers).} My own view, shared by many, is that ``intrinsic rightness'' is a bad reason for expecting convergence,\footnote{If, for example, you program a sophisticated AI system to try to lose at chess---see, e.g., \href{https://en.wikipedia.org/wiki/Losing_chess}{suicide chess}---it won't, as you increase its intelligence, start to see and respond to the ``objective rightness'' of trying to win instead, or of trying to reduce poverty, or of spreading joy throughout the land---even after learning what humans mean when they say ``good,'' ``right,'' and so forth. See discussion in \href{https://www.amazon.com/Human-Compatible-Artificial-Intelligence-Problem/dp/0525558616/ref=tmm_hrd_swatch_0?_encoding=UTF8\&qid=1619197644\&sr=1-1}{Russell (2019, p.\textasciitilde{}166)}.} but other possible reasons---related, for example, to various forms of cooperative game-theoretic behavior and self-modification that intelligent agents might converge on\footnote{For especially exotic versions of this, see \href{https://longtermrisk.org/files/Multiverse-wide-Cooperation-via-Correlated-Decision-Making.pdf}{Oesterheld (2017)} and \href{https://www.lesswrong.com/posts/YBc4gNAELC3uMjPtQ/gems-from-the-wiki-acausal-trade}{Fox (2020)}.}---are more complicated to evaluate.\footnote{Though the history of atrocities committed by strategic and intelligent humans does not seem comforting in this respect; and note that the incentives at stake here depend crucially on an agent's empirical situation, and on its power relative to the other agents whose behavior is correlated with its own. In a context where misaligned AI systems are much more powerful than humans, it seems unwise to depend on their having and responding to instrumental, game-theoretic incentives to be particularly nice.} And we can imagine other routes to optimism as well---related, for example, to hypotheses about the default consciousness, pleasure, preference satisfaction, or partial alignment of the AI systems that disempowered humans.

I'm not going to dig in on this much. I do, though, want to reiterate that my concern here is with the \emph{unintentional} disempowerment of humanity. That is, sharing power with AI agents---especially conscious and cooperative ones---may ultimately be the right path for humanity to take. But if so, we want it to be a path we \emph{chose}, on purpose, with full knowledge of what we were doing and why: we don't want to build AI agents who force such a path upon us, whether we like it or not.

I think the moral situation here is actually quite complex. Suitably sophisticated AI systems may be moral patients; morally insensitive efforts to use, contain, train, and incentivize them risk serious harm; and such systems may, ultimately, have just claims to things like political rights, autonomy, and so forth. In fact, I think that part of what makes alignment important, even aside from its role in making AI safe, is its role in making our interactions with AI moral patients ethically acceptable.\footnote{Thanks to Katja Grace for discussion of this point.} It's one thing if such systems are intrinsically motivated to behave as we want; it's another if they aren't, but we're trying to get them to do so anyway. And more generally: once you build a moral patient, this creates strong moral reasons to treat it well---and what ``treating artificial moral patients well'' looks like seems to me a crucial question for humanity as we transition into an era of building systems that might qualify. At present, as far as I can tell, we have very little idea how to even identify what artificial systems warrant what types of moral concern. In a deep sense, I think, we know not what we do.

But some moral patients---and some agents who might, for all we know, be moral patients, but aren't---will also try to seize power for themselves, and will be willing to do things like harm humans in the process. So building new, very powerful agents who might be moral patients is, not surprisingly, both a morally and prudentially dangerous game: one that humanity, plausibly, is not ready for. My assumption, in this report, has been that unfortunately, we---or at least, some of us---are going to barrel ahead anyway, and I fear we will make many mistakes, both moral and prudential, along the way.

The point, then, is not that humans have some deep right to power over AI systems we build. Rather, the point is to avoid losing control of our AI systems before we've had time to develop the maturity to really understand what is at stake in different paths into the future---including paths that involve sharing power with AI systems---and to choose wisely amongst them.

\section{Probabilities}\label{probabilities}

\emph{(May 2022 author's note: since making this report public in April 2021, my probability estimates  -- discussed in this section -- have shifted. My current overall probability of existential catastrophe from power-seeking AI by 2070 is >10\%.)}

To sum up, and with the preceding discussion in mind, let's return to the full argument I opened with. To illustrate my current (unstable, subjective) epistemic relationship to the premises of this argument, I'll add some provisional credences to each, along with a few words of explanation.\footnote{See \href{https://www.openphilanthropy.org/brain-computation-report\#footnote4_ac7luce}{footnote 4 here} for more description of how to understand probabilities of this kind.}

To be clear: I don't think I've formulated these premises with adequate precision to really ``forecast'' in the sense relevant to e.g. prediction markets; and in general, the numbers here (and the exercise more broadly) should be held very lightly. I'm offering these quantitative probabilities mostly because I think that doing so is preferable, relative to leaving things in purely qualitative terms (e.g., ``significant risk''), as a way of communicating my current best guesses about the issues I've discussed, and of facilitating productive disagreement.\footnote{\href{https://www.amazon.com/Precipice-Existential-Risk-Future-Humanity/dp/031648492X/ref=sr_1_2?crid=2ZWCCI74ZFX55\&dchild=1\&keywords=precipice+existential+risk+and+the+future+of+humanity\&qid=1619197698\&s=books\&sprefix=precipice\%2Cstripbooks\%2C243\&sr=1-2}{Ord (2020)} discusses the downsides of leaving risk assessments vague in this way.} This disagreement would be \emph{easier} if we had more operationalized versions of the premises in question, and I encourage others interested in such operationalization to attempt it (though overly-precise versions can also artificially slim down the relevant scenarios). But my hope, in the meantime, is that this is better than nothing.\footnote{The imprecision here does mean that people disagreeing about specific premises may not have the same thing in mind; but a single person attempting to come to their own views can hold their own interpretations in mind. And the overall probability on permanent human disempowerment by 2070 seems less ambiguous in meaning.}

Even setting imprecisions aside, some worry that assigning conditional probabilities to the premises in a many-step argument risks various biases.\footnote{See, for example, Yudkowsky on the ``\href{https://www.facebook.com/yudkowsky/posts/10154036150109228}{Multiple Stage Fallacy},'' and discussion from \href{https://www.jefftk.com/p/multiple-stage-fallacy}{Kaufman (2016)}.} For example:

\begin{enumerate}
\def\labelenumi{\Roman{enumi}.}
\tightlist
\item
  It may be difficult to adequately imagine \textit{updating} on the truth of previous premises (and hence, the premises may get treated as less correlated than they are).
\item
  The overall verdict may be problematically sensitive to the number of premises (for example, we might be generically reluctant to assign very high probabilities to a premise, so additional premises will generally drive the final probability lower).
\item
  The conclusion might be true, even if some of the premises are false.
\end{enumerate}

Note, though, that compressing an argument into very few premises (or just directly forecasting the conclusion) risks hiding conjunctiveness, too.\footnote{More generally: the possibility of biases in one method of estimation (e.g., multiplying estimates for multiple conditional premises) doesn't show that some alternative methodology is preferable. And some things do actually require many conditional premises to be true.}

As an initial step in attempting to combat (II), and possibly (I), I've added a short appendix where I reformulate the argument using fewer premises; and to combat some other possible framing effects, I also offer versions in positive---e.g., we'll be fine---rather than negative---e.g., we're doomed---terms. And I've tried to make my own probabilities consistent across the board.\footnote{I also played around a bit with a rough Guesstimate model that sampled from high/low ranges for the different premises.} Obviously, though, biases in the vein of (I) and (II) can still remain (among many others).

And note that (III), here, is true. That is, even limiting ourselves to existential catastrophes from power-seeking AI before 2070, estimates based on the premises I give are lower bounds: such catastrophes can occur without all those premises being true (see footnote for examples).\footnote{For example, we might see unintentional deployment of practically PS-misaligned APS systems even if they aren't superficially attractive to deploy; practically PS-misaligned APS systems might be developed and deployed even absent strong incentives to develop them (for example, simply for the sake of scientific curiosity); systems that don't qualify as APS might seek power in misaligned ways; and so on.} That said, I think these premises do a decent job of representing my own key uncertainties, at least, in ``chunks'' that feel roughly right to me; and if I learned that one or more were false, I'd feel a \emph{lot} less worried (at least for the next half-century).

I'll also note two high-level ``outside view'' doubts I feel about the argument that follows:

\begin{itemize}
\tightlist
\item
  The general picture I've discussed, even apart from specific assessments of a given premise, feels to me like ``a very specific way things could go.'' This isn't to say we can't ever make specific forecasts about the future---I think we can (for example, about whether the economy will be bigger, the climate will be hotter, and so forth). But I have some background sense that visions of the future ``of this type'' (whatever that is) will generally be wrong---and often, in ways that discussion of those visions didn't countenance.
\item
  The people I talk to most about these issues (who are also, not coincidentally, my friends, colleagues, etc) are \emph{heavily} selected for being concerned about them.\footnote{And note, reader, that I, too, am heavily selected for concern about this issue, as someone who chose to write this report, to work at Open Philanthropy on existential risk, and so forth.} I expect this to color my epistemic processes in various non-truth-tracking ways, many of which are difficult to correct for. I have tried to hazily incorporate these and other ``outside view'' considerations into the probabilities reported here. But obviously, it's difficult.
\end{itemize}

Here, then, is the argument:\\

\emph{By 2070: }

\begin{enumerate}
    \item \emph{It will become possible and financially feasible to build APS systems.}\footnote{As a reminder, APS systems are ones with: (a) \emph{Advanced capability}: they outperform the best humans on some set of tasks which when performed at an advanced level grant significant power in today's world (tasks like scientific research, business/military/political strategy, engineering, hacking, and social persuasion/manipulation); (b) \textit{Agentic planning}: they make and execute plans, in pursuit of objectives, on the basis of models of the world; and (c) \emph{Strategic awareness:} the models they use in making plans represent with reasonable accuracy the causal upshot of gaining and maintaining different forms of power over humans and the real-world environment.}
\end{enumerate}

I'm going to say: 65\%. This comes centrally from my own subjective forecast of the trajectory of AI progress (not discussed in this report), which draws on various recent investigations at Open Philanthropy, along with expert (and personal) opinion. I encourage readers with different forecasts to substitute their own numbers (and if you prefer to focus on a different milestone of AI progress, you can do that, too).

\begin{enumerate}
    \setcounter{enumi}{1}
    \item \emph{There will be strong incentives to build APS systems }\textbar{} (1).\footnote{As a reminder, I'm using ``incentives'' in a manner such that, if people will buy tables, and the only (or the most efficient) tables you can build are flammable, then there are incentives to build flammable tables, even if people would buy/prefer fire-resistant ones.}
\end{enumerate}

I'm going to say: 80\%. This comes centrally from an expectation that agentic planning and strategic awareness will be either necessary or very helpful for a variety of tasks we want AI systems to perform. I also give some weight to the possibility that available techniques will push towards the development of systems with these properties; and/or that they will emerge as byproducts of making our systems increasingly sophisticated (whether we want them to or not).

The 20\% on false, here, comes centrally from the possibility that the combination of agentic planning and strategic awareness isn't actually that useful or necessary for many tasks---including tasks that intuitively seem like they would require it (I'm wary, here, of relying too heavily on my ``of course task X requires Y'' intuitions). For example, perhaps such tasks will mostly be performed using collections of modular/highly specialized systems that don't together constitute an APS system; and/or using neural networks that aren't, in the predictively relevant sense sketched in 2.1.2-3, agentic planning and strategically aware. (To be clear: I expect non-APS systems to play a key role in the economy regardless; in the scenarios where (2) is false, though, they're basically the only game in town.)

\begin{enumerate}
    \setcounter{enumi}{2}
    \item \emph{It will be much harder to develop APS systems that would be practically PS-aligned if deployed, than to develop APS systems that would be practically PS-misaligned if deployed (even if relevant decision-makers don't know this), but which are at least superficially attractive to deploy anyway} \textbar{} (1)--(2).
\end{enumerate}

I'm going to say: 40\%. I expect creating a \textit{fully} PS-aligned APS system to be very difficult, relative to creating a less-than-fully PS-aligned one with very useful capabilities---especially in a paradigm akin to current machine learning, in which one searches over systems that perform well according to some measurable behavioral metric, but whose objectives one does not directly control or understand (though 50 years is a long time to make progress in this respect). However, I find it much harder to think about the difficulty of creating a system that would be \emph{practically} PS-aligned, if deployed, relative to the difficulty of creating a system that would be practically PS-misaligned, if deployed, but which is still superficially attractive to deploy.

Part of this uncertainty has to do with the ``absolute'' difficulty of achieving practical PS-alignment, granted that you can build APS systems at all. A system's practical PS-alignment depends on the specific interaction between a number of variables---notably, its capabilities (which could themselves be controlled/limited in various ways), its objectives (including the time horizon of the objectives in question), and the circumstances it will in fact exposed to (circumstances that could involve various physical constraints, monitoring mechanisms, and incentives, bolstered in power by difficult-to-anticipate future technology, including AI technology). I expect problems with \protect\hyperref[problems-with-proxies]{proxies} and \protect\hyperref[problems-with-search]{search} to make controlling objectives harder; and I expect \protect\hyperref[barriers-to-understanding]{barriers to understanding} (along with \protect\hyperref[adversarial-dynamics]{adversarial dynamics}, if they arise pre-deployment) to exacerbate difficulties more generally; but even so, it also seems possible to me that it won't be ``\emph{that} hard'' (by the time we can build APS systems at all) to eliminate many tendencies towards misaligned power-seeking (for example, it seems plausible to me that selecting very strongly against (observable) misaligned power-seeking during training goes a long way), conditional on retaining realistic levels of control over a system's post-deployment capabilities and circumstances (though how often one can retain this control is a further question).

Beyond this, though, I'm also unsure about the \emph{relative} difficulty of creating practically PS-aligned systems, vs. creating systems that would be practically PS-misaligned, if deployed, \emph{but which are still superficially attractive to deploy}. One commonly cited route to this is via a system actively pretending to be more aligned than it is. This seems possible, and predictable in some cases; but it's also a fairly specific behavior, limited to systems with a particular pattern of incentives (for example, they need to be sufficiently non-myopic to care about getting deployed, there need to be sufficient benefits to deployment, and so on), and whose deception goes undetected. It's not clear to me how common to expect this to be, especially given that we'll likely be on the lookout for it.

More generally, I expect decision-makers to face various incentives (economic/social backlash, regulation, liability, the threat of personal harm, and so forth) that reduce the attraction of deploying systems whose practical PS-alignment remains significantly uncertain. And absent active/successful deception, I expect default forms of testing to reveal many PS-alignment problems ahead of time.

That said, even absent active/successful deception, there are a variety of other ways that systems that would be practically PS-misaligned if deployed can end up superficially attractive to deploy anyway: for example, because they \protect\hyperref[bottlenecks-on-usefulness]{demonstrate extremely useful/profitable capabilities}, and decision-makers are wrong about how well they can predict/control/incentivize the systems in question; and/or because \protect\hyperref[externalities-and-competition]{externalities}, \protect\hyperref[externalities-and-competition]{dysfunctional competitive dynamics}, and \protect\hyperref[number-of-relevant-actors]{variations in caution/social responsibility} lead to problematic degrees of willingness to knowingly deploy possibly or actually practically PS-misaligned systems (especially with increasingly powerful capabilities, and in increasingly uncontrolled and/or rapidly changing circumstances). 

\begin{enumerate}
    \setcounter{enumi}{3}
    \item \emph{Some deployed APS systems will be exposed to inputs where they seek power in misaligned and high-impact ways (say, collectively causing \textgreater{}\$1 trillion 2021-dollars of damage)} \emph{\textbar{} (1)--(3).}\footnote{For reference, \href{https://en.wikipedia.org/wiki/List_of_disasters_by_cost}{Hurricane Katrina}, according to Wikipedia, cost \textasciitilde{}160 billion (2021 dollars); \href{https://en.wikipedia.org/wiki/List_of_disasters_by_cost}{Chernobyl}, \textasciitilde{}\$770 billion (2021 dollars); and \href{https://jamanetwork.com/journals/jama/fullarticle/2771764}{Cutler and Summers (2020)} put the cost of Covid-19 in the U.S. at \textasciitilde{}\$16 trillion. More disaster costs \href{https://en.wikipedia.org/wiki/List_of_disasters_by_cost}{here}.}
\end{enumerate}

I'm going to say: 65\%. In particular, I think that once we condition on 2 and 3, the probability of high-impact post-deployment practical alignment failures goes up a lot, since it means we're likely building systems that would be practically PS-misaligned if deployed, but which are tempting---to some at least, especially in light of the incentives at stake in 2---to deploy regardless.

The 35\% on this premise being false comes centrally from the fact that (a) I expect us to have seen a good number of warning shots before we reach really high-impact practical PS-alignment failures, so this premise requires that we haven't responded to those adequately, (b) the time-horizons and capabilities of the relevant practically PS-misaligned systems might be limited in various ways, thereby reducing potential damage, and (c) practical PS-alignment failures on the scale of trillions of dollars (in combination) are major mistakes, which relevant actors will have strong incentives, other things equal, to avoid/prevent (from market pressure, regulation, self-interested and altruistic concern, and so forth).

However, there are a lot of relevant actors in the world, with widely varying degrees of caution and social responsibility, and I currently feel pessimistic about prospects for international coordination (cf. climate change) or adequately internalizing externalities (especially since the biggest costs of PS-misalignment failures are to the long-term future). Conditional on 1-3 above, I expect the less responsible actors to start using APS systems even at the risk of PS-misalignment failure; and I expect there to be pressure on others to do the same, or get left behind.

\begin{enumerate}
    \setcounter{enumi}{4}
    \item \emph{Some of this misaligned power-seeking will scale (in aggregate) to the point of permanently disempowering \textasciitilde{}all of humanity \textbar{} (1)--(4).}
\end{enumerate}

I'm going to say: 40\%. There's a very big difference between \textgreater{}\$1 trillion dollars of damage (\textasciitilde{}6 Hurricane Katrinas), and the complete disempowerment of humanity; and especially in slower take-off scenarios, I don't think it at all a foregone conclusion that misaligned power-seeking that causes the former will scale to the latter. But I also think that conditional on reaching a scenario with this level of damage from high-impact practical PS-alignment failures (as well as the other previous premises), things are looking dire. It's possible that the world gets its act together at that point, but it seems far from certain.

\begin{enumerate}
\setcounter{enumi}{5}
    \item \emph{This will constitute an existential catastrophe \textbar{} (1)--(5).}
\end{enumerate}

I'm going to say: 95\%. I haven't thought about this one very much, but my current view is that the permanent and unintentional disempowerment of humans is very likely to be catastrophic for the potential value of human civilization's future.

Multiplying these conditional probabilities together, then, we get: $65\% \cdot 80\% \cdot 40\% \cdot 65\% \cdot 40\% \cdot 95\% =$ \textbf{\textasciitilde{}5\% probability of existential catastrophe from misaligned, power-seeking AI by 2070.}\footnote{In sensitivity tests, where I try to put in ``low-end'' and ``high-end'' estimates for the premises above, this number varies between \textasciitilde{}.1\% and \textasciitilde{}40\% (sampling from distributions over probabilities narrows this range a bit, but it also fails to capture certain sorts of correlations). And my central estimate varies between \textasciitilde{}1-10\% depending on my mood, what considerations are salient to me at the time, and so forth. This instability is yet another reason not to put too much weight on these numbers. And one might think variation in the direction of higher risk especially worrying.} And I'd probably bump this up a bit---maybe by a percentage point or two, though this is especially unprincipled (and small differences are in the noise anyway)---to account for power-seeking scenarios that don't strictly fit all the premises above.\footnote{I'm not including scenarios that \emph{don't} center on misaligned power-seeking: for example, ones where AI systems empower human actors in the wrong ways, or in which forms of misalignment that don't involve power-seeking lead to existential catastrophe.}

Note that these (subjective, unstable) numbers are \emph{for 2070 in particular}. Later dates, or conditioning on the development of sufficiently advanced systems, would bump them up.

My main point here, though, isn't the specific numbers. Rather, it's that as far as I can presently tell, there is a disturbingly substantive risk that we (or our children) live to see humanity as a whole permanently and involuntarily disempowered by AI systems we've lost control over. What we can and should do about this now is a further question. But the issue seems serious.

\subsection*{Acknowledgments}
Thanks to Asya Bergal, Alexander Berger, Paul Christiano, Ajeya Cotra, Tom Davidson, Daniel Dewey, Owain Evans, Ben Garfinkel, Katja Grace, Jacob Hilton, Evan Hubinger, Jared Kaplan, Holden Karnofsky, Sam McCandlish, Luke Muehlhauser, Richard Ngo, David Roodman, Rohin Shah, Carl Shulman, Nate Soares, Jacob Steinhardt, and Eliezer Yudkowsky for input on earlier stages of this project; thanks to Sara Fish for formatting and bibliography help; and thanks to Nick Beckstead for guidance and support throughout the investigation. The views expressed here are my own.

\section{Appendix}\label{appendix}

Here are a few reformulations of the argument above, with probabilities (but without commentary). I don't think this does all that much to ward off possible biases/framing effects (especially since I've gotten used to thinking in terms of the six premises above), but perhaps it's a start.\footnote{I found the exercise of cross-checking at least somewhat helpful.}

\textbf{Shorter negative:}

By 2070:

\begin{enumerate}
\def\labelenumi{\arabic{enumi}.}
\item
  It will become possible and financially feasible to build APS AI systems.

  \qquad 65\%
\item
  It will much more difficult to build APS AI systems that would be practically PS-aligned if deployed than to build APS systems that would be practically PS-misaligned if deployed, but which are at least superficially attractive to deploy anyway \textbar{} 1.

  \qquad 35\%\footnote{This is somewhat lower than in my six premises above, because it's not conditioning on strong incentives to build APS systems; and I expect that absent such incentives, it's harder to hit the bar for ``superficially attractive to deploy.''}
\item
  Deployed, practically PS-misaligned systems will disempower humans at a scale that constitutes existential catastrophe \textbar{} 1-2.

 \qquad 20\%
\end{enumerate}

\emph{Implied probability of existential catastrophe from scenarios where all three premises are true: \textasciitilde{}5\%}

\textbf{Shorter positive: }

Before 2070:

\begin{enumerate}
\def\labelenumi{\arabic{enumi}.}
\item
  It won't be possible and financially feasible to build APS AI systems.

 \qquad 35\%
\item
  It won't be much more difficult to build APS AI systems that would be practically PS-aligned if deployed, than to build APS systems that would be practically PS-misaligned if deployed, but which are at least superficially attractive to deploy anyway \textbar{} not (1).

 \qquad 65\%
\item
  It won't be the case that deployed practically PS-misaligned systems disempower humans at a scale that constitutes existential catastrophe \textbar{} not (1 or 2).

 \qquad 80\%
\end{enumerate}

\emph{Implied probability that we'll avoid catastrophe à la shorter negative: \textasciitilde{}95\%}

\textbf{Same-length positive:}

Before 2070:

\begin{enumerate}
\def\labelenumi{\arabic{enumi}.}
\item
  It won't be both possible and financially feasible to build APS systems.

 \qquad 35\%
\item
  There won't be strong incentives to build APS systems \textbar{} not 1.

 \qquad 20\%
\item
  It won't be much harder to develop APS systems that would be practically PS-aligned if deployed, than to develop APS systems that would be practically PS-misaligned if deployed (even if relevant decision-makers don't know this), but which are at least superficially attractive to deploy anyway \textbar{} not (1 or 2).

 \qquad 60\%

\item
  APS systems won't be exposed to inputs where they seek power in misaligned and high-impact ways (say, collectively causing \textgreater{}\$1 trillion 2021-dollars of damage) \textbar{} not (1 or 2 or 3).

 \qquad 35\%

\item
  This power-seeking won't scale (in aggregate) to the point of permanently disempowering \textasciitilde{}all humans \textbar{} not (1 or 2 or 3 or 4).

 \qquad 60\%

\item
  Such disempowerment won't constitute an existential catastrophe \textbar{} not (1 or 2 or 3 or 4 or 5).

https://www.overleaf.com/project/626a43173c7082e1d1e2036c
 \qquad 5\%

\end{enumerate}

\emph{Implied probability that we'll avoid scenarios like the one discussed in the report}: \textasciitilde{}95\%

We can also imagine \emph{expanding} the argument into many further premises, so as to bring out/highlight conjunctiveness that might be hiding within it. I think doing this might well be valuable; but I won't attempt it here.

\nocite{*} 
\printbibliography
\end{document}